\documentclass[11pt]{article}
\usepackage{jheppub}

\usepackage{amssymb}
\usepackage{lipsum}
\usepackage{bm}

\usepackage[T2A,T1]{fontenc}
\usepackage[utf8]{inputenc}
\usepackage[russian,english]{babel}
\usepackage{upgreek}
\usepackage[utf8]{inputenc}
\usepackage{subcaption}
\usepackage{graphicx}
\usepackage{amsmath}
\usepackage{amssymb}
\usepackage{physics}
\usepackage{feynmp-auto}
\usepackage{hyperref}
\DeclareMathOperator{\sign}{sign}
\usepackage{listings}
\usepackage{pythonhighlight}
\usepackage{tikz}
\usetikzlibrary{calc,patterns,angles,quotes}
\usepackage{scrextend}
\usepackage[]{mdframed}
\usepackage{tcolorbox}
\usepackage{minted}
\usepackage[svgnames]{xcolor}
\tcbuselibrary{listings}

\definecolor{mygreen}{rgb}{0,0.6,0}
\definecolor{mygray}{rgb}{0.5,0.5,0.5}
\definecolor{mymauve}{rgb}{0.58,0,0.82}

\newcommand{\mg}{{\sc MG5aMC}\ }

\title{Automated computation of spin-density matrices and quantum observables for collider physics}

\author[a]{Valentin Durupt,}
\author[a,b,c,d]{Fabio Maltoni,}
\author[a]{Olivier Mattelaer}

\affiliation[a]{
Centre for Cosmology, Particle Physics and Phenomenology (CP3),
Universit\'e Catholique de Louvain, 
Louvain-la-Neuve, 
B-1348, 
Belgium
}

\affiliation[b]{
Dipartimento di Fisica e Astronomia, Universit\`a di Bologna, 
Via Irnerio 46, 
Bologna,
 I-40126, 
 Italy
 } 

\affiliation[c]{
INFN, Sezione di Bologna,
Via Irnerio 46, 
 Bologna,
 I-40126, 
 Italy
 }

\affiliation[d]{European Organisation for Nuclear Research (CERN), 1211 Geneva, Switzerland}
\preprint{IRMP-CP3-25-34}
\abstract{
We present a fully automated framework to compute production spin-density matrices for generic collider processes at tree level within \textsc{MadGraph5\_aMC@NLO}. The method assembles helicity amplitudes into event-by-event production matrices. These are written to the LHE file in a compact form, together with run metadata, enabling direct post-processing of quantum observables. The implementation supports bi- and multipartite qubit and qutrit final states, configurable reference frames, and both polarised and unpolarised initial states. A companion, easy-to-extend library provides analysis routines to determine key quantum-information measures and witnesses. These include purity, concurrence, and entanglement of formation for qubits; Peres--Horodecki tests and negativity; spin-polarisation vectors and correlation matrices; $D$-coefficients; and stabiliser-based ``magic'' measures. As a result, multi-particle quantum correlations can be quantified systematically. We validate the implementation against known results for $t\bar t$ and $VV$ ($V=W^\pm,Z$) production in $pp$ and $e^+e^-$ collisions and in heavy-resonance decays. We then consider new applications and study quantum correlations in several LHC final states: $t\bar t W^\pm$, $tW^-$ vs.\ $t(\bar t\to W^- \bar b)$, and $t\bar t t$ vs. $t\bar t t\bar t$.}

\begin{document}
\small
\maketitle

\section{Introduction}

Quantum information theory (QI) has recently emerged as a promising ground for exploring the quantum nature of high-energy processes at colliders. Concepts such as entanglement, purity, and Bell non-locality,  traditionally studied in low-energy quantum systems, have found new relevance in particle physics, where the reconstruction of quantum states from scattering events can provide novel insights into both Standard Model (SM) dynamics and Beyond-the-Standard Model (BSM) scenarios. Recent measurements at the LHC~\cite{ATLAS:2014aus, CMS:2015cal, ATLAS:2016bac, CMS:2016piu, CMS:2019nrx, ATLAS:2019zrq, CMS:2024pts, CMS:2024zkc} and many phenomenological studies, from $t\bar t$ \cite{2003.02280,Fabbrichesi:2021npl,Severi:2021cnj,Severi:2022qjy,Aoude:2022imd,Afik:2022kwm,Aguilar-Saavedra:2022uye, Afik:2022dgh, Cheng:2023qmz, Han:2023fci, Dong:2023xiw, Cheng:2024btk, Aguilar-Saavedra:2024hwd,Aguilar-Saavedra:2023lwb,Aguilar-Saavedra:2024fig,Cheng:2024rxi, Aguilar-Saavedra:2024vpd, Han:2024ugl, Maltoni:2024tul, Maltoni:2024csn, Aoude:2025jzc, Fabbrichesi:2025psr} to $\tau^+\tau^-$~\cite{Altakach:2022ywa,Fabbrichesi:2022ovb,Ehataht:2023zzt,Fabbrichesi:2024xtq,Fabbrichesi:2024wcd,Zhang:2025mmm,Han:2025ewp}, to hadron-hadron  \cite{Tornqvist:1980af,Bertlmann:2001iw,Takubo:2021sdk,Gong:2021bcp,Fabbrichesi:2023idl,Afik:2024uif,Gabrielli:2024kbz,Fabbrichesi:2024rec,Cheng:2025cuv,Chen:2024drt,Fabbrichesi:2025rqa,Wu:2025dds,Afik:2025grr,Aoude:2025jzc}, to EW diboson ~\cite{Barr:2021zcp,Barr:2022wyq,Ashby-Pickering:2022umy,Aguilar-Saavedra:2022wam,Morales:2023gow,Aguilar-Saavedra:2023hss,Aoude:2023hxv,Fabbrichesi:2023cev,Fabbrichesi:2023jep,Subba:2024aut,Bernal:2024xhm,Sullivan:2024wzl,Bi:2023uop,Bernal:2023ruk,Fabbri:2023ncz,Grossi:2024jae,Aguilar-Saavedra:2024jkj,Wu:2024ovc,Ruzi:2024cbt,Ding:2025mzj,Fabbrichesi:2025ifv} final states, have established the feasibility and utility of quantum observables for collider processes. 
As a result, further refinements and explorations have been initiated that include the study of higher-order effects~\cite{Grossi:2024jae,DelGratta:2025qyp,Aoude:2025ovu,Aguilar-Saavedra:2025byk,Goncalves:2025qem,DelGratta:2025xjp,Gu:2025ijz}, the use of more elaborated entanglement measures \cite{Hollands:2017dov,Jozsa:1994qja,PhysRevLett.95.090503,Ollivier:2001fdq,Mintert_2005,PhysRevLett.98.140505,Mintert_2007,Vidal:2002zz,White:2024nuc,Fabbrichesi:2025aqp,Fabbrichesi:2025ywl,Fabbrichesi:2025rsg,Gargalionis:2025iqs,Liu:2025frx,Liu:2025qfl}, the study of multipartite entanglement~\cite{Morales:2024jhj,Subba:2024mnl,Aguilar-Saavedra:2024whi,Sakurai:2023nsc,Horodecki:2024bgc,Horodecki:2025tpn}, the connection with non-locality~\cite{Abel:1992kz,Tu:2019ouv,Eckstein:2021pgm,Altomonte:2023mug,Altomonte:2024upf,Hentschinski:2024gaa,Low:2025aqq,Fabbrichesi:2025aqp,Bechtle:2025ugc,Abel:2025skj} as well as the exciting exploration of entanglement as a dynamical principle~\cite{Cervera-Lierta:2017tdt,Beane:2018oxh,Low:2021ufv,Liu:2022grf,Carena:2023vjc,Liu:2023bnr,Hu:2024hex,Aoude:2024xpx,Low:2024mrk,Thaler:2024anb,Chernyshev:2024pqy,Hu:2025lua,Busoni:2025dns,Carena:2025wyh,Liu:2025bgw}. For a recent status and prospects overview, see Ref.~\cite{Afik:2025ejh}.

At the core of current collider studies lies the spin-density matrix, which encapsulates the quantum spin state of a system produced in high-energy collisions. It serves as the starting point for computing a wide range of observables: polarisation, spin correlations, entanglement measures (e.g.\ concurrence, entanglement of formation), and other markers of non-classical behaviour such as negativity or magic. The computation of spin-density matrices has been carried out so far on a process-by-process basis, often relying on custom calculations or laborious numerical reconstructions. This lack of a general framework has limited the widespread adoption of quantum information analyses in collider physics.

In this work, we present the first automated and general-purpose tool for computing spin-density matrices for arbitrary scattering processes at the tree level. Based on the \mg  ({\sc MG5aMC}) framework~\cite{Alwall:2014hca} and the {\sc MadSpin} package~\cite{Artoisenet:2012st}, our implementation leverages the helicity amplitude machinery to construct production density matrices directly from the matrix elements, and outputs them event by event in the Les Houches Event (LHE) file format~\cite{Alwall:2006yp}. The resulting matrices are fully normalised and stored in a compact form, ready for post-processing and the extraction of quantum observables.

Our implementation supports a broad range of final states, including qubit--qubit, qubit--qutrit, and qutrit--qutrit systems, as well as multipartite systems, and accommodates user-defined reference frames and helicity bases. It is capable of handling polarised or unpolarised initial states and different lab or helicity frames for density matrix evaluation. A dedicated interface within the \mg ecosystem enables a flexible configuration with the relevant information embedded into the LHE files in a transparent and user-friendly manner.

We validate our tool by reproducing known results for $t\bar t$ and $VV$ ($V=W^\pm,Z$) production at hadron and lepton colliders and in the decay of hypothetical heavy scalar and vector resonances. We then  demonstrate its versatility by applying it to novel processes, such as $t \bar t W^\pm$ and comparing $tW^-(b)$ vs. $t(\bar t \to W^-\bar b)$, and analysing $t\bar t t$ vs. $t\bar t t \bar t$ final states. 
We do this by implementing in a library a number of quantum observables that can be computed from the density matrices, including measures such as purity, concurrence, and bounds on the entanglement for higher-spin systems. These applications illustrate the wide scope of the implementation and its potential to support precision studies and BSM searches in current and future experiments.

The paper is organised as follows. In Sect.~\ref{sec:framework}, we describe the theoretical basis for constructing spin-density matrices from helicity amplitudes.  In Sect.~\ref{sec:validation} we present the validation studies performed by comparing our numerical results to known expressions in several cases relevant for SM and BSM collider  phenomenology. In Sect.~\ref{sec:applications} we provide novel and non-trivial applications to several final states involving the top quark at the LHC. We conclude in Sect.~\ref{sec:conclusions} with an outlook on future extensions of the code.  Three appendices provide more technical details: the first lists  the quantum observables available in our library (App.~\ref{app:QO}), while the second contains  how to run the code, including  explicit examples (App.~\ref{app:code}). The third provides information on the conventions used relate them to other conventions found in the literature (App.~\ref{app:conventions}). 

\section{Spin-density matrices}
\label{sec:framework}

The spin-density matrix is the most general descriptor of a quantum system, encoding both classical mixtures and genuinely quantum features, such as single-particle polarisations and inter-particle spin-spin coherences. In collider processes, the relevant object is the production spin-density matrix obtained in a given spin basis. Given their widespread use in perturbative calculations for high-energy processes, helicity amplitudes provide the ideal quantities to calculate the spin-density matrix: its diagonal elements determine population probabilities while its off-diagonal elements, the so-called coherences, capture phase information that is essential to quantify spin correlations, entanglement, and even potential CP-odd effects. Determining this matrix from experimentally accessible variables is the task of quantum tomography: by exploiting the factorisation of production and decay in the on-shell approximation, the angular and energy distributions of the decay products provide a complete set of projectors from which the elements of the density matrix can be inferred. This programme has so far been performed for $t\bar t$  production~\cite{2003.02280} and extended to $VV$ ($V=W^\pm,Z$) final states~\cite{Aoude:2023hxv}, where the interplay of polar and azimuthal modulations isolates definite spin-interference structures. In our framework, these tomographic inputs are mapped to a comprehensive suite of QI diagnostics, listed in Table \ref{tab:table_observables_implemented} and explained in more detail in App.~\ref{app:QO}, including purity, polarisation vectors, spin-correlation matrices, entanglement witnesses, and negativity, enabling a systematic and basis-controlled characterisation of the quantum state produced at colliders. While easily defined at leading order, quantum tomography becomes more involved at next-to-leading order. The relevance of higher-order effects in quantum tomography has started to be explored only recently, e.g.\ \cite{Grossi:2024jae,DelGratta:2025qyp,Goncalves:2025qem,DelGratta:2025xjp}

\subsection{Basics}
A (spin) quantum state on a Hilbert space $\mathcal{H}$ is fully described by a {density operator} $\rho$, i.e.\ a linear operator on $\mathcal{H}$ that is Hermitian, positive semidefinite, and of unit-trace:
\begin{align}
\rho^\dagger=\rho,\qquad \rho\ge 0,\qquad \Tr[\rho]=1.
\end{align}
Expectation values follow from the Born rule,
\begin{align}
\langle O\rangle \equiv \Tr[\rho\,O],
\end{align}
for any observable $O$. A {pure} state corresponds to a rank-one projector,
\begin{align}
\rho_{\text{pure}}=\ket{\psi}\bra{\psi},\qquad \Tr[\rho_{\text{pure}}^2]=1,
\end{align}
while a {mixed} state satisfies $0<\Tr[\rho^2]<1$ and admits a convex decomposition
\begin{align}
\rho_{\text{mixed}}=\sum_i p_i\,\ket{\psi_i}\bra{\psi_i},\qquad p_i\ge 0,\quad \sum_i p_i=1,
\label{eq:rho mixed pure}
\end{align}
(not unique in general). The degree of mixedness can be quantified by the {purity} $\Tr[\rho^2]$ or by the von Neumann entropy $S(\rho)=-\Tr[\rho\log\rho]$.

For composite systems $\mathcal{H}=\mathcal{H}_A\otimes\mathcal{H}_B$, the state of a subsystem is obtained by the partial trace,
\begin{align}
\rho_A=\Tr_B[\rho_{AB}],\qquad \rho_B=\Tr_A[\rho_{AB}],
\end{align}
and correlations between $A$ and $B$ are encoded in off-diagonal blocks of $\rho_{AB}$ in product bases. A state is separable if $\rho_{AB}=\sum_k p_k\, \rho_A^{(k)}\!\otimes \rho_B^{(k)}$ with $0\le p_k\le 1$; otherwise it is entangled. In collider applications, the production spin-density matrix encodes single-particle polarisations and pairwise spin coherences, which are probed through angular modulations in the decay distributions.

\par
The simplest non-trivial setting for this formalism is a pair of qubits (e.g.\ two spin-$\tfrac12$ fermions with two available spin states). Each single-particle Hilbert space has dimension two, so the composite system lives in a four-dimensional space and is described by a $4\times4$ density matrix $\rho$. A generic Hermitian, unit-trace $4\times4$ matrix contains $16-1=15$ independent real parameters. Among the many equivalent parametrisations, we adopt the conventional Pauli (Fano-Bloch) expansion because of its transparent physical interpretation:
\begin{equation}
    \rho = \frac{1}{4} \left( \mathbb{I}_4 + \sum_i B_{1i} (\sigma_i \otimes I_2) + \sum_j B_{2j} (I_2 \otimes \sigma_j) + \sum_{i,j} C_{ij} (\sigma_i \otimes \sigma_j) \right),
    \label{eq: rho 2 tops}
\end{equation}
where $\sigma_i$ ($i\in\{1,2,3\}$) are the Pauli matrices and $\mathbb{I}_4$ is the $4\times4$ identity. The parameters $B_1$ and $B_2$ are the single-particle {polarisation} (Bloch) vectors, whose components give the average spin projections of each qubit along the three Cartesian axes (3 parameters each). The $3\times3$ matrix $C$ collects the {spin correlations} between the two qubits along all axis pairs (nine parameters). Together, these $3+3+9$ numbers account for the expected total of $15$ degrees of freedom.

These coefficients are directly observable as expectation values of local and bilocal Pauli operators and can be computed theoretically or reconstructed experimentally (via tomography) as
\begin{align}
    \vec B_1 = \Tr[(\vec \sigma \otimes \mathbb{I}_2) \rho], && \vec B_2 = \Tr[(\mathbb{I}_2 \otimes \vec \sigma) \rho], && C_{ij} = \Tr[(\sigma_i \otimes \sigma_j) \rho].
    \label{eq:definition Fano coefficients}
\end{align}
In this language, $\vec B_{1,2}=\vec0$ characterises locally unpolarised subsystems, while non-zero entries of $C$ encode genuine two-qubit correlations (classical and quantum). Physical states further satisfy positivity constraints (e.g.\ $|\vec B_{1,2}|\le1$ and $C$ bounded such that $\rho\ge0$), which are automatically enforced when $\rho$ is constructed from amplitudes or from consistent tomographic data.

This formalism extends straightforwardly to bipartite systems with arbitrary (and possibly different) Hilbert-space dimensions. This is particularly relevant for collider applications involving massive vector bosons (qutrits with three spin states), mixed fermion-boson final states, or higher-spin particles (e.g.\ a massive spin-2 resonance with $d=2s+1=5$). A convenient and  physically transparent parametrisation is
\begin{equation}
    \rho = \frac{1}{d_A d_B}\,\mathbb{I}_{d_A}\otimes\mathbb{I}_{d_B}
    + \frac{1}{d_B}\sum_{i=1}^{d_A^2-1} B_{1i}\,(\tau^A_i\otimes\mathbb{I}_{d_B})
    + \frac{1}{d_A}\sum_{j=1}^{d_B^2-1} B_{2j}\,(\mathbb{I}_{d_A}\otimes\tau^B_j)
    + \sum_{i=1}^{d_A^2-1}\sum_{j=1}^{d_B^2-1} C_{ij}\,(\tau^A_i\otimes\tau^B_j),
    \label{eq:rho_2d}
\end{equation}
where $d_A$ and $d_B$ are the dimensions of $\mathcal{H}_A$ and $\mathcal{H}_B$, respectively, and $\{\tau^A_i\}$, $\{\tau^B_j\}$ form traceless Hermitian operator bases (generators of $\mathrm{SU}(d_A)$ and $\mathrm{SU}(d_B)$). With this choice, the real vectors $\vec B_1$ and $\vec B_2$ generalise single-particle polarisation (Bloch) vectors to $d_A$- and $d_B$-level systems, while the real matrix $C$ collects all inter-system spin (or helicity) correlations. The total number of free real parameters is $(d_A^2-1)+(d_B^2-1)+(d_A^2-1)(d_B^2-1)=d_A^2 d_B^2-1$, as required by hermiticity and unit trace of a $d_A d_B\times d_A d_B$ density matrix.

In practice, one typically adopts orthonormal generator bases, e.g.\ Pauli matrices for $d=2$ and Gell-Mann matrices for $d=3$, with a normalisation such as $\Tr[\tau^X_i \tau^X_j]=2\,\delta_{ij}$ ($X=A,B$). In that convention the coefficients $B_{1i}$, $B_{2j}$, and $C_{ij}$ are directly proportional to measurable expectation values of local and bilocal operators, enabling tomographic reconstruction from decay-angle distributions. As in the qubit case, positivity of $\rho$ imposes non-trivial constraints on $(\vec B_1,\vec B_2,C)$; these are automatically satisfied when $\rho$ is derived from physical amplitudes, and they provide useful consistency checks when reconstructing the state from data. Finally, while the numerical values of $(\vec B_1,\vec B_2,C)$ depend on the chosen operator bases, basis-independent features, such as singular values of $C$ or entanglement witnesses, capture the invariant physics of spin alignment and correlations across arbitrary spins.

For systems more complex than a qubit pair, the simple interpretation of $\vec B_{1,2}$ as three-component polarisations of each particle does not directly apply; higher-spin or higher-dimensional subsystems require a richer set of local spin multipoles (dipole, quadrupole, etc.). Nevertheless, one can adopt parametrisations that make the spin content manifest by expanding on irreducible tensor operators adapted to the particle spin (or, equivalently, on the generators of $\mathrm{SU}(d)$ with $d=2s+1$), which makes the role of each spin multipole explicit~\cite{Aoude:2023hxv}.

The coefficients $B_1$, $B_2$, and $C$, often called {Fano coefficients}, control the angular dependence of the decay products. In the on-shell, narrow-width approximation, production and decay factorise, and the decay distributions act as projectors onto these moments so that single-particle moments isolate $B_{1,2}$, while bilocal angular modulations determine the entries of $C$. Consequently, the coefficients can be measured experimentally through quantum tomography, thereby enabling the experimental reconstruction of the density matrix.

\par
Thus far we have described $\rho$ and its derived observables at the level of a single phase-space point (event). In many applications it is also useful to consider the {average} (incoherent) state over an ensemble of events. Let $\{i\}$ denote the selected events, each characterised by a matrix element $|\mathcal{M}|^2_i$ and a density matrix $\rho_i$. The ensemble-averaged density matrix $\bar\rho$ is then defined as the weighted sum
\begin{align}
    \bar \rho = \sum_i w_i \rho_i,&& w_i = \frac{|\mathcal{M}|^2_i}{\sum_j |\mathcal{M}|^2_j}.
    \label{eq:sum rho}
\end{align}
In a Monte Carlo setting, $w_i$ should be understood as the full event weight (including flux factors, PDFs, phase-space Jacobians, and selection cuts), normalised to unity.

Since the trace is linear, the Fano coefficients $B_1$, $B_2$, and $C$ of the averaged state are obtained by the same weighted sum as in Eq.~\eqref{eq:sum rho}. Operationally, this means one may either average the density matrices and extract $(\vec B_1,\vec B_2,C)$ once, or equivalently average the event-level coefficients directly-provided a common spin-quantisation convention and reference frame are used across events. Note that ensemble averaging produces an incoherent mixture over kinematics; while linear observables (e.g.\ polarisations and correlations) commute with this averaging, nonlinear functionals of $\rho$ (such as entanglement monotones) must be evaluated {after} forming $\bar\rho$.

\subsection{Quantum correlations and observables}

One of the principal motivations for the density-matrix formalism is the systematic study of {quantum} correlations between subsystems. While classical correlations arise from statistical mixtures (separable states), genuinely quantum correlations, {entanglement} and alike, cannot be captured by a single pure product state or by classical mixing alone. Determining whether subsystems of a globally pure state are entangled is straightforward. By contrast, establishing and quantifying entanglement among subsystems of a system in a mixed state is, in general, NP-hard, and definitive solutions/algorithms are available only for very simple composite quantum systems.

\par
For bipartite systems, a powerful diagnostic is the Peres-Horodecki or Positive Partial Transpose (PPT) criterion~\cite{Peres:1996dw,Horodecki:1996nc}. It distinguishes {separable} from {entangled} states. The test is based on the {partial transpose}: if $\rho$ is separable, then its partial transpose with respect to either subsystem is again a physical state-Hermitian, unit-trace, and {positive semidefinite}. Thus, a negative eigenvalue of the partially transposed matrix certifies entanglement (so-called NPT entanglement). The PPT criterion is {necessary and sufficient} for $2\!\times\!2$ (qubit-qubit) and $2\!\times\!3$ (qubit-qutrit) systems, but only {necessary} in higher dimensions, including the $3\!\times\!3$ (qutrit-qutrit) case relevant to massive vector-boson pairs, where PPT states can nevertheless be entangled (``bound'' entanglement).

\par
A well-known limitation of the PPT criterion is that, while it efficiently detects entanglement, it does not provide a quantitative measure of its magnitude. To characterise the {strength} of quantum correlations, one must instead resort to dedicated entanglement measures or witnesses specifically tailored to the system under study. Several such observables are listed in Table \ref{tab:table_observables_implemented} and defined in more detail in App.~\ref{app:QO}. 

For two-qubit systems, a standard monotone is the {concurrence}, which vanishes identically for separable states and attains unity for maximally entangled Bell states. Concurrence is invariant under local unitary transformations and monotonically non-increasing under Local Operations and Classical Communication (LOCC). Moreover, it admits a closed relation to the {entanglement of formation}, thereby providing an information-theoretic interpretation of the physical resources required to prepare a given state~\cite{Wootters:1997id}. 

In higher-dimensional systems, one typically employs alternative quantifiers such as the (logarithmic) {negativity}---derived from the spectrum of the partially transposed density matrix---or suitably constructed {entanglement witnesses}, each offering complementary sensitivity across different classes of mixed states. Beyond entanglement, a hierarchy of quantum correlations can be probed through measures of {steering} and {Bell non-locality} (stronger forms of quantumness), or {quantum discord} (a weaker but still non-classical form), many of which can also be evaluated directly from the density matrix~$\rho$ in a chosen basis. 

Finally, measures originating from the {resource theory of non-stabiliserness}---such as {magic}, which identifies the non-Clifford resources enabling universal quantum computation, and {mana}, which quantitatively captures this property---can likewise be computed from~$\rho$. These provide a bridge between quantum information theoretic notions of computational power and the structure of quantum correlations present in the system.

The full list of the QI observables currently implemented in the code (see \texttt{Density\_functions.py}) is given in Table \ref{tab:table_observables_implemented}. More information on the theoretical formula, meaning and applications of each observable can be found in App.~\ref{app:QO}. For each quantity we provide: (i) a  definition; (ii) the class of systems it applies to (e.g.\  qubit-qubit, qubit-qutrit, general $d_A\!\times\! d_B$, \dots); and (iii) its physical interpretation (polarisation, correlations, entanglement certification/quantification, non-classicality, {etc.}). While not comprehensive, the catalogue includes the essential diagnostics for baseline collider analyses and is straightforward to extend. Where appropriate, we also spell out normalisation conventions, invariance under local basis changes, and practical guidance for tomographic reconstruction from decay-angle distributions.

\begin{table}[H]
    \centering
    \begin{tabular}{ |p{2.2cm}|p{2cm}|p{8.2cm}|p{0.8cm}|  }
 \hline
 \multicolumn{4}{|c|}{QI  observables} \\
 \hline\hline
 Name& Systems available & Meaning &Refs.\\
 \hline\hline
 Purity   & all systems    &Quantifies how pure/mixed a quantum system is. & \\
 \hline
 Concurrence &  $2\times 2$  & Quantifies how entangled a pair of qubits is.   & \cite{Wootters:1997id}\\
 \hline
 Entanglement of formation & $2\times 2$ 
 & Quantifies how entangled a pair of qubits; also linked to Shannon entropy.&  \cite{Wootters:1997id}
 \cite{PhysRevLett.78.5022}\\
 \hline
 Peres-Horodekci criterion & $2\times 2$, $2 \times 3$&Main criterion to determine if a system is separable. It is necessary for all systems but necessary and sufficient only for $2\times 2$ and $2 \times 3$. & 
 \cite{Peres:1996dw} 
 \cite{Horodecki:1996nc} 
 \cite{Horodecki:2009zz}\\
 \hline
Fano coefficients & $d\times d'$& Represent the individual polarisations and spin correlations of the system.& 
\cite{RevModPhys.29.74}\\
\hline
Upper and lower concurrence & $3\times 3$ & For the $3\times 3$ system, concurrence is not exactly calculable. Upper/lower  bounds can be used to estimate it. & \cite{Aoude:2023hxv}\\
\hline
D coefficients & $2\times 2$, $\vec B = \vec 0$, $C$ diagonal & Entanglement measure used for $p\;p\to t\;\bar t$. Useful to separate singlet/triplet states when $\rho$ is pure and maximally entangled. & \cite{PhysRevD.53.4886}
\\
\hline
Magic & $2\times2$ & Non-classicality indicator that quantifies the non-stabiliserness of the system. & \cite{PhysRevA.71.022316}  \cite{Leone_2022}
\cite{white2024} \\\hline
Mana & $d\times d, d$ odd & Observable that extends magic for higher-dimension systems. & \cite{Jain_2020} \cite{Prakash_2020}\\
\hline
Negativity & all systems & Quantifies the negativity of the eigenvalues of the partially transposed $\rho$ to quantify entanglement. & \cite{Vidal:2002zz}  \cite{plenio2006}\\
\hline
Distances & all systems & Quantify how different two density matrices are.& \cite{Fabbrichesi:2025ywl}\\
\hline
\end{tabular}
    \caption{QI observables implemented in the analysis routine.}
\label{tab:table_observables_implemented}
\end{table}

\subsection{Implementation}
\label{sec:implementation}
The automated computation of production spin-density matrices is implemented natively within the \mg framework, leveraging its event generation, reweighting, and metadata infrastructure. For each {unweighted} event, the code assembles the spin-density matrix from helicity amplitudes and writes the result, together with the relevant run information, into the LHE record, making it directly available for downstream analyses. A companion {\tt Python} library provides high-level routines to operate on these objects, including (but not limited to) extraction of single-particle and pairwise spin polarisations, computation of entanglement diagnostics (e.g.\ concurrence for qubits) and non-classicality indicators (e.g.\ ``magic''), as well as utilities for frame changes, axis conventions, and ensemble averaging.

The interface is designed to be minimally intrusive for  \mg users: the study configuration is specified in the \texttt{reweight\_card.dat} (spin quantisation axes, frames, particles whose spin correlations are studied, etc.), after which event generation and reweighting yield LHE files augmented with per-event spin information. Typical workflows then load the LHE file into the accompanying library to perform tomographic reconstructions, build distributions of the Fano coefficients, or evaluate basis-independent quantities (e.g.\ singular values of the correlation tensor). Several worked examples in this paper illustrate recommended settings and validate the implementation across representative processes. Instructions on how to run the code are provided in App. \ref{app:code}

The primary objective of the implementation is, for each event, to construct the production matrix $R$, i.e.\ the yet-to-be-normalised spin-density matrix of the selected particles. For a two-particle system, we define
\begin{equation}
    R_{h_1 h_2} \equiv \frac{1}{N}\,
    \sum_{\text{helicities, colours}}
    \mathcal{M}_{h_1}^{*}\,\mathcal{M}_{h_2},
    \label{eq:production matrix definition}
\end{equation}
where the sum runs over the helicities of all particles not retained in the density matrix and over colour degrees of freedom (when present). Here $h_1$ and $h_2$ label the helicity configuration of the two particles whose spin state is retained, for instance, for two fermions one may take
$h\in\{\ket{+-},\ket{++},\ket{--},\ket{-+}\}$, yielding a $4\times4$ matrix.
The factor $N$ averages over the unpolarised initial-state quantum numbers (spin and colour), i.e.\ $N$ equals the product of the corresponding initial-state multiplicities. 

By construction, $R$ is a Gram matrix of (final -state) helicity amplitudes: it is Hermitian and  positive semidefinite, and its diagonal elements give the (averaged/summed) probabilities for each helicity state, while the off-diagonal elements encode relative phases responsible for spin interference and CP-odd effects. The physical, normalised spin-density matrix is then
$\rho = R/\Tr[R]$, so the explicit value of $N$ cancels once the trace normalisation is applied. The choice of helicity basis (quantisation axes and reference frame) fixes the interpretation of the matrix entries; basis changes correspond to local unitary rotations on $\rho$ and leave basis-independent observables (e.g.\  entanglement measures) unchanged.

\par
Operationally, the code evaluates the helicity amplitudes $\mathcal{M}$ for each diagram of the selected process and then constructs the production matrix via Eq.~(\ref{eq:production matrix definition}), from which the normalised density matrix $\rho$ is obtained. The computation of helicity amplitudes is native to \mg and fully automated~\cite{Alwall:2014hca}. The summation in Eq.~(\ref{eq:production matrix definition}) is performed by the {\tt Fortran} subroutine {\tt get\_density} in {\tt matrix.f}, which forms all interference products $\mathcal{M}_{h_1}^\star \mathcal{M}_{h_2}$ across helicity and colour configurations and accumulates them to yield the requested components of $R$.\footnote{A proof of concept of this type of computation within \mg was performed in Ref.~\cite{QuentinH}.}

The resulting production matrix is then passed to a {\tt Python} layer through {\tt f2py} for post-processing. 
The {\tt Python} interface serves two purposes: (i) it parses the user inputs (process, spin quantisation axes, reference frame, requested observables) and configures the {\tt Fortran} back-end accordingly; (ii) it writes the per-event spin information into the LHE record for downstream analysis. In particular, it handles frame and basis choices that affect the physical interpretation of $\rho$:

\begin{itemize}
  \item \textbf{Frame changes.} The kinematics can be boosted from the laboratory frame to the user-selected analysis frame (e.g.\  the ZMF or pair c.m.\ frame, the rest frame of one particle, or a Gottfried-Jackson frame). For massive particles, helicity is not invariant under general boosts; the code therefore applies rotation-free boosts. 

  \item \textbf{Spin basis (quantisation axis).} The density matrix is conventionally expressed in the helicity basis in a given frame. In principle, basis changes acting locally on each subsystem as unitary similarities, $\rho \to (U_A\!\otimes U_B)\,\rho\,(U_A^\dagger\!\otimes U_B^\dagger)$,  where $U_{A,B}$ are built from the appropriate Wigner $D$-matrices for the particle spins, could be used. Such transformations preserve basis-independent quantities (e.g.\ eigenvalues, entanglement monotones) while reshuffling polarisations and correlations.

  \item \textbf{Metadata and normalisation.} The interface records the chosen frame and axes, the ordering of helicity states, and the normalisation convention used to obtain $\rho = R/\Tr[R]$, ensuring reproducibility and unambiguous downstream use.
\end{itemize}

\par
To compute the density matrix of a process, one needs information of the events (namely the momenta of each external particle, their PDG codes, their status) but also information on which particles to put in the density matrix, in which reference frame, etc. All of these inputs are not unique, the momenta can be expressed in any reference frame, the spin basis of the density matrix is conventional, the phases on each coefficient of the spinors of fermions is conventional, and so on. Because of this, the components of the density matrix are convention-dependent even though the matrix itself is a fundamental object containing all information about a quantum system. Since the density matrix is not a Lorentz-invariant quantity, care must be taken when comparing its components to expressions, ensuring that the same conventions are used in both cases. The QI observables discussed in App.~\ref{app:QO}, however, are independent of the specific convention adopted.

\par
It is also important to note that conventions used in \mg for the fermion spinors \textit{via} ALOHA \cite{deAquino:2011ub} and defined in the library HELAS \cite{aMurayama} are different from conventions used by popular calculation software like {\sc FeynCalc} in {\sc Mathematica}. 
The transformations used in this paper to go from one convention to another are provided in App. \ref{app:diff}.

\section{Validation}
\label{sec:validation}
To establish that the implementation reliably computes production density matrices and their derived observables, we validate it against benchmarks from the literature. The test suite spans Standard Model processes at the LHC and at $e^+e^-$ colliders, and representative BSM scenarios at leading order (LO). Unless stated otherwise, all comparisons are performed at fixed kinematic configurations with matched spin quantisation axes and phase conventions. We verify, element by element, the agreement of the production matrix, its trace normalisation and Hermiticity, and the positivity of the resulting density matrix. We also 
compare basis-independent quantities, such as eigenvalues, singular values of the correlation tensor, and entanglement diagnostics (e.g.\ concurrence/negativity, where applicable), and confirm their invariance under local basis changes.

To validate the implementation across the relevant spin dimensionalities, we design benchmark studies covering  three cases: qubit-qubit ($2\!\times\!2$), qubit--qutrit ($2\!\times\!3$), and qutrit-qutrit ($3\!\times\!3$). For each class we choose processes with well-established control, fix frames and spin bases, and compare the production matrix $R$ (and the normalised $\rho$) element by element, as well as derived, basis-independent quantities (eigenvalues, singular values of the correlation tensor, entanglement witnesses/monotones, where applicable). Unless stated otherwise, comparisons are performed at LO, at fixed phase-space points and in fiducial regions, using consistent phase and axis conventions. The statistics associated to each of the plots corresponds to $10^6$ unweighted events.

\subsection{$t \bar t$ production at the LHC}
\label{sec:tt}

\par
\begin{figure}
    \centering
    \includegraphics[width=0.85\linewidth]{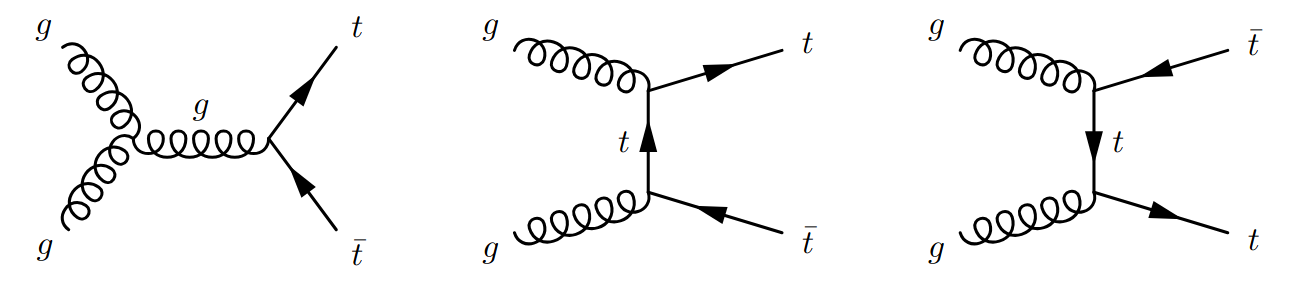}
    \caption{Diagrams for $g\;g \to t \;\bar t$ at LO in QCD.}
    \label{fig:diagrams_gg_ttbar}
\end{figure}
\begin{figure}
    \centering
    \includegraphics[width=0.25\linewidth]{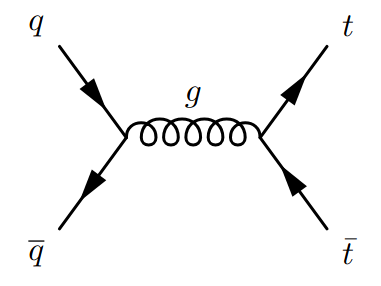}
    \caption{Diagram for $q\;\bar q \to t \;\bar t$ at LO in QCD.}
    \label{fig:diagrams_qqbar_ttbar}
\end{figure}

One of the most extensively studied processes at the LHC, and a standard benchmark process for validation, is top-antitop production,
\begin{equation}
    p \, p \to t \, \overline{t}\,.
\end{equation}
For this final state, results are available for the production density matrix, the spin--correlation matrix, and several quantum--information observables (such as the concurrence and the magic), making it an ideal reference process. Since the density matrix is constructed at the parton level, we validate our implementation separately for the two dominant production channels,

\begin{align*}
    q \; \overline{q} \to t \; \overline{t} \,,\\
    g \; g \to t \; \overline{t}\,,
\end{align*}
before combining them with parton-distribution weights. Figures \ref{fig:diagrams_gg_ttbar} and \ref{fig:diagrams_qqbar_ttbar} show the Feynman diagrams in the Standard Model at LO for the gluon-initiated and the quark-initiated channels, respectively. In this simple case, the phase space can be parametrised by centre-of-mass (c.m.) angle $\theta$ and velocity $\beta$, defined as
\begin{equation}
    \beta = \sqrt{1 - 4m_t^2 / M_{t \bar t}^2}.
    \label{eq: beta_ttbar}
\end{equation}

\par
Channel-by-channel comparisons are performed at fixed kinematics and in specified spin bases (e.g.\ helicity, beamline), checking the individual entries of the production matrix and the derived quantities (polarisations, correlation tensor, eigenvalues). The combined hadronic prediction is then obtained by PDF convolution. We verify that any residual PDF dependence in basis-independent observables is negligible within uncertainties and that expected limiting behaviours (e.g.\ $\vec B_{t,\bar t}\!\to\!0$ for unpolarised QCD production) are reproduced.

All formulas are expressed in the helicity frame $\{\hat{n},\hat{r},\hat{k}\}$ defined in App.~\ref{app:hel}, following standard conventions. Unless stated otherwise, results are given in the $t\bar t$ centre-of-mass frame, which at LO coincides with the partonic centre-of-mass frame of the initial state. In this basis the density matrix takes the form
\begin{equation}
\label{rho pp > ttbar}
    \rho = \frac{1}{4} \left( \mathbb{I}_2 \otimes \mathbb{I}_2 +  \sum_{i=1}^3 B_{1i} (\sigma_i \otimes \mathbb{I}_2) + \sum_{j=1}^3 B_{2j} (\mathbb{I}_2 \otimes \sigma_j) + \sum_{i=1}^3\sum_{j=1}^3 C_{ij} (\sigma_i \otimes \sigma_j)\right)\,,
\end{equation}
where $\vec B_1$ and $\vec B_2$ denote the polarisations of the top and antitop quarks, respectively, and $C$ is the spin-correlation matrix. To emphasise the helicity basis, the component labels $\{1,2,3\}$ may equivalently be written as $\{\hat{n},\hat{r},\hat{k}\}$; we adopt this notation hereafter.

\par
In QCD, P and CP are conserved and in the LHC the initial state is unpolarised, this implies that the final-state particles are  also unpolarised and that the spin-correlation matrix $C$ is symmetric \cite{Aoude:2022imd} so that
\begin{equation}
    \vec B_1 = \vec{0} = \vec B_2.
\end{equation}
The expressions for the spin-correlation matrix coefficients for each parton-parton process at LO are well known \cite{2003.02280, Aoude:2022imd}.
Given the full two-qubit density matrix in Eq.~\eqref{rho pp > ttbar}, any bipartite entanglement measure can be evaluated. For fermion-antifermion pairs, the standard choice is the {concurrence}, which quantifies entanglement for arbitrary (pure or mixed) two-qubit states. A detailed discussion is provided in App.~\ref{app:concurrence}; here we simply record the formula used for its computation. The concurrence is 
\begin{equation}
    \mathcal{C}(\rho) = \max(0, \lambda_1 - \lambda_2 - \lambda_3 - \lambda_4),
\end{equation}
where the $\lambda_i$ are the eigenvalues, ordered in decreasing value, of the matrix
\begin{equation}
    \sqrt{\sqrt{\rho} (\sigma_y \otimes \sigma_y) \rho^\star (\sigma_y \otimes \sigma_y) \sqrt{\rho}},
\end{equation}
with $\rho^\star$ the complex-conjugate of the density matrix and $\sigma_y = \sigma_2$.
The expression for the concurrence is thus, $ \forall \beta \in [0, 1[, \theta \in [0, 2\pi[$:
\begin{equation}
    \mathcal{C}(\rho^{q \bar q}) = \max \left(0, \frac{\beta^2 \sin^2\theta}{2 - \beta^2 \sin^2\theta} \right),
\end{equation}
\begin{align}
    \mathcal{C}(\rho^{gg}) = 
    \begin{cases}
      \frac{1}{2}\max \left( 0, \frac{2 - 4\beta^2(1 + \sin^2\theta) + 2 \beta^4 (1 + \sin^4\theta)}{1 + 2\beta^2\sin^2\theta - \beta^4(1 + \sin^4\theta)}\right) \;\; \text{if} \;\; \beta^2(1+ \sin^2\theta) < 1\\
      \frac{1}{2} \max \left(0, \frac{2 \beta^4(1 + \sin^4\theta) - 2}{1 + 2\beta^2\sin^2\theta - \beta^4(1 + \sin^4\theta)} \right) \;\; \text{if} \;\; \beta^2(1+ \sin^2\theta) \geq 1
    \end{cases}\,.
\end{align}

One can also remark that for the gluon case there is a zone of the phase space where the concurrence is $0$. The boundaries of this region are defined by the following functions:
\begin{align}
    &\beta_{\text{low}}(\theta) = \sqrt{\frac{1 +\sin^2\theta - \sqrt{2}\sin\theta}{1 + \sin^4\theta}} \,,\\
    &\beta_{\text{up}}(\theta) = \frac{1}{(1 + \sin^4\theta)^{1/4}}\,.
\end{align}

Another observable that can be compared with analytical results from the literature is magic, which is detailed in section (\ref{app:magic}) and whose expression for a mixed (i.e.\ non pure) pair of fermions is:
\begin{equation}
    \tilde{M}_2(\rho^I) = -\log_2 \left( \frac{(\tilde{A}^I)^4 + \sum_{i = 1}^3 \left(( \tilde{B}_{1i}^{I})^4 + (\tilde{B}_{2i}^{I})^4\right) + \sum_{i,j = 1}^3 (\tilde{C}^I_{ij})^4}{(\tilde{A}^I)^2\left[(\tilde{A}^I)^2 + \sum_{i = 1}^3 \left(( \tilde{B}_{1i}^{I})^2 + (\tilde{B}_{2i}^{I})^2\right) + \sum_{i,j = 1}^3 (\tilde{C}^I_{ij})^2\right]}\right),
     \label{eq: magic formula}
\end{equation}
with $I = q\bar q, g g$ and $\tilde X$ the non-normalised Fano coefficients:
\begin{align}
    \tilde A^I = \frac{\Tr[R^I]}{4}, &&  B_i^I = \frac{\tilde B_i^I}{\tilde A^I}, && C_{ij}^I = \frac{\tilde C_{ij}^I}{\tilde A^I}.
\end{align}
Using the Fano coefficients computed in the literature, we can determine the expression for magic using equation (\ref{eq: magic formula}). For compactness, we define $z\equiv \cos\theta$ to simplify the formula. For the quark-antiquark initiated channel, the expression reads
\begin{equation}
    \tilde{M}_2(\rho^{q \bar  q}) = - \log_2 \left(\frac{c_0^{q \bar q}(z) + c_2^{q \bar q}(z)\beta^2 + c_4^{q \bar q}(z)\beta^4 + c_6^{q \bar q}(z) \beta^6 + c_8^{q \bar q}(z) \beta^8}{\left(\beta^2 \left(z^2-1\right)+2\right)^2 \left(\beta^4 \left(z^2-1\right)^2+2 \beta^2 \left(z^2-1\right)+2\right)} \right),
    \label{eq:magic start}
\end{equation}
with 
\begin{align}
    &c_0^{q \bar q}(z) = 8 \left(2 z^8-4 z^6+4 z^4-2 z^2+1\right),\\
    &c_2^{q \bar q}(z)  = -8 \left(z^2-1\right)^2 \left(4 z^4-z^2+2\right),\\
    &c_4^{q \bar q}(z)  = 4 \left(z^2-1\right)^2 \left(5 z^4-3 z^2+3\right),\\
    &c_6^{q \bar q}(z)  = -4(z^2 - 1),\\
    &c_8^{q \bar q}(z)  = \left(z^2-1\right)^4.
\end{align}
For the gluon-gluon-initiated channel, the expression is
\begin{equation}
    \tilde{M}_2(\rho^{gg}) = -\log_2 \left(\frac{N^{gg}}{D^{gg}} \right),
\end{equation}
with
\begin{align}
        D^{gg} =& \left(\beta^4 \left(z^4-2 z^2+2\right)+2 \beta^2 \left(z^2-1\right)-1\right)^2 (\beta^8 \left(z^4-2 z^2+2\right)^2\nonumber \\ &+ 2 \beta^6 \left(z^6-5 z^4+7 z^2-4\right)+\beta^4 \left(4 z^4-8 z^2+6\right)-2 \beta^2+1)\,,\\
    N^{gg} =& 1 + N_2^{gg}(z) \beta^2 + N_4^{gg}(z) \beta^4 + N_6^{gg}(z) \beta^6 + N_8^{gg}(z) \beta^8 + N_{10}^{gg}(z) \beta^{10}\nonumber \\&+ N_{12}^{gg}(z) \beta^{12} + N_{14}^{gg}(z) \beta^{14} + N_{16}^{gg}(z) \beta^{16}\,,
\end{align}
and
\begin{align}
    N_2^{gg}(z) =& -4,\\
    N_4^{gg}(z) =& 12 \left(z^8-3 z^6+5 z^4-5 z^2+3\right),\\
    N_6^{gg}(z) =& 4\left(6 z^{10}-33 z^8+72 z^6-90 z^4+66 z^2-28\right),\\
    N_8^{gg}(z) =& 2 (8 z^{16}-48 z^{14}+136 z^{12}-260 z^{10}+399 z^8-496 z^6\nonumber \\&+464 z^4-280 z^2+96),\\
    N_{10}^{gg}(z) =& -4(8 z^{16}-50 z^{14}+146 z^{12}-268 z^{10}+361 z^8-382 z^6\nonumber\\
    &+312 z^4-172 z^2+52),\\
    N_{12}^{gg}(z) =& 4(5 z^{16}-33 z^{14}+105 z^{12}-211 z^{10}+303 z^8-324 z^6\nonumber\\
    &+254 z^4-132 z^2+36),\\
    N_{14}^{gg}(z) =& -4 \left(z^4-2 z^2+2\right)^4,\\
    N_{16}^{gg}(z) =& \left(z^4-2 z^2+2\right)^4\,. 
    \label{eq:magic end}
\end{align}

\par
Two complementary checks are performed to validate the results of the simulation. 
First, one can examine the distributions of {concurrence} and {magic} across the phase space for each individual process to verify that the expected patterns of quantum correlations are correctly reproduced. 
Second, one can evaluate the relative error in the density matrix elements to assess the numerical precision of the computation and ensure the internal consistency of the simulation.

\begin{figure}[!t]\centering
\subfloat[]{\label{}\includegraphics[width=.49\linewidth]{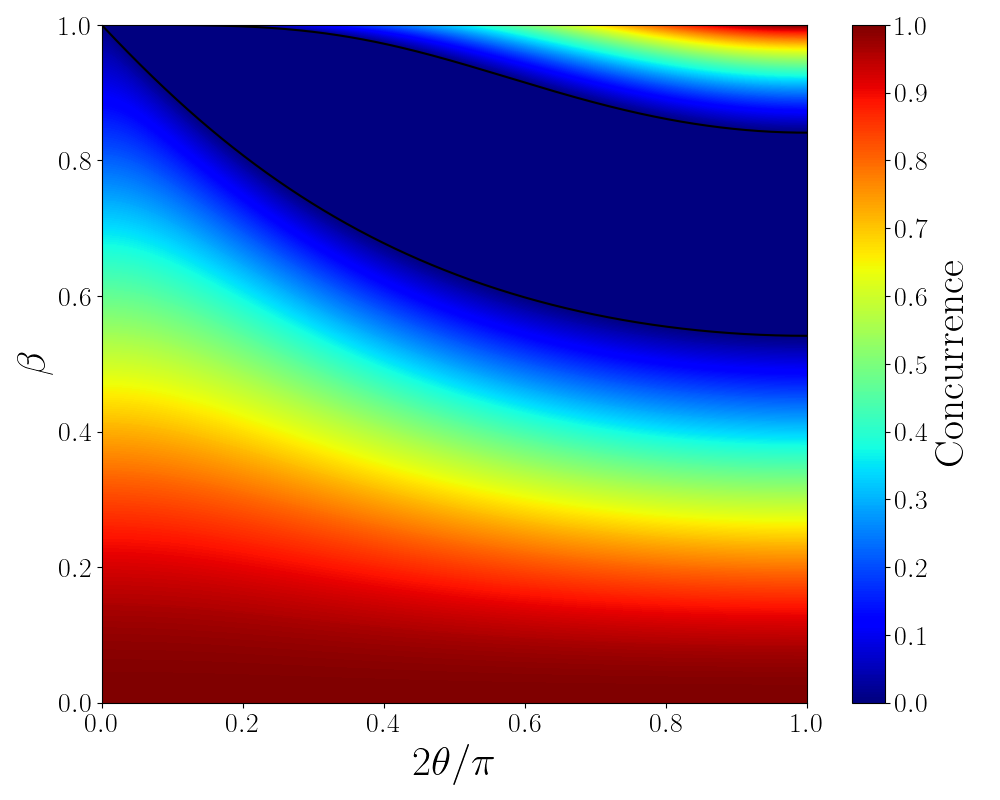}}\hfill
\subfloat[]{\label{}\includegraphics[width=.49\linewidth]{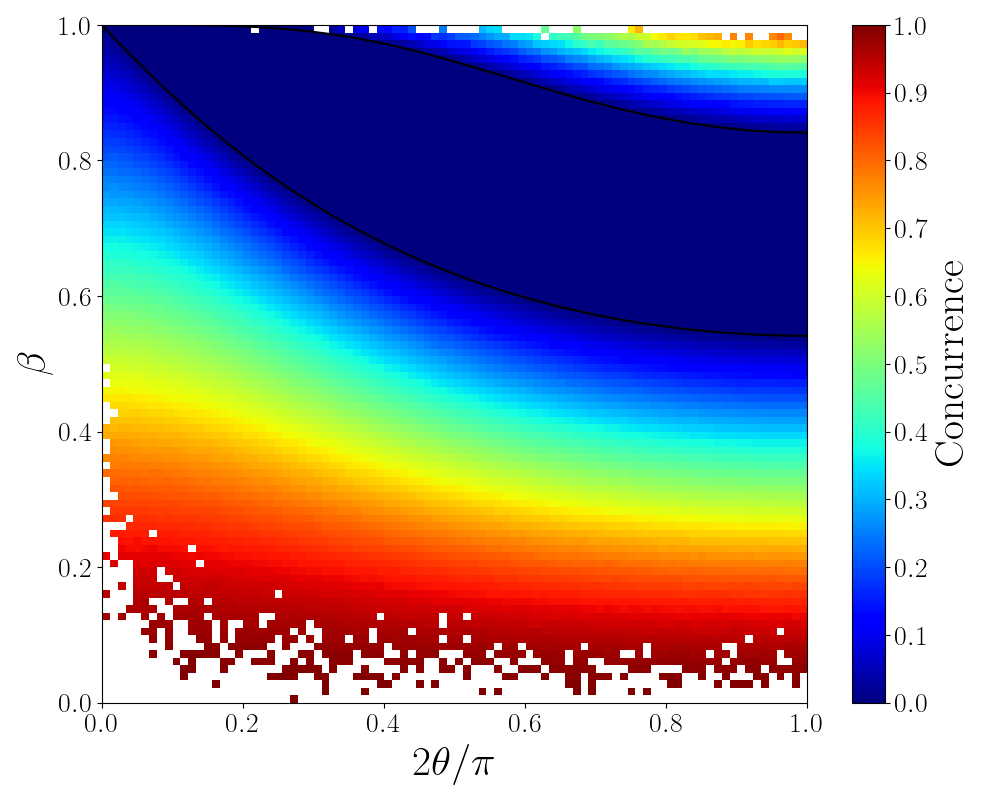}}
\caption{Concurrence as a function of $\theta$ (\ref{eq: theta1}) and $\beta$ (\ref{eq: beta_ttbar}),  in the centre-of-mass of the pair $t \bar t$ for the process $gg \to t\bar t$ at $\sqrt{s} = 13$~TeV. (a) (b) simulation results.  The coarse graining on the right is related to the available statistics in the corresponding phase space region (1M of unweighted events).}
\label{fig: Concurrence g g > t tbar}
\end{figure}
\par
For comparison purposes, parton-level events have been generated for $pp$ collisions at $\sqrt{s}=13$ TeV with the default \texttt{NNPDF23 NLO PDF} set~\cite{Ball:2012cx}. 
The plots in Fig.~\ref{fig: Concurrence g g > t tbar} display the concurrence over the phase space for the gluon-initiated process, comparing the prediction with the simulation output. The two results exhibit excellent agreement, with a relative bin-by-bin deviation below $1\%$ across most of the phase space, and in particular in the region of highest cross-section. Notably, the region of strong spin correlation at large values of~$\beta$ approaches a pure triplet configuration, corresponding to one of the maximally entangled Bell states.
Indeed, one can explicitly calculate that, in the limit $\beta \to 1$ and for $\theta = \pi/2$, the density matrix takes the form
\begin{equation}
    \rho = \frac{1}{2} \begin{pmatrix}
        1 & 0 & 0 & 1 \\ 0 & 0 & 0 & 0 \\ 0 & 0 & 0 & 0 \\ 1 & 0 & 0 & 1
    \end{pmatrix},
\end{equation}
expressed in the basis $\{ \ket{+-}, \ket{++}, \ket{--}, \ket{-+}\}$. 
It follows that the system described by this density matrix is a pure state generated by the Bell state $\ket{\Psi^+}$,
\begin{align}
    \ket{\Psi^+} = \frac{\ket{+-} + \ket{-+}}{\sqrt{2}}, &&\rho = \ket{\Psi^+} \bra{\Psi^+}.
\end{align}
This state is known to be maximally entangled: the spins of the two particles are fully correlated with opposite orientations. 
Correspondingly, the spin-correlation matrix of this state is given by
\begin{equation}
    C_{\Psi^+} = \begin{pmatrix}
        1 & 0 & 0 \\ 0 & -1 & 0 \\ 0 & 0 & 1
    \end{pmatrix}.
\end{equation}
At the opposite limit of the phase space, at threshold ($\beta \to 0$), the density matrix of the system becomes
\begin{equation}
    \rho = \begin{pmatrix}
        0 & 0 & 0 & 0 \\ 0 & 1 & 1 & 0 \\ 0 & 1 & 1 & 0 \\ 0 & 0 & 0 & 0
    \end{pmatrix}.
\end{equation}
The system is again found to be in a pure state, corresponding to another Bell state, $\ket{\Phi^+}$,
\begin{align}
    \rho = \ket{\Phi^+} \bra{\Phi^+}, && \ket{\Phi^+} = \frac{\ket{--} + \ket{++}}{\sqrt{2}}.
\end{align}
This state is likewise maximally entangled: the spins of the two particles are anti-aligned. 
The associated spin-correlation matrix is therefore
\begin{equation}
    C_{\Phi^+} = \begin{pmatrix}
        1 & 0 & 0 \\ 0 & 1 & 0 \\ 0 & 0 & -1
    \end{pmatrix}.
\end{equation}
The plots discussed above demonstrate that the simulation is able to accurately reproduce several QI observables across the entire phase space. 
However, in addition to these global comparisons, we also wish to assess the precision of the computation on an event-by-event basis. 
To this end, the density matrix must be evaluated in exactly the same reference frame and conventions as those used in the literature. 
This requires adopting the same definition of the four-momenta (\textit{ie.} the helicity frame), the same basis for the density matrix, and identical conventions for the phases of the helicity states. 
Such details are rarely specified explicitly in published works, and thus had to be carefully reconstructed for the purpose of this comparison.
 
\par
The four-momenta are expressed in the helicity frame defined in App.~\ref{app:hel}. 
The basis used for the density matrix is $\{\ket{+-}, \ket{++}, \ket{--}, \ket{-+}\}$, where the first helicity label refers to the top quark and the second to the anti-top quark. 
The conventions for the definition of spinors and helicities are detailed in App.~\ref{app:hel}. 
It should be noted, however, that Ref.~\cite{2003.02280} employed {\sc FeynCalc} to obtain the expressions, whereas our simulation relies on \mg. 
Since these two frameworks use different phase conventions for fermion spinors, the corresponding density matrices differ by a global phase for each helicity component. 
To account for this difference, a unitary transformation $U$ is applied such that $\rho \to U^\dagger \rho U$, thereby restoring the correct relative phases between the spinor components. 
The explicit form and parameters of the transformation are given in App.~\ref{app:diff}.

Once these definitions are given, we can compare the density matrix to the one computed by the \mg code.  We have tested several phase-space points, yet for illustration and reference we choose one  in the high cross-section zone where most events are generated, with a bin centred at $\beta = 0.827$ and $2\theta/\pi = 0.414$. The and simulated density matrices are found to be:
\begin{equation}
    \rho_{\rm TH} = \begin{pmatrix}
        0.21101789&0.0349677i&0.0349677i&-0.04731302\\
        -0.0349677i&0.28898211&-0.03643422&-0.0349677i\\
        -0.0349677i&-0.03643422&0.28898211&-0.0349677i \\
        -0.04731302&0.0349677i&0.0349677i&0.21101789 
    \end{pmatrix}\,,
\end{equation}

\begin{equation}
    \rho_{\rm MG} = \begin{pmatrix}
        0.211017999&         \varepsilon+0.03496734i & \varepsilon+0.03496734i & -0.0473125707 \\
        \varepsilon-0.03496734i&  0.288982001& -0.0364339178&      \varepsilon-0.03496734i \\
        \varepsilon-0.03496734i& -0.0364339178& 0.288982001&        \varepsilon-0.03496734i \\
        -0.0473125707&        \varepsilon+0.03496734i & \varepsilon+0.03496734i& 0.211017999    
    \end{pmatrix},
\end{equation}
with $\varepsilon = -1.90673940.10^{-4}$ which is two orders of magnitude smaller than the imaginary part of each component of the matrix. We have checked other several other points which we omit here for brevity. Our comparison demonstrates that the code is able to accurately compute the density matrix for the gluon-initiated process.

An analogous study has been performed for the case of an initial state composed of light quarks. 
Since results are available at leading order (LO), a direct comparison with the simulation can be carried out. 
The parameters used to generate the data are identical to those adopted for the gluon-initiated process at $\sqrt{s} = 13~\text{TeV}$.

\begin{figure}[!htp]\centering
\subfloat[]{\includegraphics[width=.49\linewidth]{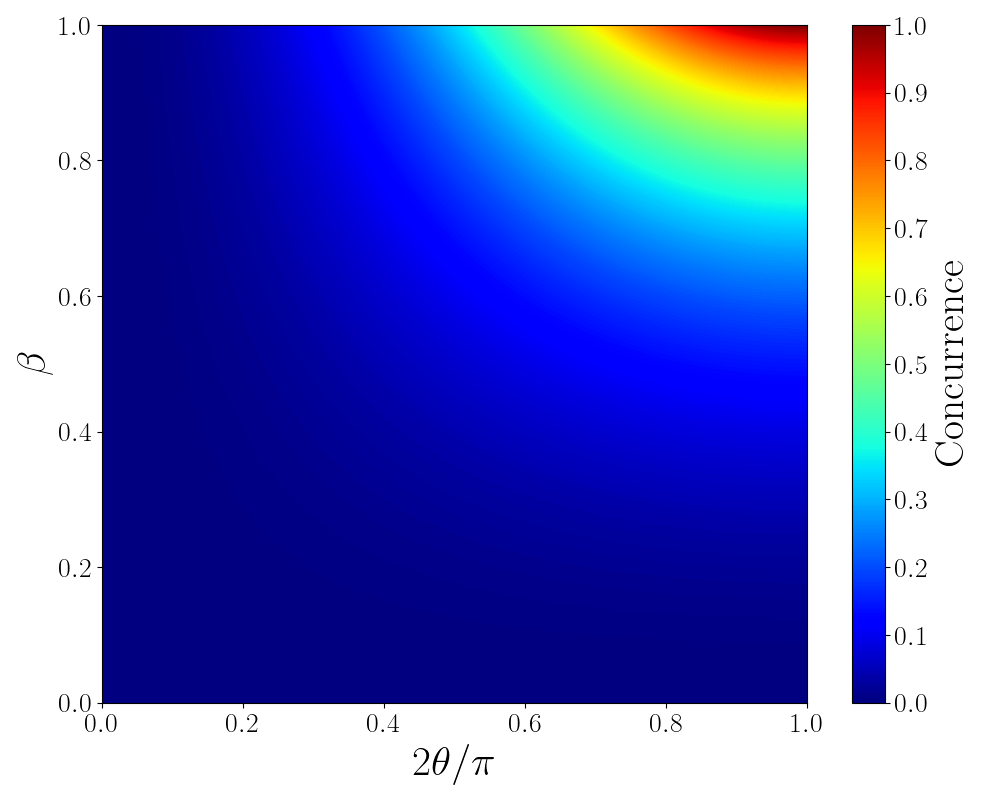}}\hfill
\subfloat[]{\includegraphics[width=.49\linewidth]{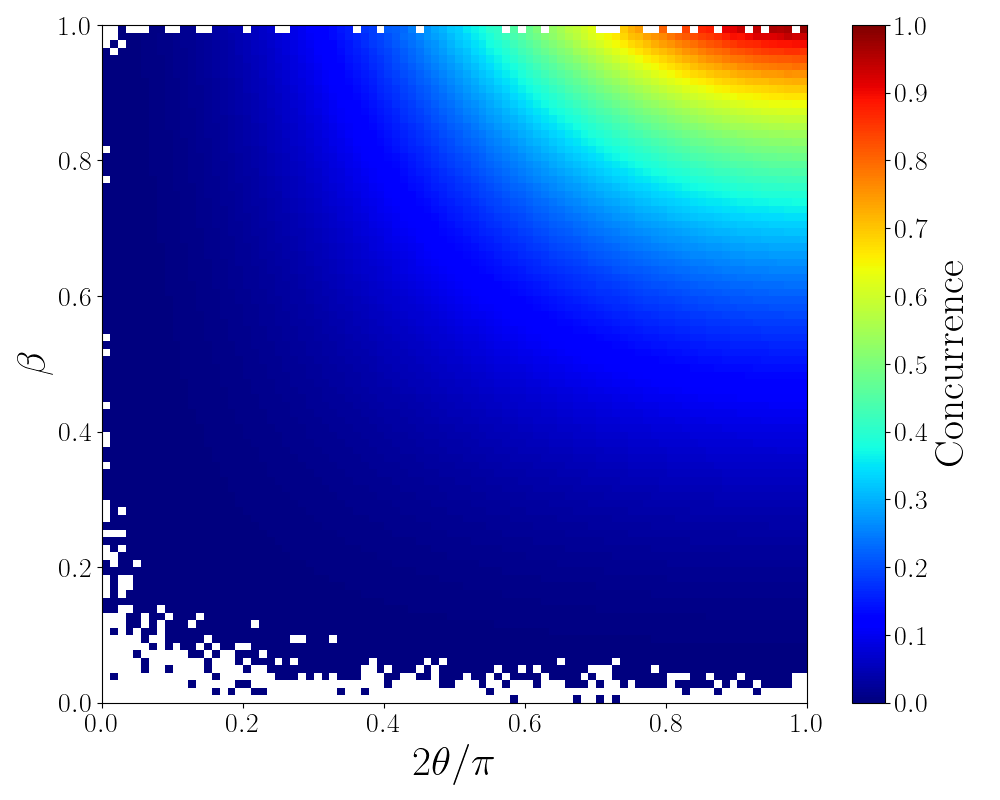}}
\caption{Same as Fig.~\ref{fig: Concurrence g g > t tbar} for the $q \bar q \to t\bar t$ production channel. }
\label{fig:Concurrence_qqbar_ttbar}
\end{figure}

The plots in Fig.~\ref{fig:Concurrence_qqbar_ttbar} display the concurrence for the light-quark--initiated channel, with the prediction shown on the left and the simulation result on the right. 
The first observation is the excellent agreement between the simulated data and the expression, with a relative deviation that we calculated to be below $1\%$ in all regions of the phase space where the statistical precision is sufficient. 
Notably, the region of strong spin correlation, i.e.\ $\beta \to 1$ and  $\theta = \pi/2$, corresponds to a pure state, namely the Bell state $\ket{\Psi^+}$, identical to the configuration found in the same kinematic region for the gluon-initiated process. 
This implies that the same conclusion regarding the Bell-state structure applies to the total process $p\,p \to t\,\bar t$. 
In contrast, no high-entanglement region is observed near threshold for the quark-initiated channel. 
This behaviour arises because, in this case, only opposite-helicity quarks contribute to the interaction, whereas in the gluon-initiated process both like-helicity and opposite-helicity configurations are allowed. 
The implications of this difference will be discussed in more detail later in this section.

\par
Until here we have computed the density matrix and concurrence for both partonic channels of $t\bar t$ production at LHC because the density matrix is an object that is calculated at the parton level. To enable the code to predict results directly comparable with collider data, it needs to be able to compute the density matrix for the full $p\;p\to t\;\bar t$ process. As discussed in the introduction, this is done by summing the density matrices with a weight which corresponds to probability weight associated with each partonic subprocess

\begin{equation}
    \rho_{\text{tot}}(\sqrt{s}, \hat{k}) \equiv \sum_{I} w_{I}(\sqrt{s}) \rho^{I} (\sqrt{s}, \hat{k}).
\end{equation}
where $w_I$ is the probability of finding the subprocess $I$. In the case of the process $p\, p \to t\, \bar t$, we have $I = q \bar q, gg$.
The weight $w_I$ can be calculated as a weighted sum of the amplitudes $\tilde{A}_I$:
\begin{equation}
    w_I (M_{t \bar t}, \hat{k}) = \frac{L^I (M_{t \bar t}) \tilde{A}^I(M_{t \bar t}, \hat{k})}{\sum_J L^J(M_{t \bar t}) \tilde{A}^J(M_{t \bar t}, \hat{k})},
\end{equation}
where the $L^I(M_{t \bar t})$ are called the luminosity functions and $\tilde A_I = \Tr[R^I]/4$. Note that compared to the density matrices, it does not depend on the choice of frame, it only depends on the total energy of the system. The luminosity functions are defined through PDFs
\begin{equation}
    L^I (M_{t \bar t}, \sqrt{s}) \equiv \frac{2\tau}{\sqrt{s}} \int_{\tau}^{1/\tau} \frac{d\zeta}{\zeta} N_I(\tau \zeta) N_{\bar I}\left(\frac{\tau}{\zeta}\right),
\end{equation}
where $\tau = M_{t\bar t}/\sqrt{s}$ and $N_I(x)$ are the PDF for each particle. With these elements, we can compare the simulation data for $p \, p \to t \, \bar t$ to the results from \cite{2003.02280}. The parameters used for generating the data are the same as the one used for the partonic channels.

\begin{figure}[!htp]\centering
\subfloat[]{\label{pp-a}\includegraphics[width=.49\linewidth]{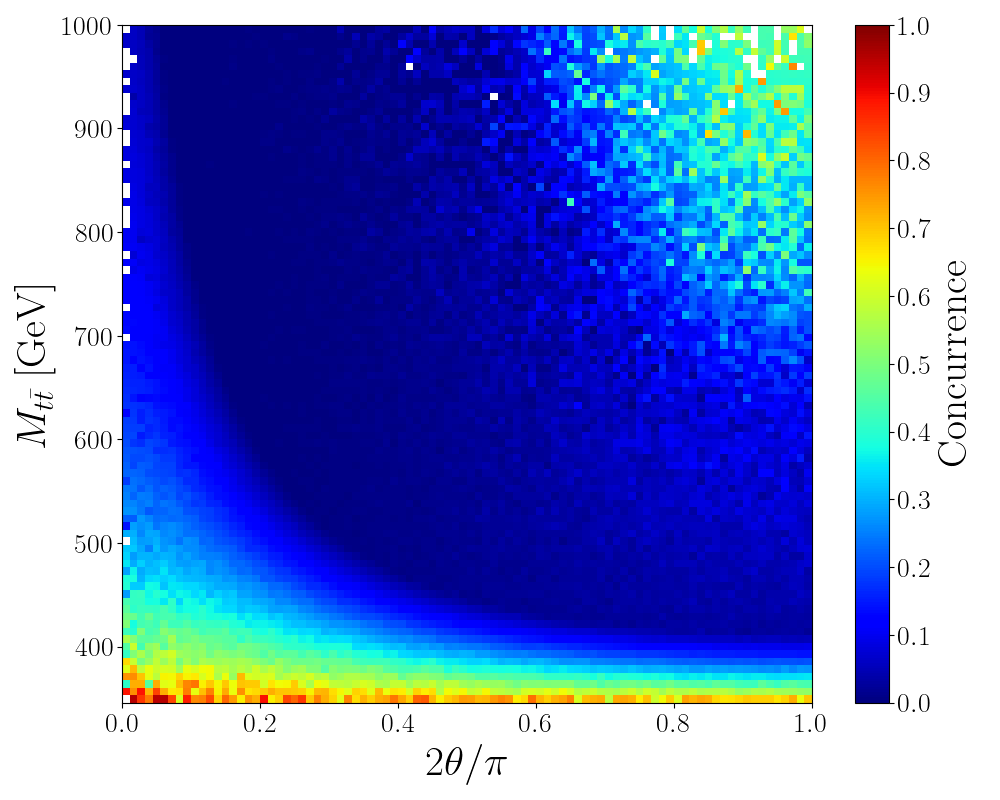}}
\subfloat[]{\label{pp-c}\includegraphics[width=.38\linewidth]{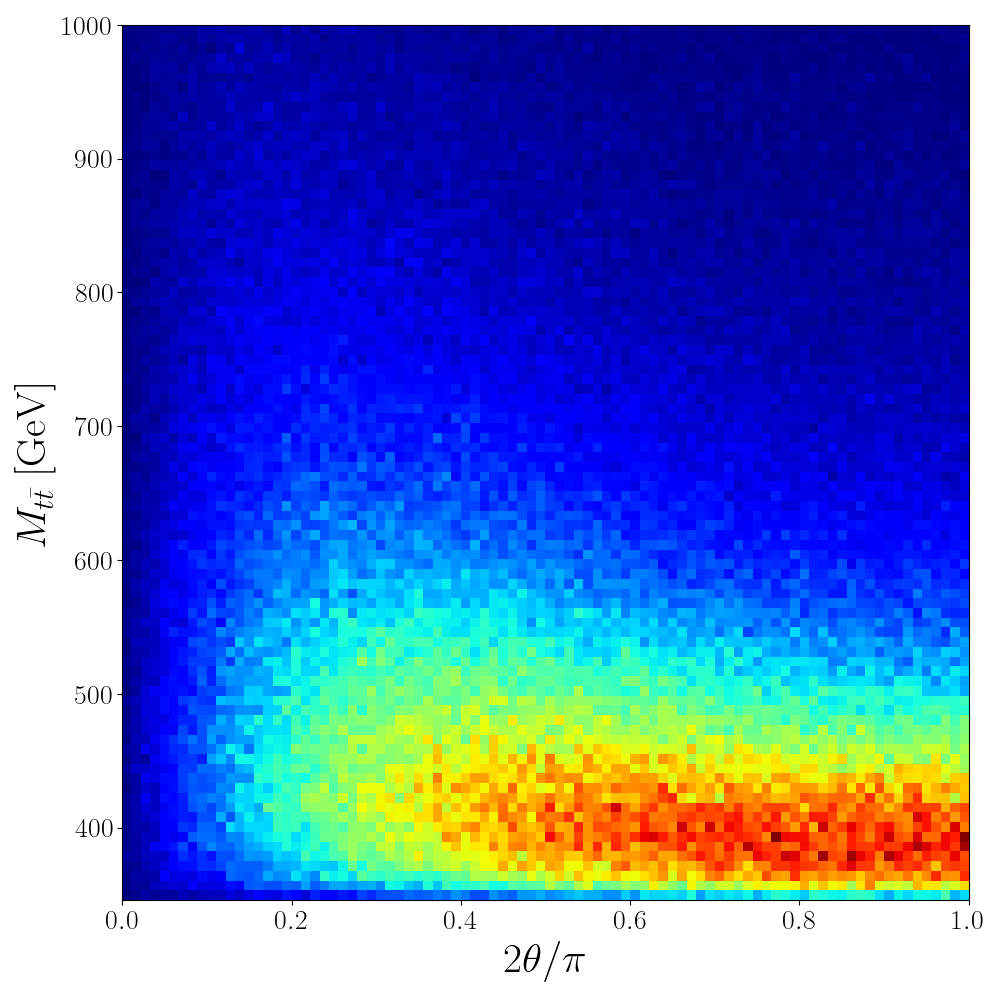}}\par 

\caption{Concurrence (\ref{pp-a}) and cross-section (\ref{pp-c}) as a function of the production angle $\theta$ (\ref{eq: theta1}) and of $M_{t\bar t}$ in the centre-of-mass of the pair $t \bar t$ at the LHC ($p\; p$  at $\sqrt{s} = 13$ TeV). They can be compared with those obtained analytically, see Fig.~5 of Ref.~\cite{2003.02280}.}
\label{fig:ttc}
\end{figure}

\begin{figure}[!htp]\centering
\subfloat[]{\label{fig:high-beta cuts}\includegraphics[width=.49\linewidth]{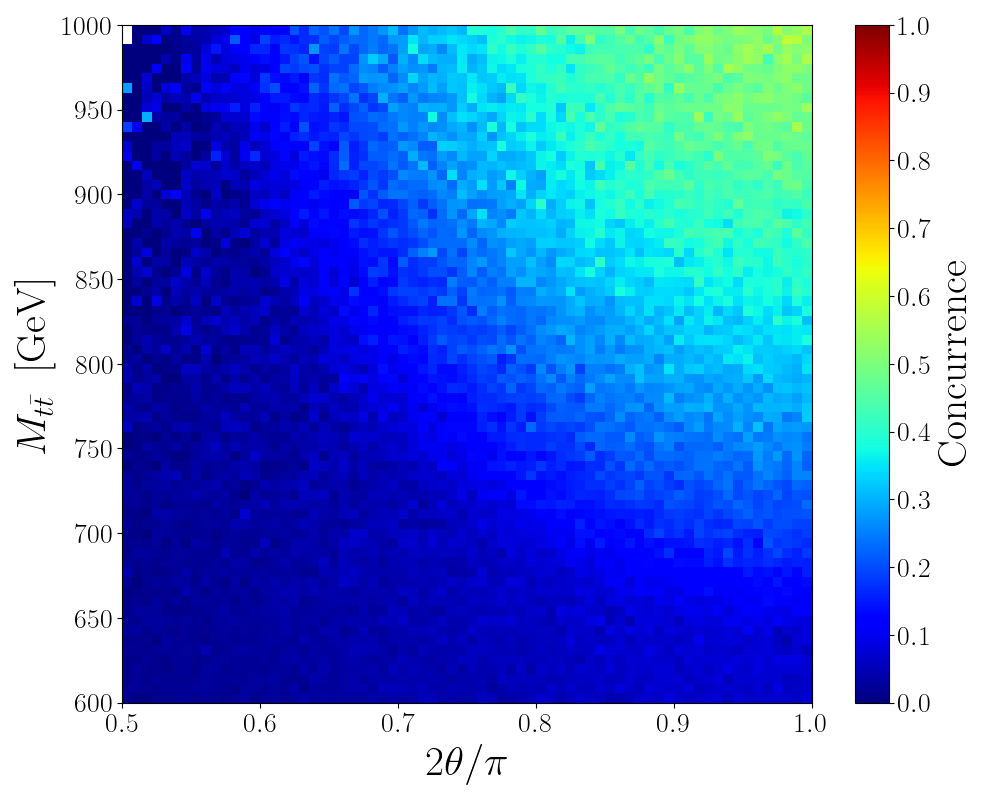}}\hfill
\subfloat[]{\label{fig:threshold cuts}\includegraphics[width=.49\linewidth]{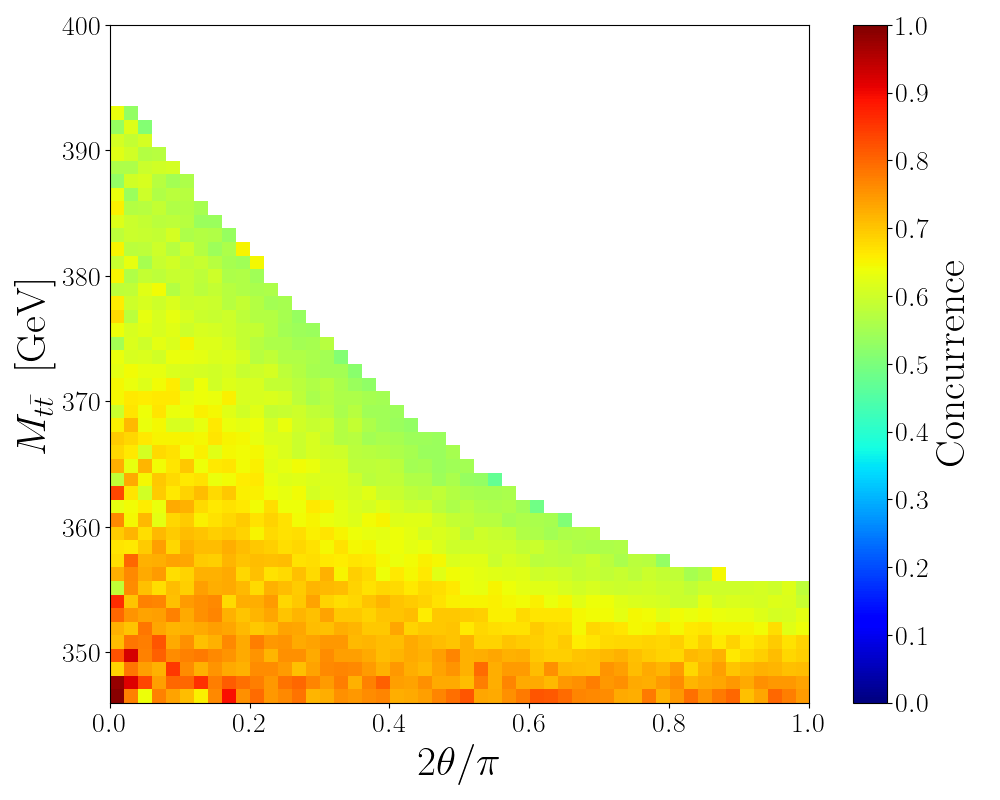}}\par
\caption{Concurrence of $p \, p \to t\, \bar t$ in the high-entanglement phase-space regions:  (\ref{fig:high-beta cuts}) corresponds to the high-$\beta$ region, generated with  $M_{t \bar t} \geq 600$~GeV and $\abs{\eta} < 0.8$. (\ref{fig:threshold cuts}) corresponds to the threshold region, generated with $M_{t \bar t} \leq 400$~GeV. }
\label{fig:tt-cuts}
\end{figure}

Figure~\ref{fig:ttc} shows the concurrence distribution over the phase space for the process $p\,p \to t\,\bar t$, together with the corresponding event density, which indicates the statistical weight of each region. Compared to the parton-level analysis, small differences can be detected in areas of phase space which are not very well populated.  A useful cross-check consists in applying a cut on the invariant mass of the $t\bar t$ system, thereby improving the statistical precision in the affected regions. Figure~\ref{fig:high-beta cuts} shows the concurrence distribution over the phase space after applying the cuts $M_{t\bar t} \geq 600~\text{GeV}$ and $|\eta_t| < 0.8$, where $\eta_t$ denotes the rapidity of the top quark. These cuts ensure a sufficient number of events in the selected region, thereby reducing the impact of statistical uncertainties. A direct comparison with the prediction in Fig.~\ref{fig:ttc} shows that the two results are in very good agreement, with only minimal differences. The same procedure has been applied to the threshold region, using the cuts $M_{t\bar t} \leq 400~\text{GeV}$.

This study on the process $p \, p \to t \, \bar t$ shows that the code allows precise computation of the concurrence over the whole phase space which serves as an observable of entanglement; moreover, its dependence on the choice of PDFs is small~\cite{2003.02280}.

A further consistency check is obtained by considering polarised initial states in the gluon-fusion channel. While in quark-antiquark annihilation the incoming fermions must have opposite helicities, gluons can be either like- or unlike-helicity. Therefore, one expects a markedly different behaviour for the unlike-helicity configuration $g_L g_R$ and for the like-helicity configuration $g_R g_R$. As shown in Ref.~\cite{Mahlon:2010gw}, when focusing on the spin correlations  between the top and the anti-top quark, the processes $g_R g_L \to t \, \bar t$ and $q \, \bar q \to t \, \bar t$ display very similar features. 

\begin{figure}[t!]\centering
\subfloat[]{\label{gRgR concurrence plot}\includegraphics[width=.45\linewidth]{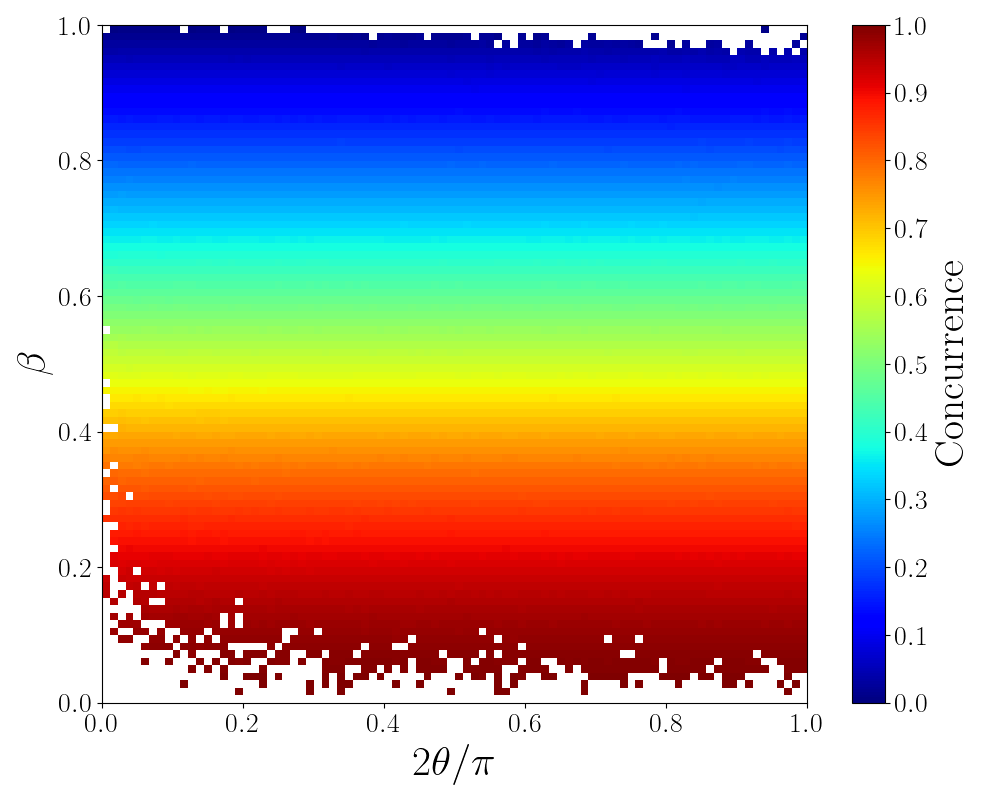}}\hfill
\subfloat[]{\label{gRgL concurrence plot}\includegraphics[width=.45\linewidth]{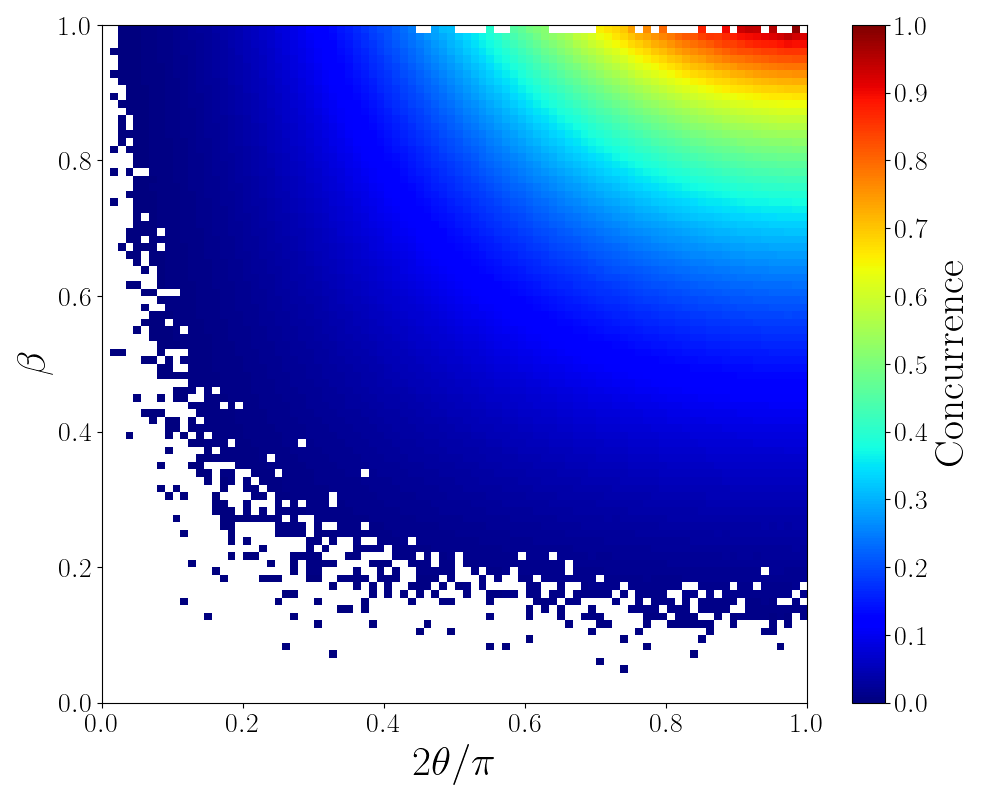}}\par
\caption{Concurrence over the phase space of the gluon-fusion process, where the initial gluons are polarised with like and unlike helicities, is shown in Figs.~\ref{gRgR concurrence plot} and~\ref{gRgL concurrence plot}, respectively. In the like-helicity configuration (Fig.~\ref{gRgR concurrence plot}), the region of strong correlation corresponds to the threshold region of the process $g \, g \to t \, \bar t$. Conversely, for the unlike-helicity configuration (Fig.~\ref{gRgL concurrence plot}), strong correlations arise in the high-energy region of the phase space. Notably, this latter concurrence distribution is identical to that obtained for the $q \, \bar q \to t \, \bar t$ channel (Fig.~\ref{fig:Concurrence_qqbar_ttbar}), as also discussed in Ref.~\cite{Mahlon:2010gw}.
}
\label{fig: Concurrence gRL gRL > t tbar}
\end{figure}

In the low-$\beta$ region, corresponding to the threshold limit where the top quarks are produced nearly at rest, the helicities of the top quarks align with those of the initial gluons. The spin of each top is then fixed independently of the other, implying the absence of entanglement and hence a vanishing concurrence. On the opposite side of the phase space, in the high-$\beta$ region ($|\vec{p}| \to \infty$), only the helicity configurations $t_L \, \bar t_R$ and $t_R \, \bar t_L$ contribute. In this limit, the spins, identical to the helicities, are perfectly correlated, leading to a concurrence that tends to unity, in agreement with the behaviour observed in Fig.~\ref{gRgL concurrence plot}.

In the case of like-helicity gluons, the situation is reversed: strong entanglement is expected in the threshold region, while no correlation arises in the high-$\beta$ limit~\cite{Mahlon:2010gw}. This pattern is precisely what we observe in Fig.~\ref{gRgR concurrence plot}.

\par
A final test for this process consists in computing a different quantum witness for which an expression is available. Earlier in this section, in Eqs.~(\ref{eq:magic start}--\ref{eq:magic end}), we derived the expression for the quantity known as ``magic'' in both production channels. Although magic is not strictly an entanglement witness, it quantifies another genuinely quantum feature, the so-called non-stabiliserness, widely used in QI yet scarcely explored in collider-physics contexts.

\par
\begin{figure}[!htp]\centering
\subfloat[]{\label{fig:magic1}\includegraphics[width=.45\linewidth]{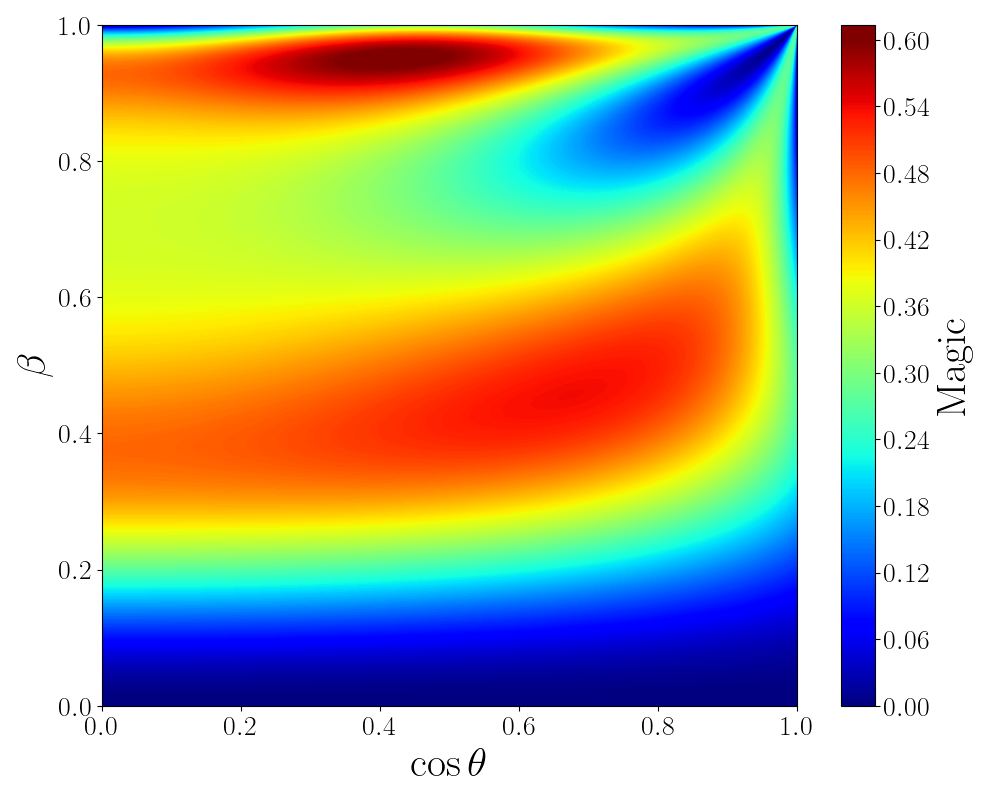}}\hfill
\subfloat[]{\label{fig:magic2}\includegraphics[width=.45\linewidth]{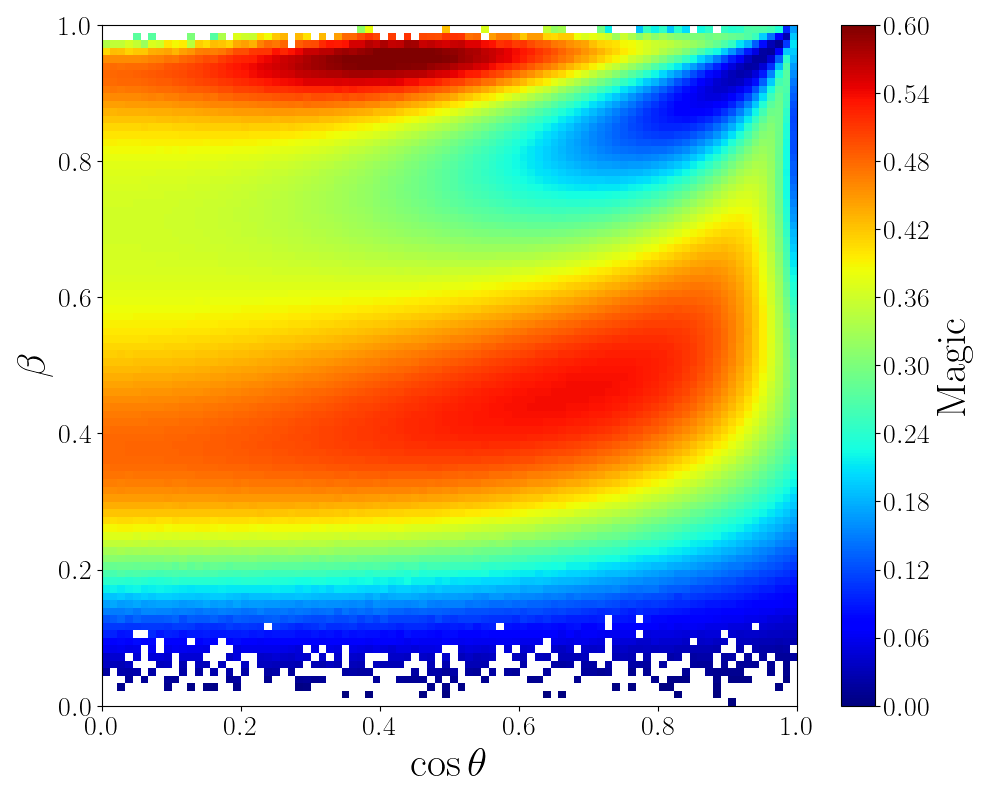}}\par
\caption{The value of  magic over  the phase-space spanned by $\cos\theta$ (\ref{eq: theta1}) and $\beta$ (\ref{eq: beta_ttbar}). (a) result. (b) Simulation result. Plots are extremely close, in most of the phase space, bins deviate from the result by a relative error of about 0.1\% in most of the phase space.}
\label{fig: Magic g g > t tbar}
\end{figure}

Figure~\ref{fig: Magic g g > t tbar} shows the value of magic for the gluon-initiated channel (left) and the corresponding result computed with \mg (right) over the phase space $(\cos\theta, \beta)$. The distribution of magic differs noticeably from that of the concurrence: regions exhibiting large values of magic often correspond to low concurrence. The agreement between the and simulated results is excellent, with relative deviations below 0.5\% in each bin across most of the phase space, and in particular in the regions with the largest cross-section. In regions of limited statistics the uncertainty increases slightly, though effect is negligible for any realistic experimental determination of the observable.

The same analysis can be carried out for the $q \bar q$ channel, yielding identical conclusions and therefore not shown here.

\par
In summary, a comprehensive set of validation tests has been performed on the well-studied process of $t\bar t$ production at the LHC, comparing the simulated results with the corresponding predictions. The excellent agreement observed confirms the robustness and reliability of the implementation for this benchmark process. The following sections extend the analysis to other processes to further demonstrate the versatility of the framework.

\subsection{$t \bar t$ production at $e^+e^-$ colliders}

A second test case that we consider is the production of a $t\bar t$ pair at an electron--positron collider, $e^+ e^- \to t \bar t$. The results for this process are well established at leading order~\cite{Maltoni:2024csn}, providing a benchmark for comparison. The main qualitative difference with respect to $t \bar t$ production at the LHC arises from the presence of  polarised weak--boson interactions.

\begin{figure}[hbt]
    \centering
    \includegraphics[width=0.2\linewidth]{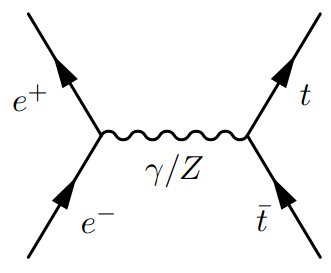}
    \caption{Feynman diagram of the process $e^+ \, e^- \to t \, \bar t$ at tree level.}
    \label{diagram e+e-}
\end{figure}

The Feynman diagram for this process is shown in Fig.~\ref{diagram e+e-}, and the QED vertices, following the notation of Ref.~\cite{Maltoni:2024tul}, are given by
\begin{align}
    (t \, \bar t \, A^\mu) = ie Q_t \gamma^\mu, && ( t \, \bar t \, Z^\mu) = \frac{ie}{c_w s_w} \gamma^\mu(g_{V_t} \mathbb{I} - g_{A_t} \gamma^5),
\end{align}
where $g_{V_t} = T^3/2 - Q s_w^2$ and $g_{A_t} = T^3/2$, with $T^3$ denoting the weak isospin of the fermion, $Q$ its electric charge, and $s_w$ the sine of the weak mixing angle.

\par
As discussed earlier, the density matrix and the spin--correlation matrix strongly depends on the chosen conventions (definition of the reference frame and  phase choices for spinors, etc.), and it is therefore essential to specify them explicitly. The reference frame adopted here is the same as in the study of $t\bar t$ production at the LHC, described in detail in App.~\ref{app:hel}. The density matrix is always expressed in the basis $\{\ket{+-}, \ket{++}, \ket{--}, \ket{-+}\}$. The results obtained are compared with the results of Ref.~\cite{Maltoni:2024csn}, which differ slightly from those implemented in {\sc MG5aMC}. 

\par
While $t\bar t$ production at the LHC is a CP--even process---which allows one to set the polarisations of the top and antitop quarks to zero---this is not the case at an electron--positron collider, where weak interactions induce CP--odd effects. Consequently, the top and antitop polarisations do not need to vanish. 
As in the LHC case, this process is characterised by two kinematic degrees of freedom: the production angle $\theta$ and the top velocity $\beta$,
\begin{equation}
    \beta = \sqrt{1 - 4m_t^2 / M_{t \bar t}^2}.
    \label{eq: beta_epem}
\end{equation}
The expressions for the spin-correlation coefficients are \cite{Maltoni:2024csn}

\begin{align}
    \begin{cases}
        A^{[0]} = F^{[0]} ( \beta^2 c_\theta^2 - \beta^2 + 2), \\
        \tilde C ^{[0]}_{kk} = F^{[0]} \left( \beta^2 - (\beta^2 - 2 \right) c_\theta^2), \\
        \tilde C ^{[0]}_{kr} = 2F^{[0]} \sqrt{1 - \beta^2} c_\theta s_\theta,\\
        \tilde C ^{[0]}_{rr} = F^{[0]} s_\theta^2 ( 2 - \beta^2), \\
        \tilde C ^{[0]}_{nn} = - F^{[0]} \beta^2 s_\theta^2,
    \end{cases} && 
    \begin{cases}
        A^{[1]} = 2 F^{[1]} c_\theta, \\
        \tilde C ^{[1]}_{kk} = 2F^{[1]} c_\theta,\\
        \tilde C ^{[1]}_{kr} = F^{[1]} \sqrt{1 - \beta^2} s_\theta, \\
        \tilde C ^{[1]}_{rr} = 0, \\
        \tilde C ^{[1]}_{nn} = 0,
    \end{cases} &&
    \begin{cases}
        A^{[2]} = F^{[2]} s_\theta^2, \\
        \tilde C ^{[2]}_{kk} = F^{[2]} s_\theta^2,\\
        \tilde C ^{[2]}_{kr} =0, \\
        \tilde C ^{[2]}_{rr} = - F^{[2]} s_\theta^2, \\
        \tilde C ^{[2]}_{nn} = F^{[2]} s_\theta^2,
    \end{cases}
\end{align}
where $C_{ij} = \tilde C_{ij}/ A$ and $s_\theta \equiv \sin\theta$, $c_\theta \equiv \cos\theta$.

The different coefficients $\tilde C_{ij}$ are separated according to the number of $\gamma^5$ insertions. This means that, to obtain the total contribution for a given element, one must sum over the three channels. For instance,
\begin{equation}
    A \equiv A^{[0]} + A^{[1]} + A^{[2]}.
\end{equation}
The prefactors $F^{[i]}$ can be found in the appendix of \cite{Maltoni:2024csn}.

\par
The first comparison we perform is based on the concurrence of the process across the phase space, since the expression of the spin--correlation matrix is available. The simulated data were generated with \mg at $\sqrt{s} = 3~\text{TeV}$ using the electron PDF \texttt{isronlyll}. The inclusion of a PDF allows one to probe different values of $\beta$. In contrast, the results used for comparison are obtained by varying the total centre--of--mass energy of the process, but involve only electron--positron initial states. Therefore, we ensured that our simulated sample does not include any event with photonic initial states.

\begin{figure}[!htp]\centering
\subfloat[]{\label{fig:e-e+1}\includegraphics[width=.45\linewidth]{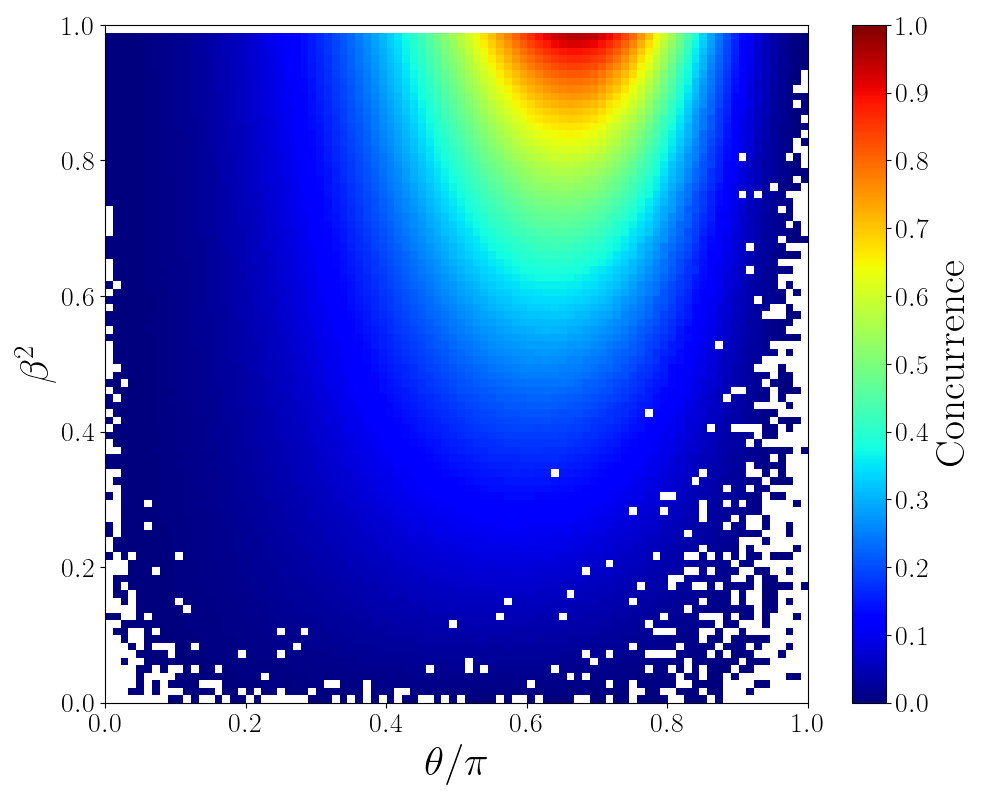}}\hfill
\subfloat[]{\label{fig:e-e+2}\includegraphics[width=.45\linewidth]{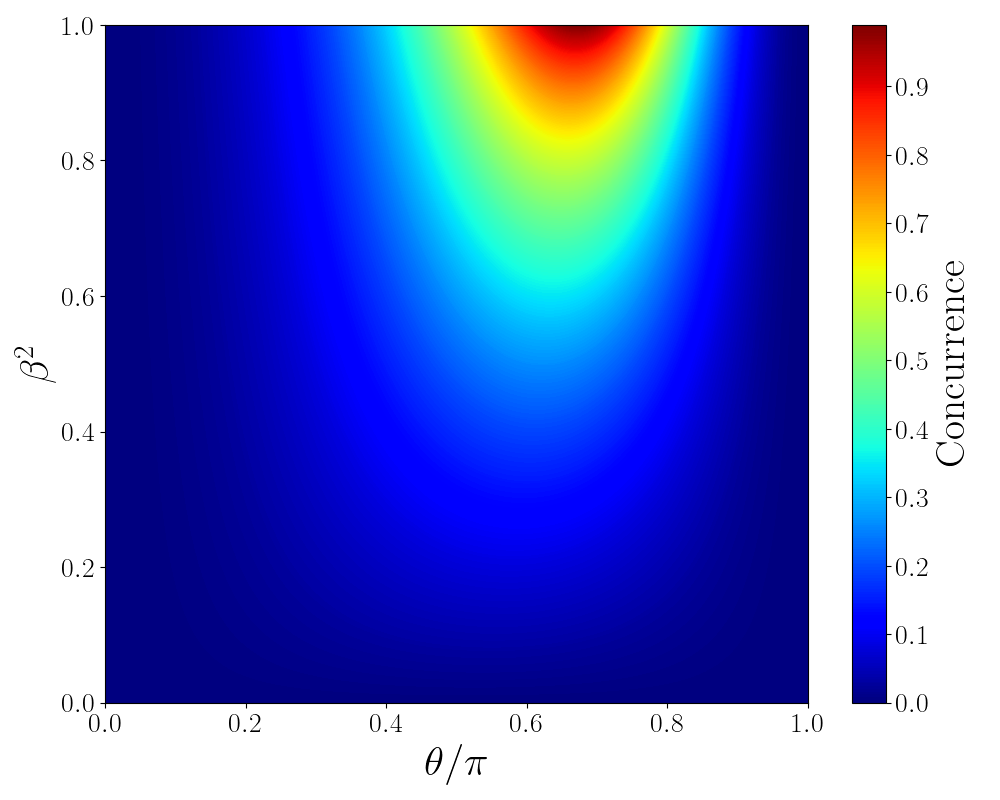}}\par
\caption{Concurrence of $e^+ \, e^- \to t\, \bar t$ over the phase space spanned by $\theta$ (\ref{eq: theta1}) and $\beta$ (\ref{eq: beta_epem}) in the centre-of-mass frame of $t\bar t$. (\ref{fig:e-e+1}) Result of \mg simulation at $\sqrt{s} = 3$TeV and with the pdf \texttt{isronlyll}. (\ref{fig:e-e+2}) analytical plot of the concurrence.}
\label{fig: Concurrence e- e+ > t tbar cuts}
\end{figure}

Figure~\ref{fig: Concurrence e- e+ > t tbar cuts} shows the concurrence over the phase space $(\theta, \beta)$. Unlike $t \bar t$ production at the LHC, the top quarks produced in lepton colliders are entangled throughout the entire phase space. Moreover, in both hadron and lepton colliders (Fig.~\ref{fig:ttc} and Fig.~\ref{fig: Concurrence e- e+ > t tbar cuts}), a region of maximal entanglement emerges at high energies ($\beta \to 1$). The corresponding production angle, however, differs between the two cases: at the LHC, maximal entanglement occurs for $\theta = \pi/2$, whereas at an electron--positron collider it is reached for $\theta \simeq 0.67\,\pi$. This difference arises from the fact that the strong interaction conserves parity, whereas the weak interaction does not~\cite{Maltoni:2024csn}.

\par
The simulation results obtained with \mg are in very good agreement with the analytical solution. The relative deviation between the simulated and analytical results is found to be of order $1\%$ or smaller across most of the phase space. Larger deviations appear only in bins with low statistics, as expected. When phase--space cuts are applied to enhance the event yield in these regions, as was done for $t\bar t$ production at the LHC, the relative errors are again reduced to about $1\%$. This confirms that the implementation accurately reproduces the concurrence at leading order for the process $e^+ e^- \to t \bar t$ over the entire phase space.

\par
Another class of quantum--information observables of interest are the entanglement markers $D^{(i)}$. These quantities are more readily accessible experimentally than the concurrence and make it possible to distinguish between entangled and separable states. Their expressions are given in Eqs.~(\ref{eq:D1}--\ref{eq:Dz}), and we recall that the condition for entanglement is that at least one of the $D$ markers be strictly smaller than $-1/3$. In the following, we compare the values of these markers with their analytical counterparts to assess the precision achievable with the current implementation. Before doing so, we also perform an event--by--event comparison.  We note that the marker $D^{(1)}$ is fixed to $+1/3$, while the others vary. Equations~(\ref{eq:varepsilon1}--\ref{eq:varepsilon2}) provide, for each marker, the corresponding average relative error $\bar{\varepsilon}$,
\begin{align}
\label{eq:varepsilon1}
   &\bar \varepsilon_{D^{(1)}} = 2.10^{-17}, && \bar \varepsilon_{D^{(k)}} = 2.10^{-6}, \\
   &\bar \varepsilon_{D^{(r)}} = 8.10^{-6}, && \bar \varepsilon_{D^{(n)}} = -5.10^{-8}.
   \label{eq:varepsilon2}
\end{align}
We find that the relative error on the different $D^{(i)}$ markers remains below $10^{-5}$, and often even smaller, demonstrating that the code can compute both the density matrix of a process and the associated observables with high precision. We then compare the values of the $D^{(i)}$ markers with their analytical counterparts. For electron colliders, these markers have been evaluated in Ref.~\cite{Maltoni:2024csn} at different centre--of--mass energies ($365~\text{GeV}$, $500~\text{GeV}$, and $3~\text{TeV}$); we therefore compute them for the same energies. Among all markers, $D^{(n)}$ is the most relevant for this process, as it is the smallest and thus the most sensitive to entanglement. Equations~(\ref{eq:Dnfirst}--\ref{eq:Dnlast}) show the comparison between our results and those reported in the literature:
\begin{align}
    \label{eq:Dnfirst}
    &\sqrt{s} = 365\; \text{GeV}: && D^{(n)}_{\text{th}} = - 0.35, && D^{(n)}_{\text{sim}} = -0.3492106, \\
    &\sqrt{s} = 500\; \text{GeV}: && D^{(n)}_{\text{th}} = - 0.42, && D^{(n)}_{\text{sim}} = -0.4222872, \\
    &\sqrt{s} = 3\; \text{TeV}: && D^{(n)}_{\text{th}} = -0.52, && D^{(n)}_{\text{sim}} = -0.5212332.
    \label{eq:Dnlast}
\end{align}
We observe that, for the first two energy points, the simulation results agree perfectly with the analytical predictions.\footnote{In Ref.~\cite{Maltoni:2024csn}, the value of $D^{(n)}$ at $\sqrt{s} = 3$ TeV is written as $-0.53$ instead of $-0.52$ as calculated in this paper. This difference may have been due to rounding effects or just a typographical error.} This confirms that the code reproduces the results with very high precision. For completeness, Eqs.~(\ref{eq:D markers e-e+ initial}--\ref{eq:D markers e-e+ final}) present the comparison for the other $D$ markers, as well as for the off--diagonal elements of the spin--correlation matrix in each case. The simulated values are found to be in excellent agreement with the ones, demonstrating the capability of the code to accurately describe processes at both hadron and lepton colliders.

\begin{align}
\label{eq:D markers e-e+ initial}
    &\sqrt{s} = 365\; \text{GeV}: && D^{(1)}_{\text{th}} = \frac{1}{3}, && D^{(1)}_{\text{sim}} = 0.333333, \\
    &                             && D^{(k)}_{\text{th}} = -0.071914, && D^{(k)}_{\text{sim}} = -0.071914, \\
    &                             && D^{(r)}_{\text{th}} = 0.087791, && D^{(r)}_{\text{sim}} = 0.087791, \\
    &                             && [(C_{kr} + C_{rk})/2]_{\text{th}} = 0.13276, && [(C_{kr} + C_{rk})/2]_{\text{sim}} = 0.13276, \\
    &                             && [(C_{rn} + C_{nr})/2]_{\text{th}} = 0, && [(C_{rn} + C_{nr})/2]_{\text{sim}} = -2.10^{-5}, \\
    &                             && [(C_{nk} + C_{kn})/2]_{\text{th}} = 0, && [(C_{nk} + C_{kn})/2]_{\text{sim}} = -2.10^{-8}.
\end{align}

\begin{align}
    &\sqrt{s} = 500\; \text{GeV}: && D^{(1)}_{\text{th}} = \frac{1}{3}, && D^{(1)}_{\text{sim}} = 0.333333, \\
    &                             && D^{(k)}_{\text{th}} = 0.103499, && D^{(k)}_{\text{sim}} = 0.103499, \\
    &                             && D^{(r)}_{\text{th}} = -0.0145456, && D^{(r)}_{\text{sim}} =-0.0145455, \\
    &                             && [(C_{kr} + C_{rk})/2]_{\text{th}} = 0.225993, && [(C_{kr} + C_{rk})/2]_{\text{sim}} = 0.225993, \\
    &                             && [(C_{rn} + C_{nr})/2]_{\text{th}} = 0, && [(C_{rn} + C_{nr})/2]_{\text{sim}} = -2.10^{-5}  , \\
    &                             && [(C_{nk} + C_{kn})/2]_{\text{th}} = 0, && [(C_{nk} + C_{kn})/2]_{\text{sim}} =1.10^{-8}.
\end{align}

\begin{align}
    &\sqrt{s} = 3\; \text{TeV}: && D^{(1)}_{\text{th}} = \frac{1}{3}, && D^{(1)}_{\text{sim}} = 0.333333, \\
    &                             && D^{(k)}_{\text{th}} = 0.326378, && D^{(k)}_{\text{sim}} = 0.326378, \\
    &                             && D^{(r)}_{\text{th}} = -0.138479, && D^{(r)}_{\text{sim}} =-0.138479, \\
    &                             && [(C_{kr} + C_{rk})/2]_{\text{th}} = 0.0547778, && [(C_{kr} + C_{rk})/2]_{\text{sim}} = 0.0547778, \\
    &                             && [(C_{rn} + C_{nr})/2]_{\text{th}} = 0, && [(C_{rn} + C_{nr})/2]_{\text{sim}} =  -1.10^{-6}, \\
    &                             && [(C_{nk} + C_{kn})/2]_{\text{th}} = 0, && [(C_{nk} + C_{kn})/2]_{\text{sim}} = 2.10^{-10}.
    \label{eq:D markers e-e+ final}
\end{align}

\subsection{$t \bar t$ production from scalar and vector resonance decays}

Up to this point, we have examined common processes whose density matrices depend on the kinematics. We now turn to simpler cases: resonant decays of scalars, pseudoscalars, vector bosons, axial vectors, etc. The advantage of these processes is that their density matrices are constant and can therefore be easily compared with results. They also provide a stringent test of the code ability to handle particles of different spin.

The first process we study is the decay $\phi \to t \bar t$, where $\phi$ denotes a mixed scalar-pseudoscalar state characterised by a mixing angle $\alpha$ and a mass $M_\phi \geq 2 m_t$.

\begin{figure}
    \centering
    \includegraphics[width=0.25\linewidth]{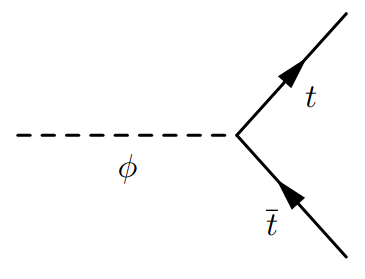}
    \caption{Diagram of the decay of a heavy particle, mixed scalar-pseudoscalar into $t\bar t$ in the SM at LO}
    \label{fig:placeholder}
\end{figure}

The interaction of this particle with the top quark is
\begin{equation}
    \mathcal{L}_{\text{int}} = -\frac{y_t}{\sqrt{2}} \phi\,  \bar t (\cos\alpha + i \sin\alpha \gamma^5)t + \text{h.c.}.
    \label{eq:Lagrangian_Higgs_Characterisation}
\end{equation}
The density matrix can be easily calculated analytically giving
\begin{equation}
    \rho = \frac{1}{2}\begin{pmatrix}
        0 &0 &0&0\\0&1&  -\frac{\beta c_\alpha - is\alpha}{\beta c_\alpha + is_\alpha} & 0\\ 0 &  -\frac{\beta c_\alpha + is\alpha}{\beta c_\alpha - is_\alpha} & 1&0\\0&0&0&0
    \end{pmatrix},
\end{equation}
where $c_\alpha = \cos\alpha$, $s_\alpha = \sin\alpha$ and $\beta = \sqrt{1 - 4 m^2_t/M_\phi^2}$. The associated spin-correlation matrix is 
\begin{equation}
    C =     \begin{pmatrix}
        -\frac{\beta^2c_\alpha^2 - s_\alpha^2}{\beta^2 c_\alpha^2 + s_\alpha^2} & \frac{\beta s_{2\alpha}}{\beta^2 c_\alpha^2 + s_\alpha^2} & 0\\ \frac{-\beta s_{2\alpha}}{\beta^2 c_\alpha^2 + s_\alpha^2} & -\frac{\beta^2c_\alpha^2 - s_\alpha^2}{\beta^2 c_\alpha^2 + s_\alpha^2} & 0\\ 0& 0 & -1
    \end{pmatrix}.
\end{equation}
One can observe that if $\alpha \neq 0$, the spin--correlation matrix is not symmetric. This reflects the fact that the presence of a pseudoscalar component in the interaction leads to CP violation. As mentioned earlier, these matrices do not depend on the kinematics of the process when $\alpha \in \{0, \pi/2\}$.

Instead of comparing the concurrence, as done previously, we consider the density matrices and the spin--correlation matrices for kinematically independent processes. We have compared the expressions with the simulated data for varying values of $(\alpha, \beta)$ and expect identical results. For $\alpha = 0$, corresponding to a pure scalar $\phi$, the system is independent of $\beta$, and the \mg matrices reproduce the ones exactly:

\begin{align}
&\rho_{\rm MG}(\alpha = 0) = \frac{1}{2}\begin{pmatrix} 0 & 0&0&0\\ 0&1&-1&0\\0&-1&1&0\\0&0&0&0 \end{pmatrix} = \rho_{\rm TH}(\alpha = 0) 
\end{align}
\begin{align}
    &C_{MG}(\alpha = 0) = \begin{pmatrix}-1 &0&0\\0&-1&0\\0&0&-1\end{pmatrix} = C_{TH}(\alpha = 0).
\end{align}

We next consider the  case $\alpha = \pi/2$, $\phi$ is then a pure pseudo-scalar, the system remains independent of $\beta$ and the matrices calculated with \mg match exactly the ones:
\begin{align}
&\rho_{\rm MG}(\alpha = \pi/2)     = \frac{1}{2}\begin{pmatrix} 0 & 0&0&0\\ 0&1&1&0\\0&1&1&0\\0&0&0&0 \end{pmatrix} = \rho_{\rm TH}(\alpha = \pi/2),
\end{align}
\begin{align}
    &C_{MG}(\alpha = \pi/2) = \begin{pmatrix}1 &0&0\\0&1&0\\0&0&-1\end{pmatrix} = C_{TH}(\alpha = \pi/2).
\end{align}

Finally, we test this process for intermediate values of $\alpha$, where the matrices depend on $\beta$. We therefore select two representative points in phase space to illustrate that the code behaves as expected. The first point corresponds to $\alpha = \pi/3$ and $\beta = 0.7257106861553025$ ($M_\phi = 500~\text{GeV}$):

\begin{align}
    &\rho_{\rm MG}  = \begin{pmatrix} 0 & 0&0&0\\ 0&0.5&0.350664199+0.356419163i&0\\0&0.350664199-0.356419163i&0.5&0\\0&0&0&0 \end{pmatrix} = \rho_{\rm TH}.
\end{align}
We compute the relative error with respect to the calculation for both the real and imaginary parts of the off--diagonal terms, in order to verify the numerical precision of the computation. The relative error is found to be of order $10^{-12}$--$10^{-13}$ for all elements, indicating that the calculation can be regarded as exact for this point.

The second point is $\alpha = \pi/6$ and $\beta = 0.5102940328869229$ ($M_\phi = 400$ GeV):
\begin{align}
    &\rho_{MG}  = \begin{pmatrix} 0 & 0&0&0\\ 0&0.5&0.061419268+0.496213335i&0\\0&0.061419268-0.496213335i&0.5&0\\0&0&0&0 \end{pmatrix}=\rho_{TH}.
\end{align}
The relative error for each element of this density matrix is also of order $10^{-12}$--$10^{-13}$, so the calculation can be regarded as exact. These comparisons demonstrate that the code can compute the density matrix of basic qubit processes with extremely high precision, achieving relative errors at the level of $10^{-12}$--$10^{-13}$.

Having verified that the code performs correctly for the decay of scalar and pseudoscalar particles, we now turn to the decay of a massive vector boson. As in the scalar case, we consider the simple process of a vector boson decaying into a top quark and an antitop quark. The vector boson is taken to be a mixture of vector and pseudovector components, with respective coupling factors $g_V$ and $g_A$. The corresponding interaction Lagrangian for this decay is
\begin{equation}
    \mathcal{L}_{\text{int}} = \frac{g}{\cos\theta_W} V_\mu \bar t (g_V - g_A \gamma^5) \gamma^\mu t + \text{h.c.},
\end{equation}
where
\begin{align}
    g_V = \frac{1}{2}T^3 - Q\sin^2\theta_W, && g_A = \frac{1}{2}T^3, && g = \frac{e}{\sin\theta_W},
\end{align}
with $e$ the electron charge, $T^3$ is  the weak-isospin projection, and $\theta_W$ the weak mixing angle.

\begin{figure}[hbt]
    \centering
    \includegraphics[width=0.25\linewidth]{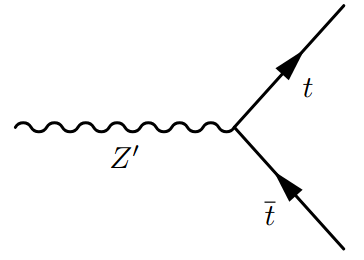}
    \caption{Diagram of the decay of a heavy $Z^\prime$ boson into $t\bar t$ with $M_Z^\prime \geq 2 m_t$}
    \label{fig:diagram_decay_z_ttbar}
\end{figure}

One can note that no vector boson in the Standard Model is massive enough to produce a $t \bar t$ pair on shell. To make the process possible, we therefore introduce a hypothetical boson $Z^\prime$ with a mass sufficiently large, i.e.\ $M_{Z^\prime} \geq 2 m_t$. 

The main difference with respect to the previous decay is that the initial state now admits three possible polarisation states, each yielding a distinct density matrix. To obtain the decay of an unpolarised massive vector boson, we therefore average over the polarisations. The density matrices corresponding to the three possible polarisation states of the $Z^\prime$ are:

\begin{align}
    \rho_- = \begin{pmatrix}
        1 & 0 & 0 & 0 \\ 0 & 0 & 0 & 0 \\ 0 & 0 & 0 & 0 \\ 0 & 0 & 0 & 0
    \end{pmatrix}, && \rho_0 = \frac{1}{2} \begin{pmatrix}
        0 & 0 & 0 & 0 \\ 0 & 1 & -1 & 0\\ 0 & -1 & 1 & 0\\ 0 & 0 & 0 & 0
    \end{pmatrix}, &&
    \rho_+ = \begin{pmatrix}
        0 & 0 & 0 & 0 \\ 0 & 0 & 0 & 0 \\ 0 & 0 & 0 & 0 \\ 0 & 0 & 0 & 1
    \end{pmatrix}.
\end{align}
The density matrix of the unpolarised boson is obtained by averaging the individual density matrices with their matrix element 
\begin{equation}
    \rho_{\rm tot} = \frac{1}{|\mathcal{M}_-|^2 + |\mathcal{M}_0|^2 + |\mathcal{M}_+|^2} \left(|\mathcal{M}_-|^2 \rho_- + |\mathcal{M}_0|^2\rho_0 + |\mathcal{M}_+|^2 \rho_+ \right).
\end{equation}
This gives 
\begin{equation}
    \rho_{\rm tot} = \frac{1}{N} \begin{pmatrix}
        K_- & 0 & 0 & 0 \\ 0 & K_0 & -K_0 & 0 \\ 0 & -K_0 & K_0 & 0 \\ 0 & 0 & 0 & K_+
    \end{pmatrix},
\end{equation}
with
\begin{align}
    &N = (1 - 4Q\sin^2\theta_W)^2(1 - \beta^2) + (1 - 4Q\sin^2\theta_W + \beta)^2 + (1 - 4Q\sin^2\theta_W - \beta)^2,\\
    &K_- = (1 - 4Q\sin^2\theta_W + \beta)^2, \\
    &K_0 = \frac{1}{2}(1 - 4Q\sin^2\theta_W)^2(1 - \beta^2),\\
    &K_+ = (1 - 4Q\sin^2\theta_W - \beta)^2,
\end{align}
where $\beta = \sqrt{1 - 4m_t^2/M_{Z^\prime}^2}$.

We can now compare the density matrix with the results obtained from \mg. When only $M_{Z^\prime}$ is varied, the density matrix depends solely on the value of $\beta$. We  perform the comparison in three regions of the phase space: at threshold ($\beta \to 0$), in the high-energy region ($\beta \to 1$), and in an intermediate region. For $\beta = 0.024037142$ ($M_{Z^\prime} = 346.1$ GeV) we have
\begin{align}
    &\rho_{\rm TH} = \begin{pmatrix}
        0.373039660 & 0 & 0 & 0 \\
        0 & 0.166216523 & -0.166216523 & 0 \\
        0 & -0.166216523 & 0.166216523 & 0 \\
        0 & 0 & 0 & 0.294527295
    \end{pmatrix} = \rho_{\rm MG},
\end{align}
For $\beta = 0.721897500$ ($M_{Z^\prime} = 500$ GeV) we have
\begin{align}
    &\rho_{\rm TH} = \begin{pmatrix}
        0.877267800 & 0 & 0 & 0 \\
        0 & 0.027331383 & -0.027331383 & 0 \\
        0 & -0.027331383 & 0.027331383 & 0 \\
        0 & 0 & 0 & 0.068069435
    \end{pmatrix} =\rho_{\rm MG},
\end{align}

For $\beta = 0.997602808$ ($M_{Z^\prime} = 5000$ GeV) we have
\begin{align}
    &\rho_{\rm TH} = \begin{pmatrix}
        0.849681224 & 0 & 0 & 0 \\
        0 & 0.000171017 & -0.000171017 & 0 \\
        0 & -0.000171017 & 0.000171017 & 0 \\
        0 & 0 & 0 & 0.149976742
    \end{pmatrix} = \rho_{\rm MG}
\end{align}
where the equalities are true up to a precision of $10^{-6}$.

These comparison points demonstrate that \mg accurately computes the elements of the density matrix for the decay of a vector or axial--vector boson across the entire phase space (provided the boson mass is allowed to vary). The relative error for this process is consistently of order $10^{-6}$ or smaller for all matrix elements. Although this uncertainty remains very small and enables highly precise computations, a slight loss of precision is observed. It may be attributed to the averaging over different density matrices required for the final result.

\subsection{$VV$:  heavy Higgs decay and pair production at the LHC}
\label{sec:vv}

 Another interesting class of systems whose spin correlations merit investigation are qutrit systems, exemplified by massive vector bosons. These systems are more challenging to study, not only because their spin structure is more intricate, but also because no observable as powerful as the concurrence is available for them. Indeed, the concurrence is defined through a minimisation procedure, which prevents its (and even numerical) evaluation except in the case of two qubits.

\begin{figure}[H]
    \centering
    \includegraphics[width=0.25\linewidth]{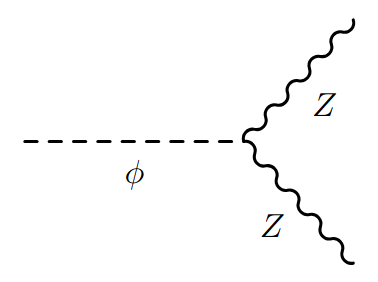}
    \caption{Feynman diagram of the process $\phi \to Z Z$ at LO where $\phi$ is heavy SM-like scalar of mass $M_\phi \geq 2 m_t$.}
    \label{fig:phi_decay}
\end{figure}

The first system we consider is the simplest one: the decay of a heavy SM--like scalar into a pair of massive vector bosons. In this study, we focus on the process $\phi \to Z Z$, although the results are identical for the $W$ boson since the $hZZ$ and $hW^+W^-$ vertices share the same tensor structure in the SM. At leading order, this process is described by a single Feynman diagram Fig. \ref{fig:phi_decay}.

The Lagrangian of interaction for this decay is simple
\begin{equation}
    \mathcal{L}_{\text{int}} = \frac{m_Z^2}{v} \phi Z^\mu Z_\mu.
\end{equation}

As for the other decays, we define the dimensionless quantity $\beta$, which is related to  vector--boson pair invariant mass.

\begin{equation}
    \beta = \sqrt{1 - \frac{4 M_Z^2}{M_\phi^2}}.
\end{equation}

Since massive vector bosons have three possible spin projections, rather than two as in the fermionic case, the density matrix for this process is a $9 \times 9$ matrix. We express it in the basis $\{\ket{--}, \ket{-0}, \ket{-+}, \ket{0-}, \ket{00}, \ket{0+}, \ket{+-}, \ket{+0}, \ket{++}\}$. The resulting density matrix for this process is given by:

\begin{equation}
    \rho = \frac{1}{3\beta^4 - 2\beta^2 + 3} \begin{pmatrix}
      (1 - \beta^2)^2 & 0 & 0 & 0 & -(1 - \beta^4) & 0 & 0 & 0 & (1 - \beta^2)^2 \\
      0 & 0 & 0 & 0 & 0 & 0 & 0 & 0 & 0 \\
      0 & 0 & 0 & 0 & 0 & 0 & 0 & 0 & 0 \\
      0 & 0 & 0 & 0 & 0 & 0 & 0 & 0 & 0 \\
      -(1 - \beta^4) & 0 & 0 & 0 & (1+\beta^2)^2 & 0 & 0 & 0 & -(1 - \beta^4) \\
      0 & 0 & 0 & 0 & 0 & 0 & 0 & 0 & 0 \\
      0 & 0 & 0 & 0 & 0 & 0 & 0 & 0 & 0 \\
      0 & 0 & 0 & 0 & 0 & 0 & 0 & 0 & 0 \\
      (1 - \beta^2)^2 & 0 & 0 & 0 & -(1- \beta^4) & 0 & 0 & 0 & (1 - \beta^2)^2
    \end{pmatrix}.
    \label{density matrix phi > Z Z}
\end{equation}

One can note that this density matrix depends only on $\beta$, and therefore on the mass of the heavy scalar. As done previously, we compare the result of Eq.~(\ref{density matrix phi > Z Z}) with the simulation results obtained using \mg in different regions of the phase space: the threshold region ($\beta \to 0$), the boosted region ($\beta \to 1$), and an intermediate region.

In the boosted regime, we consider $M_\phi = 1~\text{TeV}$, corresponding to $\beta = 0.983228862$ ($M_Z = 91.188~\text{GeV}$).

\begin{equation}
    \rho_{\rm TH} = \begin{pmatrix}
      0.000285844 & 0 & 0 & 0 &  -0.016902085 & 0 & 0 & 0 & 0.000285844 \\
      0 & 0 & 0 & 0 & 0 & 0 & 0 & 0 & 0 \\
      0 & 0 & 0 & 0 & 0 & 0 & 0 & 0 & 0 \\
      0 & 0 & 0 & 0 & 0 & 0 & 0 & 0 & 0 \\
      -0.016902085 & 0 & 0 & 0 & 0.999428312 & 0 & 0 & 0 & -0.016902085 \\
      0 & 0 & 0 & 0 & 0 & 0 & 0 & 0 & 0 \\
      0 & 0 & 0 & 0 & 0 & 0 & 0 & 0 & 0 \\
      0 & 0 & 0 & 0 & 0 & 0 & 0 & 0 & 0 \\
      0.000285844 & 0 & 0 & 0 & -0.016902085 & 0 & 0 & 0 & 0.000285844
    \end{pmatrix} =\rho_{\rm MG}.
\end{equation}

The relative error between the coefficients is extremely small: for each individual coefficient, taking the calculation as reference, the relative error is of order $10^{-12}$. The computation can therefore be regarded as exact.

In the threshold region, we take $M_\phi = 182.5$ GeV, so $\beta = 0.036857054$ ($M_Z = 91.188$ GeV).
\begin{equation}
    \rho_{\rm TH} = \begin{pmatrix}
      0.332729035 & 0 & 0 & 0 & -0.333634251 & 0 & 0 & 0 & 0.332729035 \\
      0 & 0 & 0 & 0 & 0 & 0 & 0 & 0 & 0 \\
      0 & 0 & 0 & 0 & 0 & 0 & 0 & 0 & 0 \\
      0 & 0 & 0 & 0 & 0 & 0 & 0 & 0 & 0 \\
      -0.333634251 & 0 & 0 & 0 & 0.334541930 & 0 & 0 & 0 & -0.333634251 \\
      0 & 0 & 0 & 0 & 0 & 0 & 0 & 0 & 0 \\
      0 & 0 & 0 & 0 & 0 & 0 & 0 & 0 & 0 \\
      0 & 0 & 0 & 0 & 0 & 0 & 0 & 0 & 0 \\
      0.332729035 & 0 & 0 & 0 & -0.333634251 & 0 & 0 & 0 & 0.332729035
    \end{pmatrix} = \rho_{\rm MG}.
\end{equation}

For this value of $\beta$, the relative error between the result and the value computed by \mg is even smaller, of order $10^{-16}$ for the individual elements of the density matrix.

The third and final point is in the intermediate region, we take $M_H = 250$ GeV, so $\beta = 0.683976545$.

\begin{equation}
    \rho_{\rm TH} = \begin{pmatrix}
      0.104086252 & 0 & 0 & 0 & -0.287085974 & 0 & 0 & 0 & 0.104086252 \\
      0 & 0 & 0 & 0 & 0 & 0 & 0 & 0 & 0 \\
      0 & 0 & 0 & 0 & 0 & 0 & 0 & 0 & 0 \\
      0 & 0 & 0 & 0 & 0 & 0 & 0 & 0 & 0 \\
      -0.287085974 & 0 & 0 & 0 & 0.791827496 & 0 & 0 & 0 & -0.287085974 \\
      0 & 0 & 0 & 0 & 0 & 0 & 0 & 0 & 0 \\
      0 & 0 & 0 & 0 & 0 & 0 & 0 & 0 & 0 \\
      0 & 0 & 0 & 0 & 0 & 0 & 0 & 0 & 0 \\
      0.104086252 & 0 & 0 & 0 & -0.287085974 & 0 & 0 & 0 & 0.104086252
    \end{pmatrix} = \rho_{\rm MG}.
\end{equation}
For this value of $\beta$, the relative error between the result and that computed by \mg is of order $10^{-13}$ for the individual elements of the density matrix.
The study of this simple example demonstrates that the code correctly handles systems composed of two qutrits. 

\par
The next validation step, after successfully reproducing the density matrix for decay processes, is to include the production stage. We therefore consider the production of a pair of weak bosons at the LHC. The analytical results for these processes have been derived in Ref.~\cite{Aoude:2023hxv}, and our simulated results will be compared against them. The processes under study are:
\begin{align}
\label{eq: processes VV 1}
    &p \, p \to W^+ \, W^-, && u \, \bar u \to W^+ \, W^- && d \, \bar d \to W^+ \, W^- \\
    &u \, \bar u \to Z \, Z && d \, \bar d \to Z \, Z, && u \bar d \to Z \, W^+,
\label{eq: processes VV 2}
\end{align}
where $u$ and $d$, respectively represent the up and down quark family in a five-flavour scheme: $u = \{u, c\}$, $d = \{d, s, b\}$.

The conventions adopted for the simulations closely follow those of Ref.~\cite{Aoude:2023hxv} to enable a direct comparison. Specifically, the production angle $\theta$ is defined between the antiquark $\bar q$ and the $W^+$, or between the antiquark $\bar q$ and the $Z$ in the case of $Z$--boson pair production. The reference frame used is the helicity frame defined in App.~\ref{app:hel}: it is boosted to the rest frame of the boson pair, with the reference direction for the helicity basis taken along the $W^+$ boson (or the $Z$ in the case of $ZZ$ production). The simulations employ the latest \texttt{NNPDF4.0} NNLO PDF set~\cite{Ball_2022}, which is the one used in Ref. \cite{Aoude:2023hxv}, and the input parameters are:
\begin{align}
    m_W = 80.377\, \text{GeV}, && &m_Z = 91.1876\, \text{GeV}, \\
    m_h = 125.35\, \text{GeV}, && &G_f = 1.166378 .10^{-5}\, \text{GeV}.
\end{align}

\begin{figure}[t!]  
 \begin{minipage}{0.41\textwidth}
   \includegraphics[width=\textwidth]{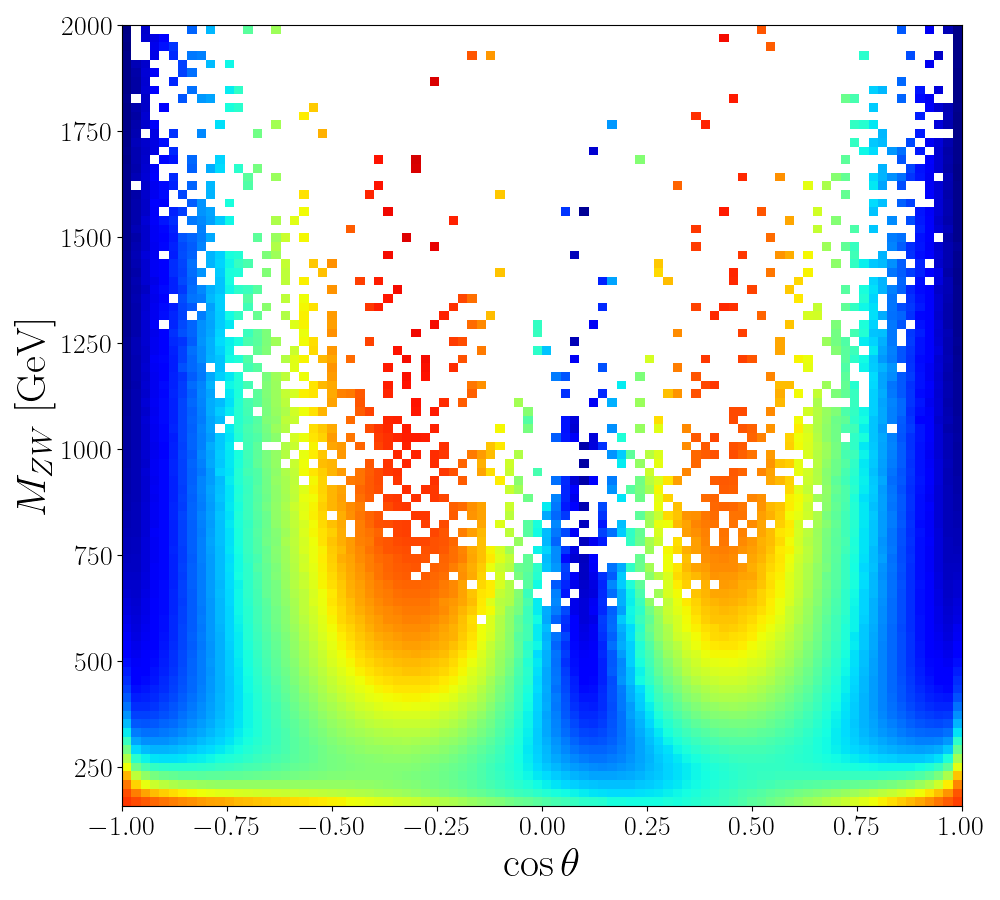}
  \end{minipage}
  \hfill
  \begin{minipage}{0.41\textwidth}
   \includegraphics[width=\textwidth]{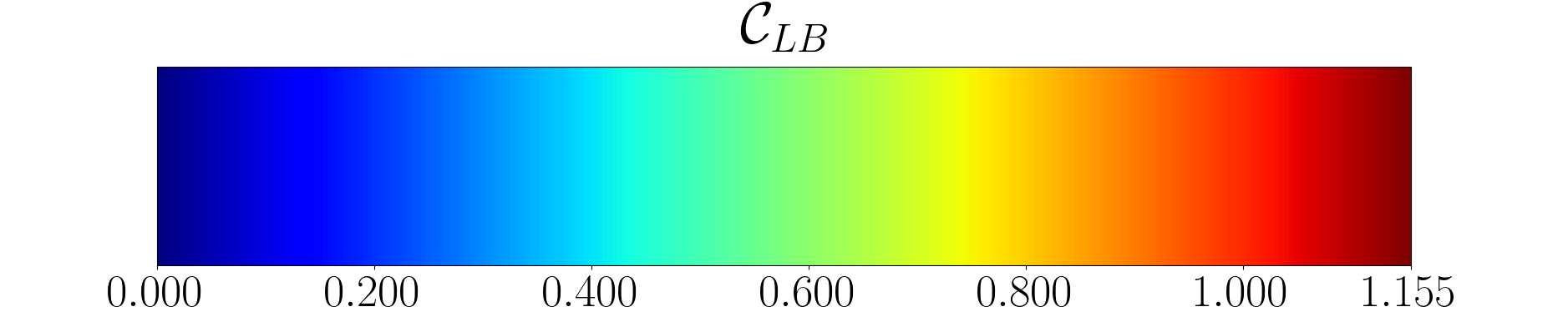}
  \end{minipage}
\end{figure}
\begin{figure}[ht]  
 \begin{minipage}{0.41\textwidth}
   \includegraphics[width=\textwidth]{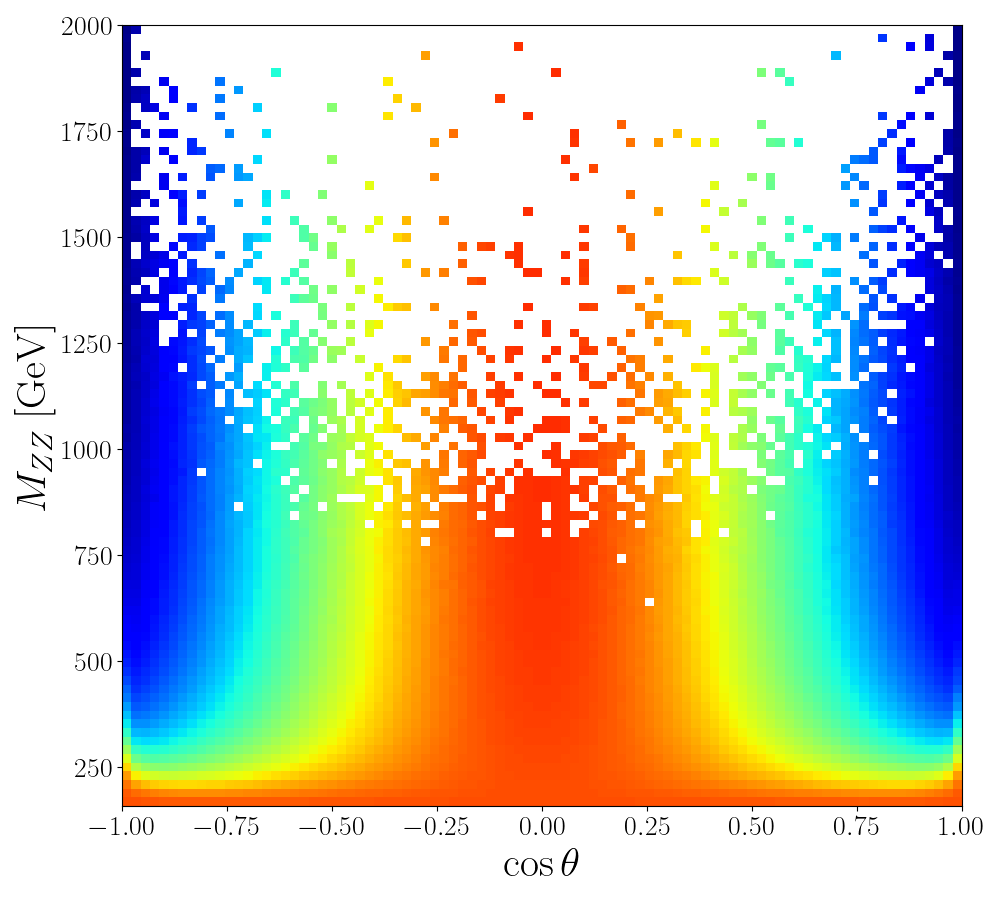}
  \end{minipage}
  \hfill
  \begin{minipage}{0.41\textwidth}
   \includegraphics[width=\textwidth]{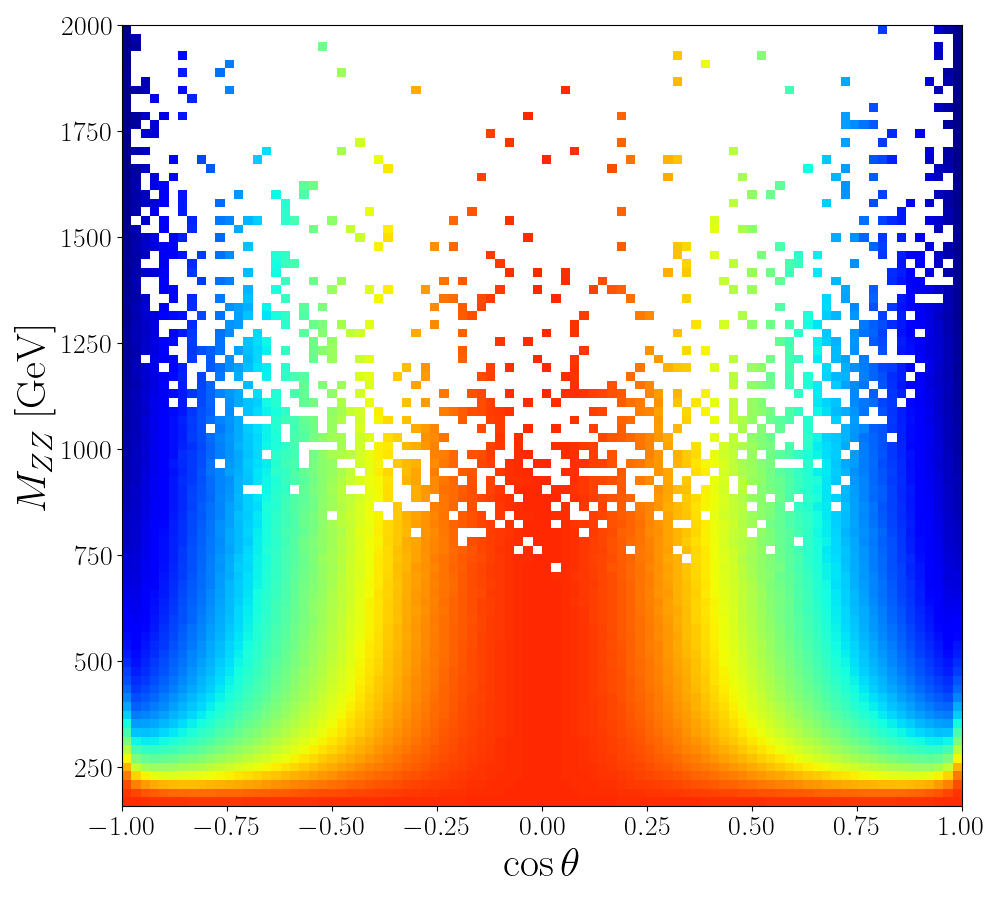}
  \end{minipage}
\end{figure}    
\begin{figure}[ht]  
 \begin{minipage}{0.41\textwidth}
   \includegraphics[width=\textwidth]{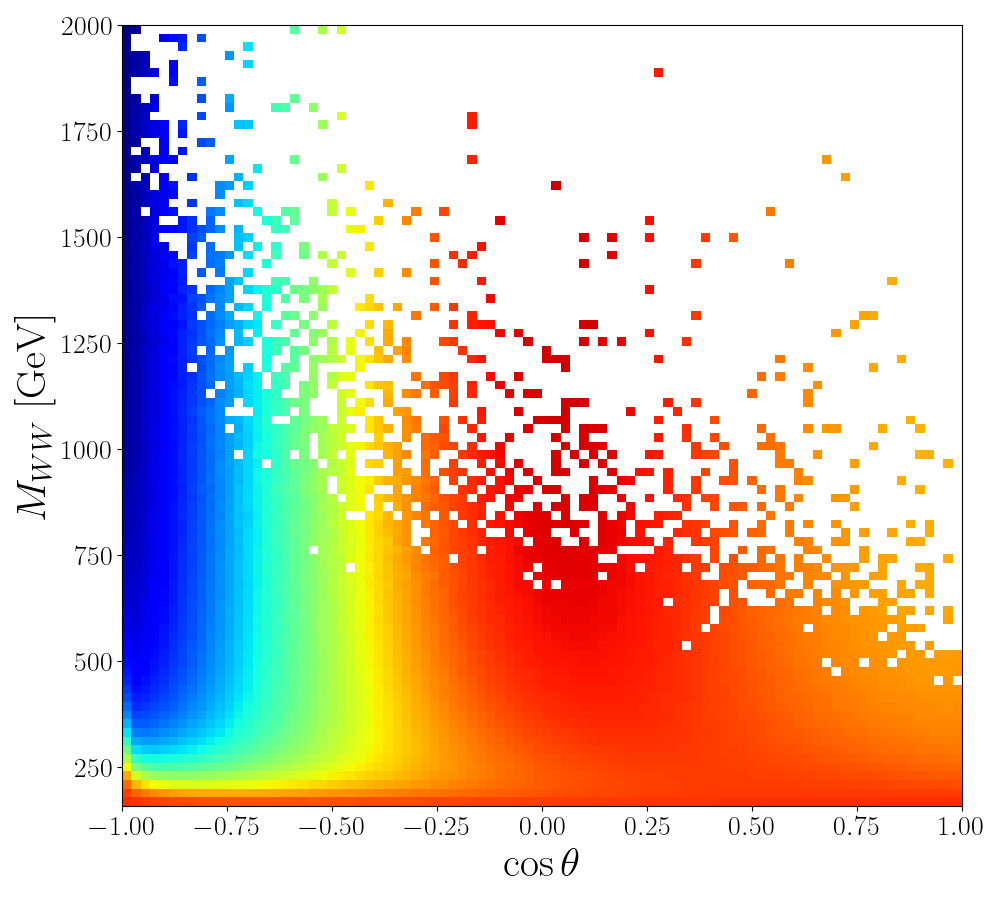}
  \end{minipage}
  \hfill
  \begin{minipage}{0.41\textwidth}
   \includegraphics[width=\textwidth]{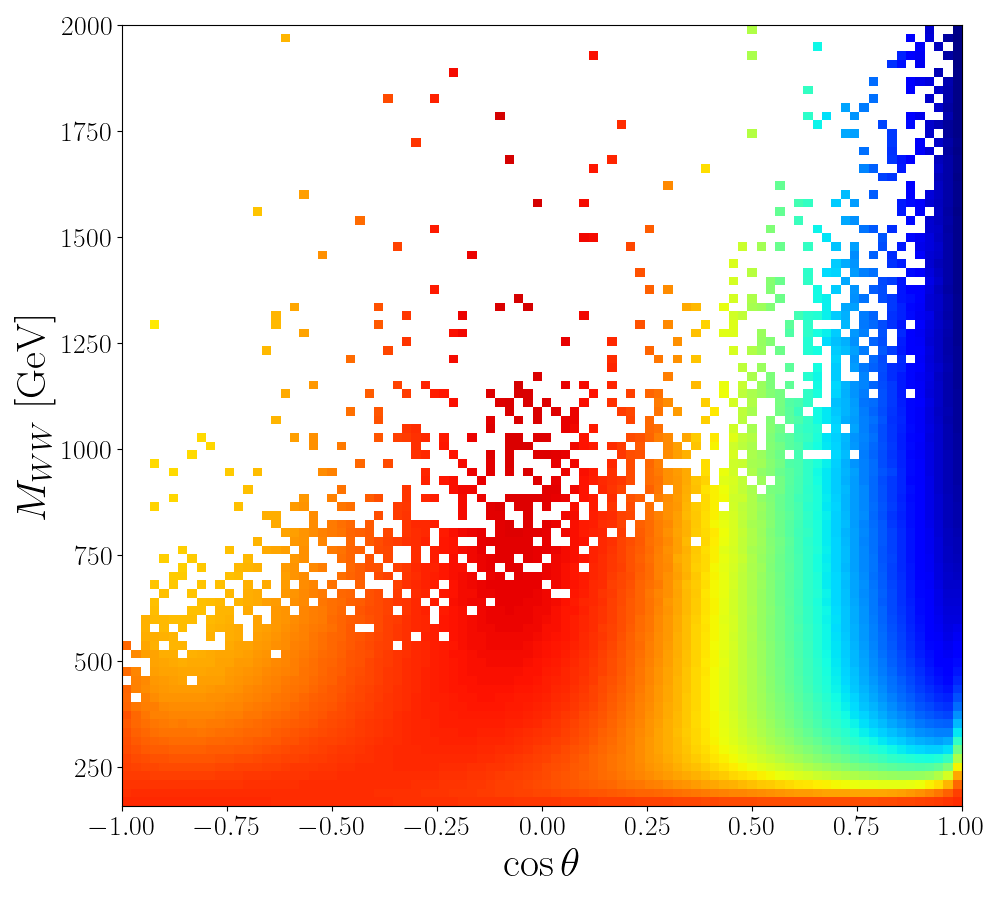}
  \end{minipage}
 \caption{Lower bound of the concurrence $\mathcal{C}_{\textrm{LB}}$ over the phase space spanned by $\theta$ (\ref{eq: theta1}) and $M_{VV}$ generated by \mg in the partonic channels of the production of weak bosons in the LHC. The angle $\theta$ is defined in the rest-frame of the weak bosons as the angle between the anti-quark $\bar q$ and $W^+$ (or $Z$ in the $ZZ$ channel). (upper-left) Channel $u\,\bar d \to W^+ \,Z$. (middle-left) Channel $u\,\bar u \to Z \,Z$. (middle-right) Channel $d\,\bar d \to Z \,Z$. (bottom-left) Channel $u\,\bar u \to W^+ \,W^-$ . (bottom-right) Channel $d\,\bar d \to W^+ \,W^-$.}
        \label{fig: partonic channels weak bosons}
\end{figure}    
\clearpage

\par
In the case of qutrits, the concurrence observable exhibits distinct properties. First, it is not known how to compute it exactly for systems composed of a pair of qutrits; however, upper and lower bounds can be evaluated to estimate its value. Second, unlike for qubits, the concurrence for qutrit pairs can exceed unity. The definitions of these bounds, following Ref.~\cite{Aoude:2023hxv}, are recalled here and discussed in more detail in App.~\ref{app:QO}:
\begin{align}
    &\mathcal{C}_{LB}^2 \equiv 2 \max (0, \Tr[\rho^2] - \Tr[\rho_A^2], \Tr[\rho^2] - \Tr[\rho_B^2]), \\
    &\mathcal{C}_{UB}^2 \equiv 2 \min (1 - \Tr[\rho_A^2], 1 - \Tr[\rho_B^2]),
\end{align}
where $\rho_A$ and $\rho_B$ are the density matrices of each particle in the pair. They are obtained by tracing out one of the Hilbert spaces, $\mathcal{H}_A$ (or $\mathcal{H}_B$), from the total density matrix $\rho$ defined on $\mathcal{H} = \mathcal{H}_A \otimes \mathcal{H}_B$.

\par
Figure~\ref{fig: partonic channels weak bosons} shows the lower bound of the concurrence over the phase space $(\theta, m_{VV})$, where $V$ denotes a weak boson, for the five partonic processes defined in Eqs.~(\ref{eq: processes VV 1}--\ref{eq: processes VV 2}). The different partonic channels are independent from one another and differ slightly in their couplings to the $Z$ boson, since the weak isospin takes different values for the down- and up-quark families. This results in small variations in the corresponding density matrices and motivates plotting both channels separately. Note that the up-quark channel is not the mirror image of the down-quark channel, in agreement with the results.  

Because of the distribution of events across the phase space, we did not scan the entire region as was done for the process $p p \to t \bar t$. This omission,  however, has negligible impact, since the unprobed regions do not contribute significantly to the total cross-section. The plots in Fig.~\ref{fig: partonic channels weak bosons} reproduce very accurately those of Fig.~11 in Ref.~\cite{Aoude:2023hxv}, demonstrating that our code accurately reproduces the theoretical predictions. Finally, as an additional validation, we compute various quantum--information observables for the full process $p p \to W^+ W^-$ to verify that the code correctly accounts for the combined contributions from different partonic channels, as previously done for $p p \to t \bar t$.

\par
For $t \bar t$ production, we have studied the concurrence directly, whereas the $ZZ$, $WW$, and $ZW$ processes we computed upper and lower bounds on the concurrence to estimate its value. In the case of $t \bar t$ production, we also evaluated the quantity known as {magic}, which, however, cannot be defined for $3\times3$ systems such as pairs of massive vector bosons. A possible analogue for magic is the quantity known as {mana}. To our knowledge, this observable has not previously been applied in collider settings; we therefore compute it for a realistic process to demonstrate its feasibility and correct implementation in our code.  

\begin{figure}[!htp]\centering
\subfloat[]{\label{}\includegraphics[width=.49\linewidth]{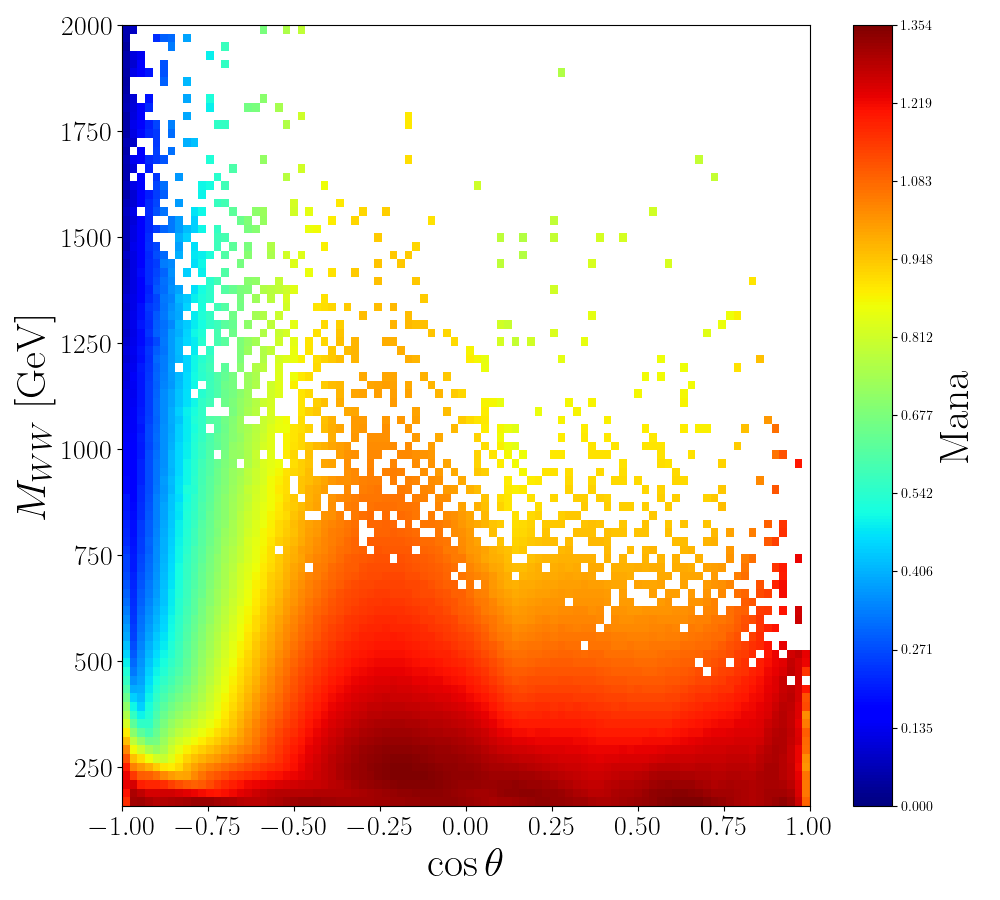}}\hfill
\subfloat[]{\label{}\includegraphics[width=.49\linewidth]{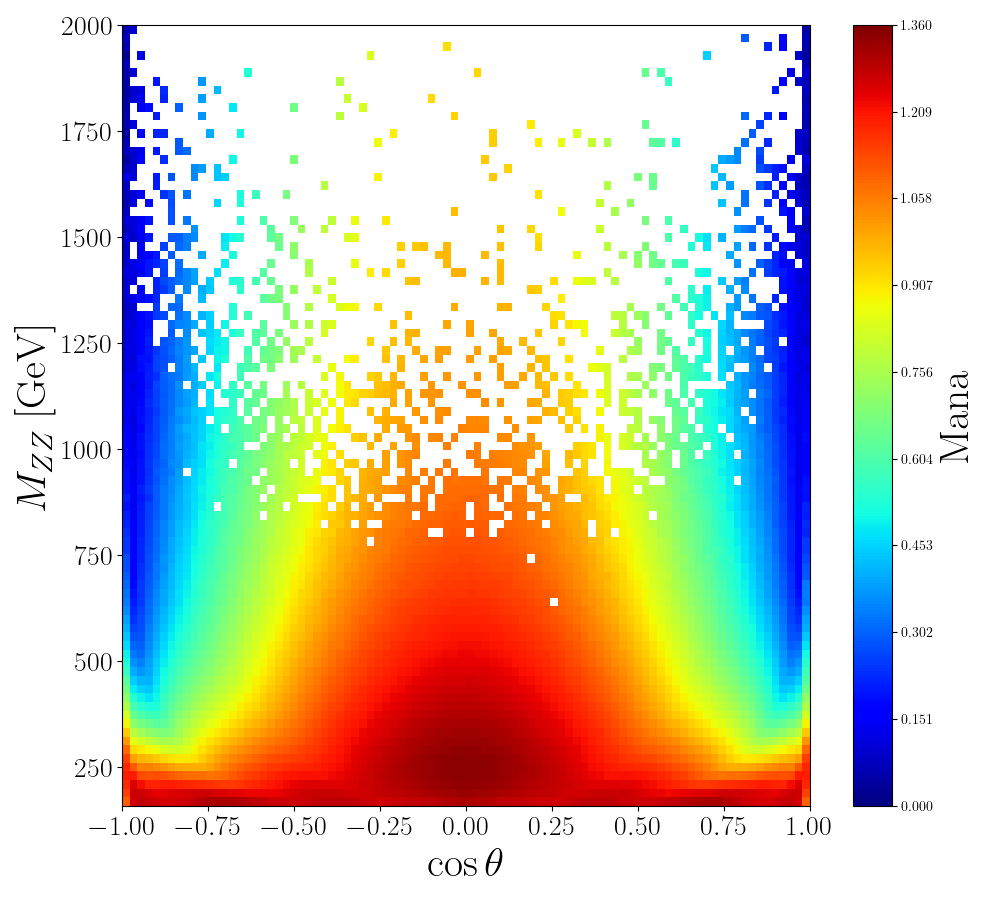}}\par
\caption{Mana over the phase space spanned by $M_{VV}$ and $\cos\theta$ (\ref{eq: theta1}) for $u\;\bar u \to W^-\;W^+$ (a) and $d\;\bar d \to Z\;Z$ (b). These plots show the same trend as the plots of the lower bound of concurrence.}
\label{fig:Mana_VV}
\end{figure}

Figure~\ref{fig:Mana_VV} shows the values of the {mana} for $WW$ and $ZZ$ production in the up--quark--initiated channel. These distributions exhibit the same overall trend as the plots of the lower bound of the concurrence: states with higher concurrence also tend to display higher mana. This is not always the case for the magic, for which separable states can exhibit maximal values. A more detailed study of collider processes using mana would be of interest to better understand its behaviour and physical interpretation.

\section{Applications}
\label{sec:applications}

Having established numerical agreement in all validation tests, we now turn to phenomenological applications at the LHC. These include fiducial predictions with realistic selections, distributions of polarisations and correlation tensors, and QI observables, together with an estimate of statistical uncertainties from finite event samples. Where relevant, we also examine the sensitivity of basis-dependent quantities and highlight basis-independent measures that provide a robust characterisation of the produced quantum states.

\subsection{$t\bar t W$ production at the LHC}
\label{sec:ttw}
\begin{figure}[t!]
    \centering
    \includegraphics[width=0.8\linewidth]{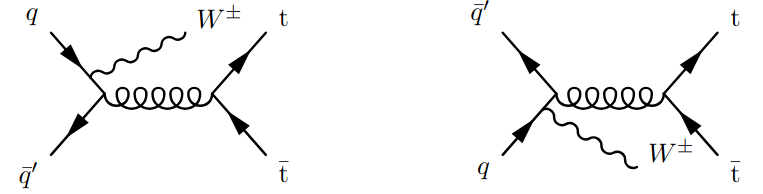}
    \caption{Feynman diagrams of the process $q \, \bar q^\prime \to t \, \bar t \, W^\pm$ at tree-level.}
    \label{fig:feynman diagrams q q' > t tbar Wpm}
\end{figure}

The first quantum system we analyse is $t\bar t$ in association with a $W$ boson:
\[
q\,\bar q^{\prime}\to t\,\bar t\,W^{\pm}.
\]
At tree level (LO in $g_s^2 g_W$) the amplitude is generated by two diagrams, shown in Fig.~\ref{fig:feynman diagrams q q' > t tbar Wpm}, with the $W$ radiated from the initial-state quark line while the $t\bar t$ pair is created via an $s$-channel gluon. This process is closely related to inclusive $t\bar t$ production (Sect.~\ref{sec:tt}), and we will systematically compare the two.
Two features are particularly relevant for the spin-quantum structure:
\begin{itemize}
  \item \textbf{Initial state:} the $gg$ channel that dominates inclusive $t\bar t$ at the LHC is {absent} at LO in $t\bar tW^{\pm}$. As a result, the strong near-threshold entanglement driven by gluon fusion in $t\bar t$ is not present here; we therefore expect sizeable entanglement primarily in the high-energy regime, where helicity selection enhances off-diagonal spin coherences.
  \item \textbf{Chiral emission:} the $W$ is emitted only from the initial-state quarks and couples through the left-handed charged current. This V-A structure induces a non-trivial polarisation of the interacting quark leg, effectively preparing a {polarised} $q\bar q'$ initial state for $t\bar t$ production. Consequently, the one-body polarisations $\vec B_{t,\bar t}$ and the correlation tensor $C_{ij}$ of the $t\bar t$ subsystem differ qualitatively from the unpolarised $q\bar q$ or $gg$ cases, with characteristic patterns in the helicity basis $\{\hat n,\hat r,\hat k\}$.
\end{itemize}

In what follows we construct the event-by-event production matrix $R$ for the $t\bar t$ subsystem within $q\,\bar q^{\prime}\to t\,\bar t\,W^{\pm}$, extract the normalised density matrix $\rho$, and study its Fano coefficients $(\vec B_t,\vec B_{\bar t},C)$. We then compare basis-independent diagnostics (eigenvalues of $\rho$, singular values of $C$, and two-qubit entanglement measures such as the concurrence and the magic) to the corresponding quantities in inclusive $t\bar t$ production, highlighting the impact of the chiral initial state and the absence of the $gg$ channel at LO.

In the study of the process $p p \to t \bar t$, we have seen that the top quarks are unpolarised. It is well known that the top quarks inherit their polarisation from the initial--state quarks, which in turn arises from the emission of the $W^\pm$ boson~\cite{Maltoni:2014zpa}. We now examine this difference in more detail, employing various quantum--information observables, such as the concurrence, to characterise the distinctions between the two processes. Moreover, since this process involves a new final--state particle, we can study the spin correlations between the top quark and the $W^\pm$ boson. This requires introducing new observables, as the concurrence is no longer computable in this case.

The data used in this study were generated with \mg with a center of mass energy of $13$ TeV in order to simulate processes from the LHC. A 5 flavour scheme is used so we consider all quarks except the top quark to be massless, the PDF used is the \texttt{NNPDF4.0 NNLO PDF} \cite{Ball_2022} and the input parameters used for the data generation are 
\begin{align*}
    m_Z =91.118\;\textrm{GeV} , && G_F=1.166390.10^{-5} \;\textrm{GeV}^{-2}, && \alpha_{EM}^{-1} = 1.325070.10^{2}.
\end{align*}

\par
The density matrix of two top quarks can be parametrised following Eq.~(\ref{eq: rho 2 tops}) with $B_1$ and $B_2$ the polarisation vectors of each of the top quarks. Because of the linearity of the trace, one can apply formula (\ref{eq:definition Fano coefficients}) on the average density density matrix or average the individual polarisation in the same way. For this process we have three external observable particles, we can thus study the spin correlations between any of the three pairs $t \bar t$, $t W^\pm$ and $\bar t W^\pm$. We will first study the system made of the pair of top quarks in order to compare it to the known process $q \, \bar q \to t \, \bar t$, in this case the $W$ boson can be seen as a way to polarise the initial state. Then we will study the spin correlations between the weak boson and any of the top quarks.

\paragraph*{Quantum spin correlations in $t \bar t$}
The goal of this first study is to compare with the process $q \bar q \to t \bar t$. We recall that, for this process, the polarisations of the top quarks vanish, as a general consequence of the vector nature of the strong interaction. In the present case, however, a weak-interaction vertex is introduced which, unlike the strong interaction, is sensitive to helicity and therefore induces a net polarisation in the final state. The average polarisations of the two top quarks are given by:

\begin{align}
    \langle S_t \rangle = \begin{pmatrix}
        -1.63 .10^{-5} \\ 0.603 \\ - 2.78 .10^{-4}
    \end{pmatrix}, && \langle S_{\bar t} \rangle = \begin{pmatrix}
        1.63 .10^{-5} \\ -0.603 \\ - 2.78 .10^{-4}
    \end{pmatrix}.
\end{align}

The spin-correlation matrix also becomes dense, as correlations now appear in all directions, unlike in the original process. 

\par
To explore this further, we compare the values of different observables, such as the concurrence and the purity, with those of the original process over the phase-space $(\cos\theta, M_{tW^-})$. However compared to all other processes studied up until now that were $2 \to 2$ processes, this process has $3$ particles in the final state. This means that there is an ambiguity on the definition of the angle $\theta$ used to parametrise the phase-space, indeed $\theta$ can be defined as the angle between the top in the centre-of-mass frame of $tW^-$ and the beam axis in the lab frame $\hat z$ or between the top in the centre-of-mass frame of $tW^-$ and the beam axis in the boosted frame. Both these definitions are the same for $2 \to 2$ processes as explained in App.~\ref{app:hel} but not for more complex processes. For the rest of this paper, we have decided to use the beam axis in the lab frame $\hat z$ as reference for the definition of $\theta$, the definition can be found in equation (\ref{eq: theta1}). Differences between the two definitions are visible but do not change the trend and thus the interpretation of the plots and will not be discussed in the rest of the paper.

\par
\begin{figure}[!htp]\centering
\subfloat[]{\label{fig:Concurrence qq_ttbarwmp1}\includegraphics[width=.49\linewidth]{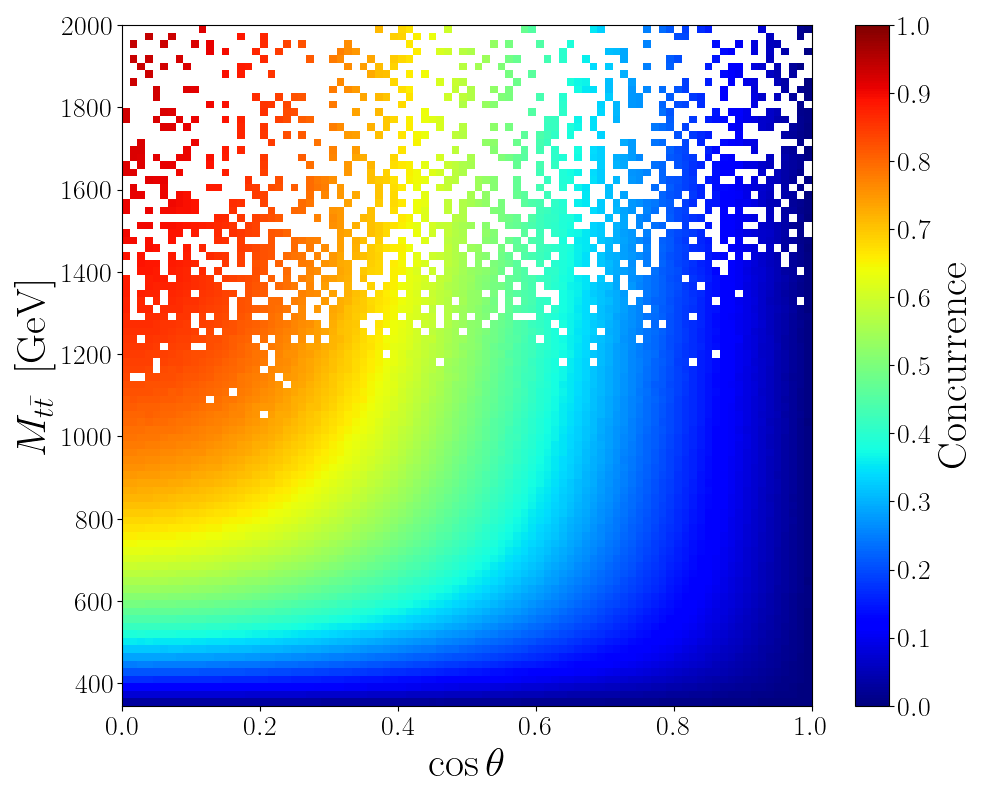}}\hfill
\subfloat[]{\label{fig:Concurrence qq_ttbarwmp2}\includegraphics[width=.49\linewidth]{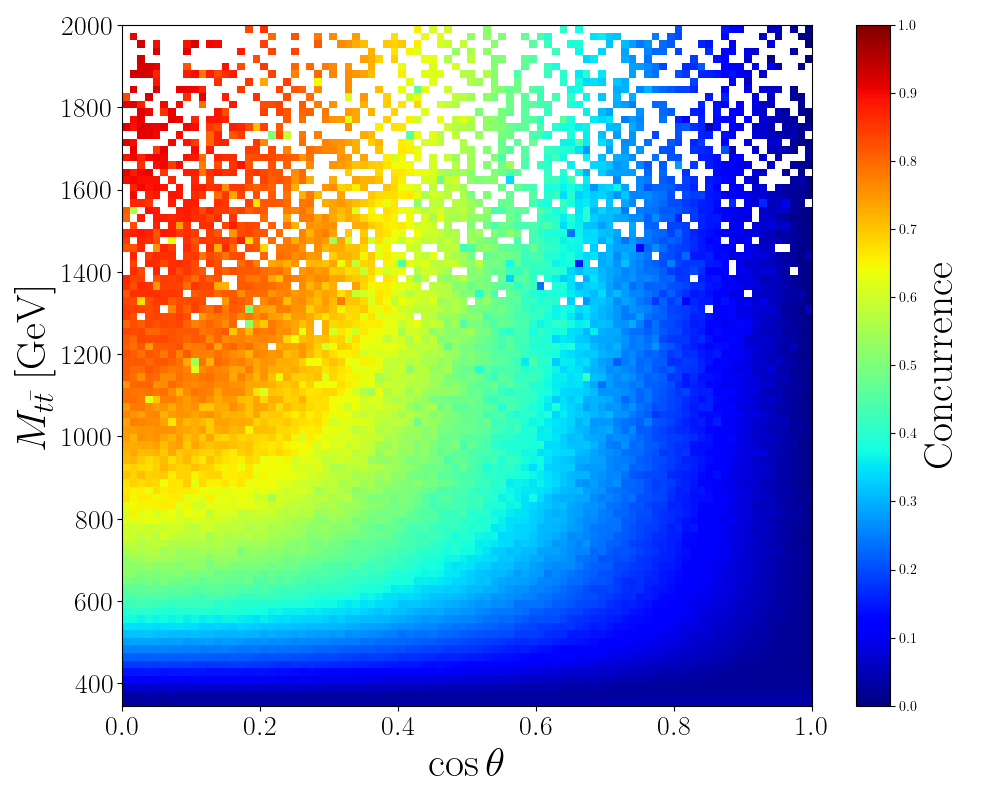}}\par
\caption{Concurrence of $q\,\bar q \to t\,\bar t$ (a) and of $q\,\bar q^\prime \to t \, \bar t \, W^\pm$ (b) over the phase space spanned by $M_{t\bar t}$ and $\cos\theta$ (\ref{eq: theta1}) in the centre-of-mass frame of $t\bar t$. The histograms are nearly identical showing that polarising the initial state does not seem to impact significantly the level of entanglement of the final state for this process.}
\label{fig:Concurrence qq_ttbarwmp}
\end{figure}

Figure~\ref{fig:Concurrence qq_ttbarwmp} shows the concurrence over the phase space for the processes $q \bar q \to t \bar t$ (left) and $q \bar q^\prime \to t \bar t W^\pm$ (right). The two histograms are nearly identical, indicating that the polarisation of the initial state does not significantly affect the level of entanglement in the final state for this process. Note that for the process $q \bar q^\prime \to t \bar t W^\pm$, the histogram appears less smooth, which is due to the presence of two contributing channels, compared with only one in the simpler $t \bar t$ process.

\begin{figure}[!htp]\centering
\subfloat[]{\label{fig:Purity qq_ttbarwmp1}\includegraphics[width=.49\linewidth]{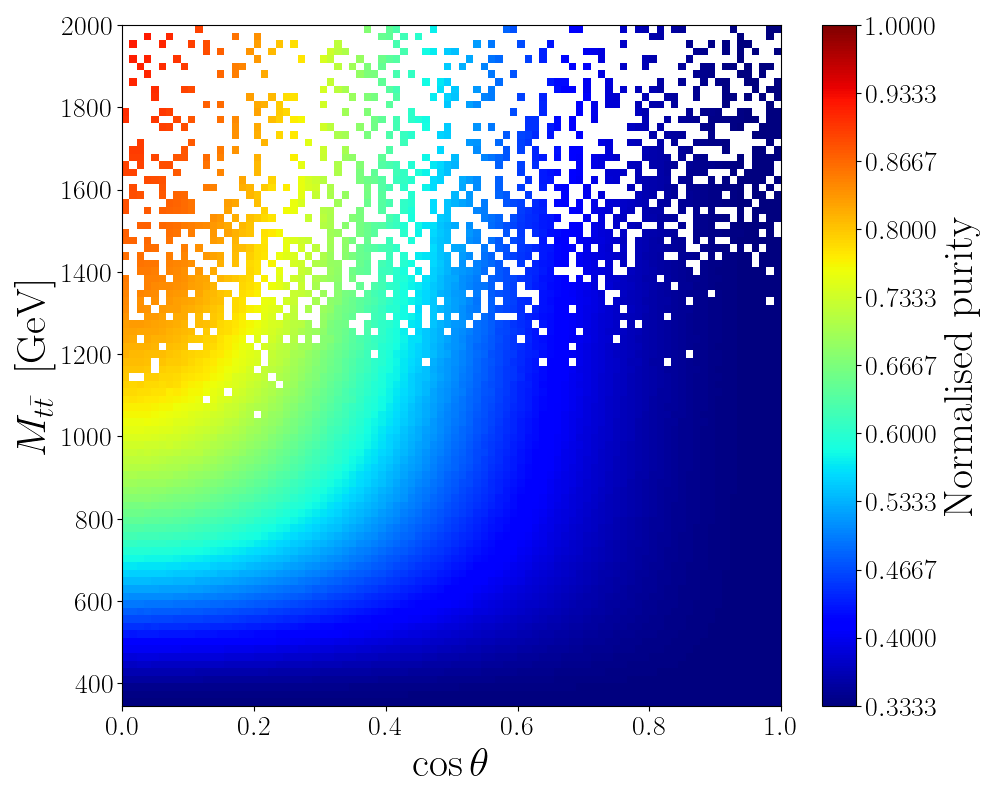}}\hfill
\subfloat[]{\label{fig:Purity qq_ttbarwmp2}\includegraphics[width=.49\linewidth]{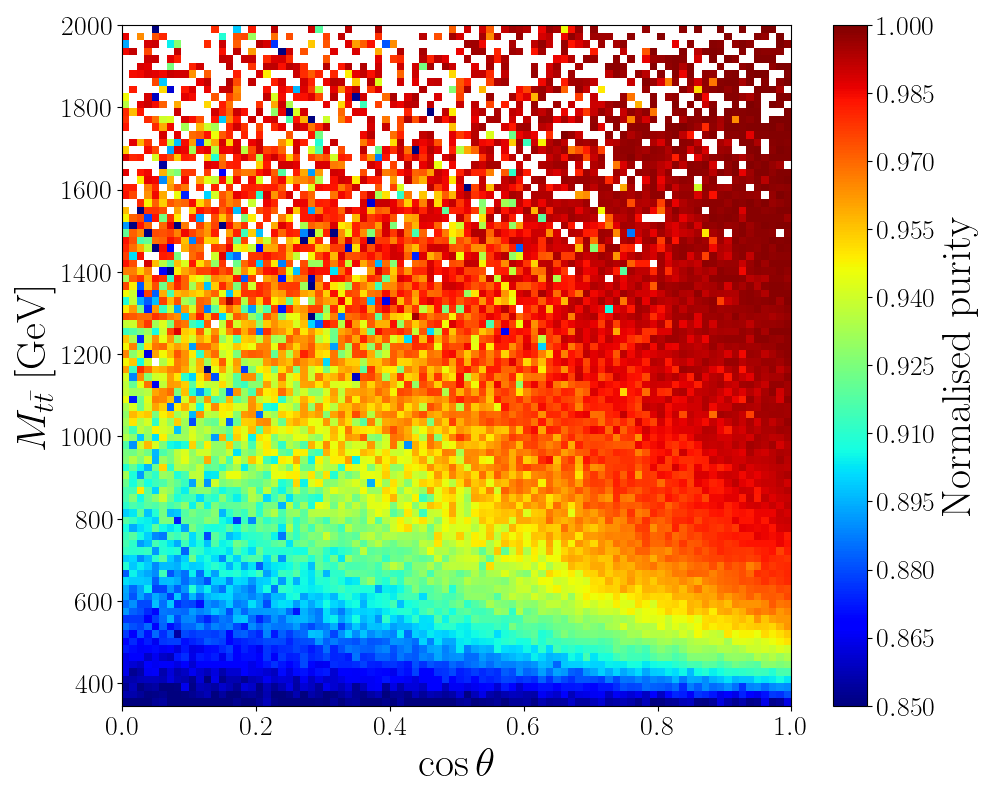}}\par
\caption{Normalised purity of $q\,\bar q \to t\,\bar t$ (a) and of $q\,\bar q^\prime \to t \, \bar t \, W^\pm$ (b) over the phase space spanned by $M_{t\bar t}$ and $\cos\theta$ (\ref{eq: theta1}) in the centre-of-mass frame of $t\bar t$. Note that the two scales are different. We observe that figure (b) shows a much higher degree of purity compared to figure (a) as well as a totally different distribution of purity.}
\label{fig:Purity qq_ttbarwmp}
\end{figure}

Another observable that can be easily computed for this process is the purity, which quantifies how mixed the quantum state is. Specifically, we plot the normalised purity $\mu$ over the same phase space. Figure~\ref{fig:Purity qq_ttbarwmp} shows the normalised purity across the phase space for $q \bar q \to t \bar t$ (left) and $q \bar q^\prime \to t \bar t W^\pm$ (right). A purity of $1$ corresponds to a pure quantum state, while a value of $0$ corresponds to a maximally mixed state.  

For $t\bar t$, the purity approaches unity in the high--energy, low--$\cos\theta$ region, where the states are entangled. This behaviour is consistent with the fact that such strongly correlated states tend toward Bell states, which are pure. In contrast, $t\bar t W^\pm$ exhibits a markedly different purity distribution: the purity is significantly higher across the entire phase space compared to the original process and is no longer correlated with regions of high entanglement. This indicates that polarising the initial state leads to the production of much purer states throughout the phase space, while having little effect on the degree of entanglement.

\paragraph*{Quantum spin correlations in $t W^-$ and $\bar t W^+$}
We have first considered the $t \bar t$ system to compare it with the initial $q \bar q \to t \bar t$ process and to understand the impact of polarising the initial state. However, the emission of the weak boson $W^\pm$ also allows us to study the quantum spin correlations between the boson and one of the top quarks. All calculations are performed in the centre--of--mass frame of the $t \bar t$ pair. In this frame, the symmetries are such that it is sufficient to study a single subsystem, chosen here to be the $t W^+$ system.

This system is no longer a $2 \times 2$ system, and the concurrence is therefore not easily computable. We thus require new observables to quantify entanglement in this case. As discussed in App.~\ref{app:QO}, when reviewing several entanglement witnesses and other quantum--information observables, the Peres--Horodecki criterion remains both necessary and sufficient for $2 \times 3$ systems, making it an excellent witness for our study. The Peres--Horodecki criterion states that a system described by a density matrix $\rho$ is entangled if and only if its partial transpose possesses at least one negative eigenvalue (for $2 \times 2$ and $2 \times 3$ systems). We are therefore interested in the smallest eigenvalue of $\rho^{T_B}$.  

The {negativity} (and its logarithmic form, the {logarithmic negativity}), defined in App.~\ref{app:QO}, is an observable derived from the Peres--Horodecki criterion, quantifying the negativity of the partial transpose of the density matrix. Specifically, for $2 \times 3$ systems, a logarithmic negativity of zero corresponds to a separable state, while a non-zero value indicates entanglement, with a maximal value of one corresponding to a maximally entangled state.

\begin{figure}[!htb]
    \centering
    \includegraphics[width=0.5\linewidth]{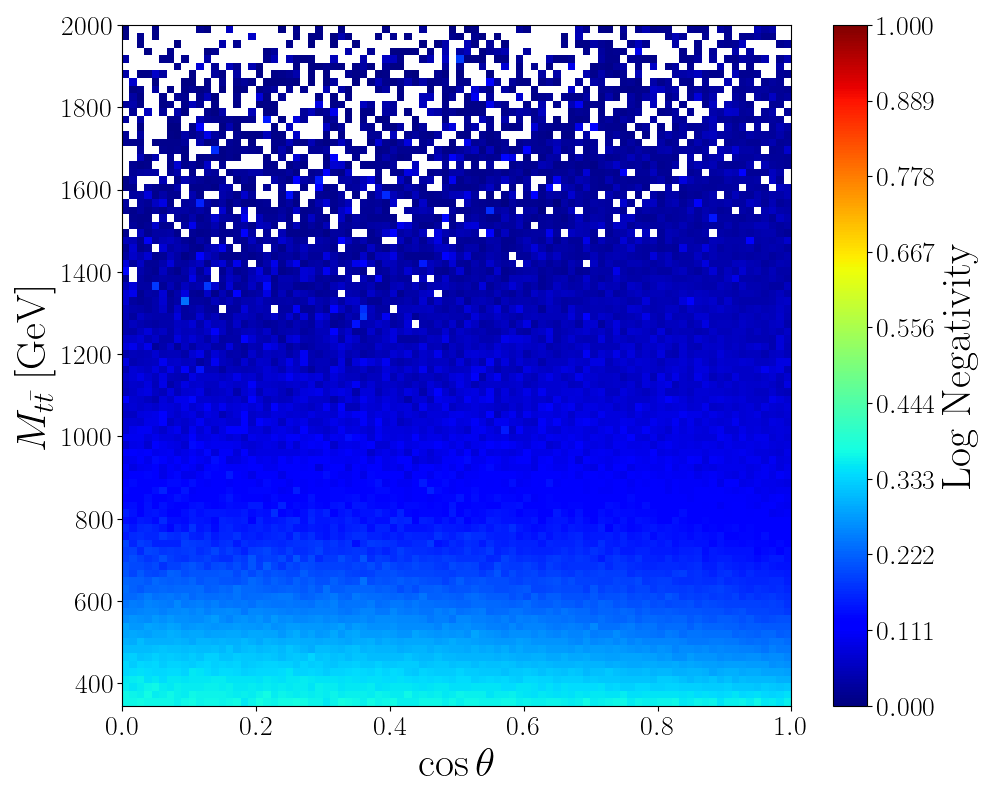}
    \caption{Value of the logarithmic negativity over the phase space spanned by $\cos\theta$ (\ref{eq: theta1}) and $M_{t\bar t}$ for $tW^+$ production. Entanglement of the system described by $\rho$ is equivalent to a non-zero value.}
    \label{fig:LogNeg_twp}
\end{figure}

Figure~\ref{fig:LogNeg_twp} shows the value of the logarithmic negativity over the phase space $(M_{tW^+}, \cos\theta)$. The observable is non-zero across the entire phase space, indicating that the system is entangled in all regions. However, the threshold region exhibits the highest level of entanglement, whereas the high--energy region shows almost vanishing logarithmic negativity, implying that the system becomes nearly separable. The dependence on the production angle $\theta$ is weak, which is consistent with the behaviour expected for entangled states in the threshold region.  

Interestingly, while no entanglement is observed in the threshold region for the $t \bar t$ subsystem, owing to the absence of a gluon initial state, the $t W^+$ subsystem does exhibit entanglement in this region. 

\par

\subsection{$t (\bar t \to W^- \bar b)$ vs.\ $t W^-(b)$}
\label{sec:tw}
Distinguishing $p p \to t W^-(b)$ and $p p \to t \bar t$, with $\bar t \to \bar b W^-$, provides a particularly interesting testbed for studying quantum correlations in top production and decay. These channels, which are known to interfere at NLO in QCD, probe the transition between entangled and effectively separable subsystems when one of the heavy quarks subsequently decays. In this context, the recent studies  on {post--entanglement} effects~\cite{Aguilar-Saavedra:2023hss,Aguilar-Saavedra:2024hwd} have shown that the pattern of spin correlations encoded in the top decay products can preserve, distort, or even revive the quantum features present in the production stage. Investigating these phenomena through quantum--information observables thus offers a complementary perspective on the interplay between production, decay, and entanglement loss in realistic collider environments.

\begin{figure}[htb]
    \centering
    \includegraphics[width=0.8\linewidth]{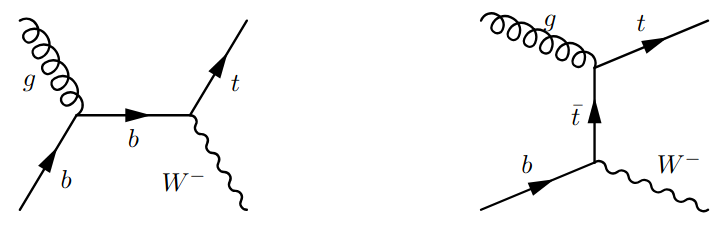}
    \caption{Feynman diagrams of the process $p  p \to t W^-$ in the Standard Model at LO in the five-flavour scheme.}
    \label{fig:diagrams_gb_twm}
\end{figure}

\begin{figure}[htb]
    \centering
    \includegraphics[width=0.8\linewidth]{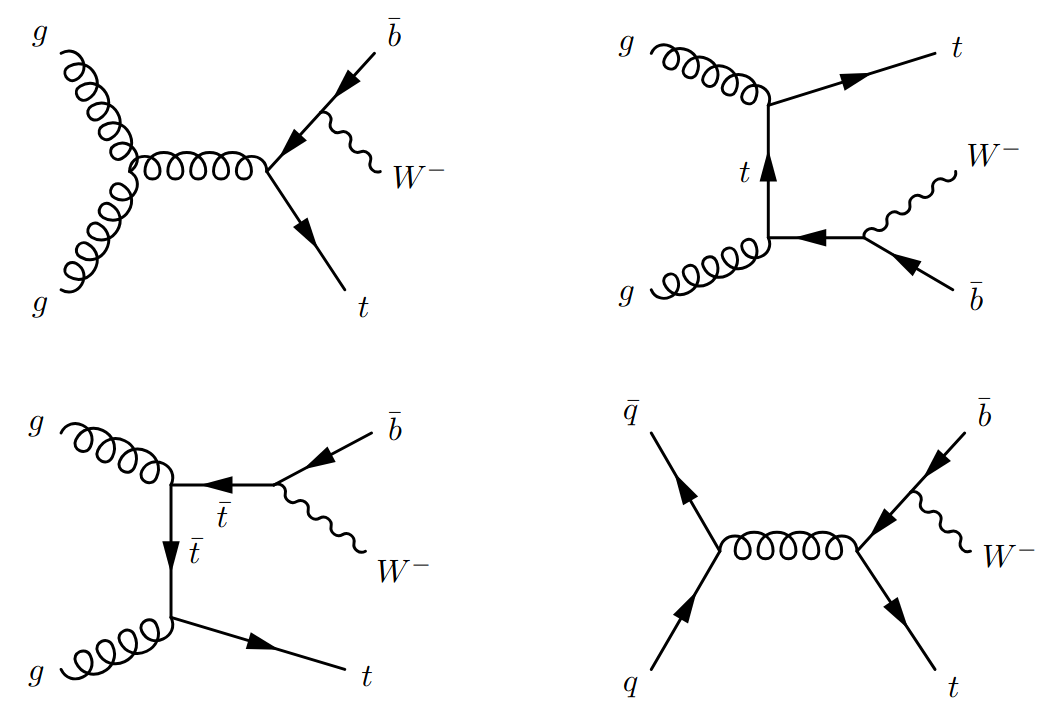}
    \caption{Feynman diagrams for the process $p p \to t \bar t,  \bar t \to \bar b  W^-$ in the Standard Model at LO.}
    \label{fig:diagrams p p > t tbar decay}
\end{figure}

As widely studied, $tW$ production at next-to-leading order (NLO) in QCD interferes with $t \bar t$ production, making their separation a non-trivial task. Techniques such as diagram removal and diagram subtraction have been developed for this purpose~\cite{Frixione:2008yi,White:2009yt,Cascioli:2013wga,Frederix:2016rdc,Jezo:2016ujg,Demartin_2017_tWH}. 
In this section, we employ QI observables to investigate how different the processes 
$p p \to t W^-$ and $p p \to t \bar t$, with $\bar t \to \bar b W^-$, are, whose Feynman diagrams are shown in Figs.~\ref{fig:diagrams_gb_twm} and~\ref{fig:diagrams p p > t tbar decay}.

In this study, we consider the quantum spin correlations between the $t$ quark and the $W^-$ boson. Since the $W^-$ is a massive vector boson, it has three possible polarisation states, making the system a qubit--qutrit configuration for which the concurrence is not directly computable. We therefore employ alternative quantum--information observables to quantify the degree of entanglement between the two particles. In a $2 \times 3$ (qubit--qutrit) system, the Peres--Horodecki criterion remains both necessary and sufficient, and can thus be applied here. We also make use of the {negativity}, an observable derived from the Peres--Horodecki criterion, to measure the degree of entanglement, and of the {purity} to assess how mixed the quantum state of the $tW^\pm$ pair is. This list of observables is not exhaustive: further quantities, such as the {mana} or other quantum--information measures listed in App.~\ref{app:QO}, could also be computed once implemented in the code.

The data used for this analysis were generated with \mg at a centre--of--mass energy of $13~\text{TeV}$, corresponding to LHC conditions. A five--flavour scheme is employed, treating all quarks except the top as massless. The simulations use the \texttt{NNPDF4.0} NNLO PDF set~\cite{Ball_2022}, and the input parameters adopted for event generation are
\begin{align*}
    m_Z =91.118\;\textrm{GeV} , && G_F=1.16639.10^{-5} \;\textrm{GeV}^{-2}, && \alpha_{EM}^{-1} = 1.32507.10^{2}, 
\end{align*}
while the value of $\alpha_S$ is set by the PDF.

We start by computing the differences in the observables introduced earlier to assess whether any of them can help discriminate between the two processes. 

\begin{figure}[!htp]\centering
\subfloat[]{\label{}\includegraphics[width=.33\linewidth]{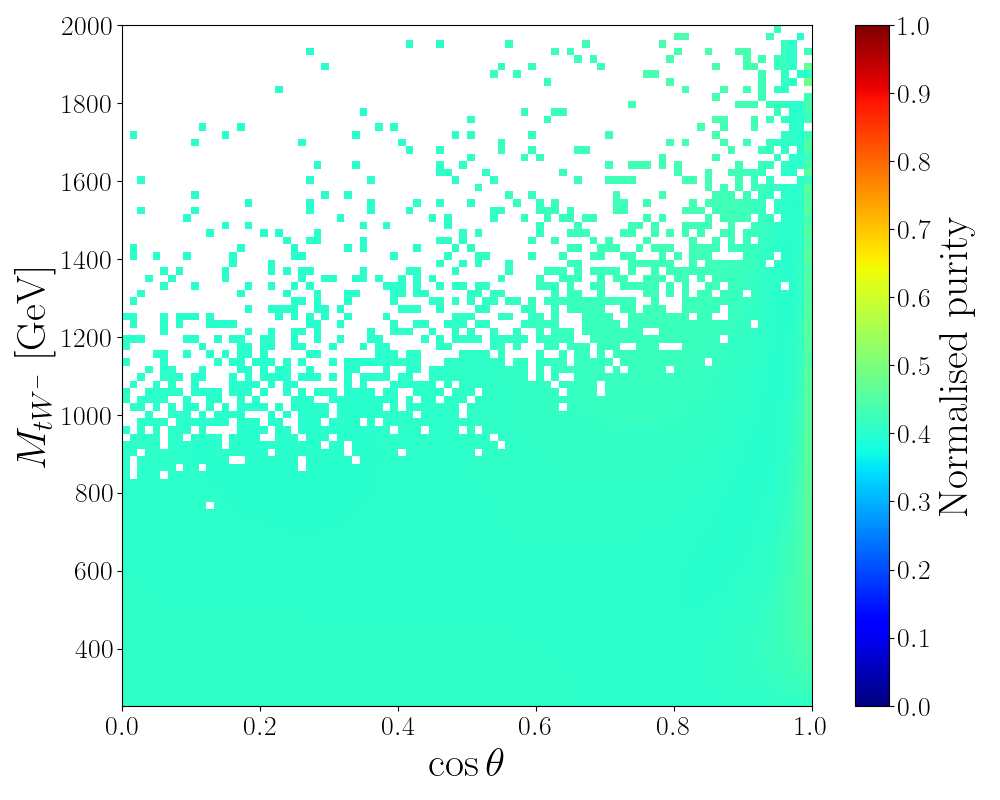}}
\subfloat[]{\label{}\includegraphics[width=.33\linewidth]{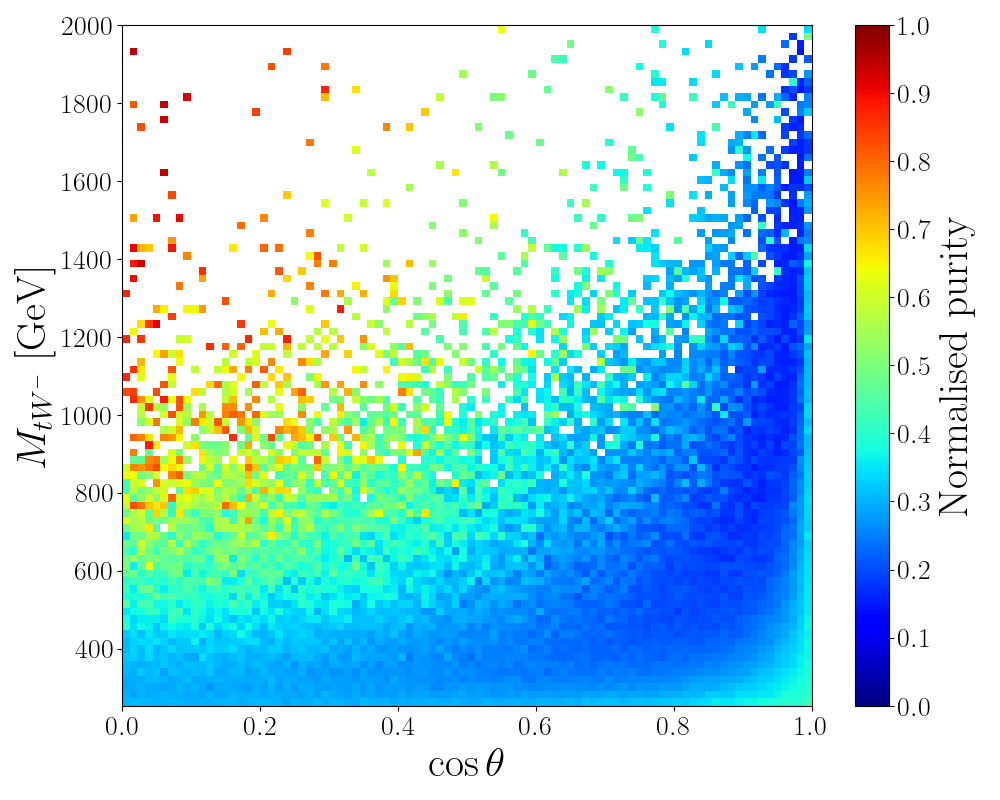}}
\subfloat[]{\label{}\includegraphics[width=.33\linewidth]{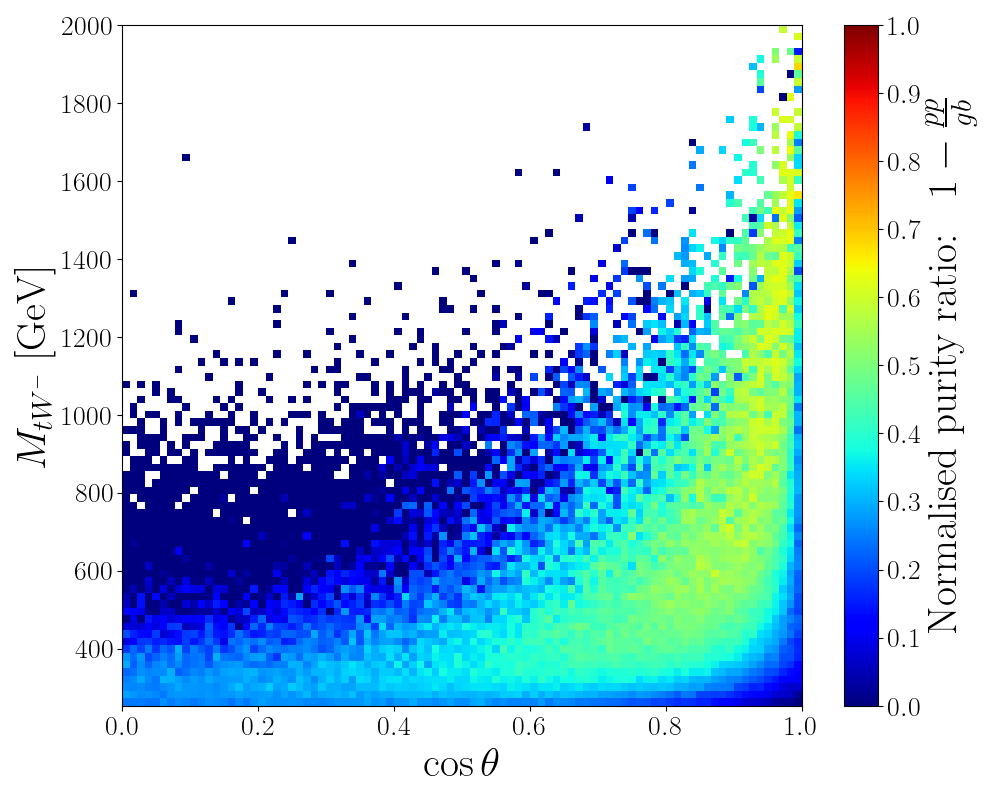}}

\caption{Normalised purity over the phase-space for (a) $g  b \to t  W^-$  and (b) $p  p \to t (\bar t  \to \bar b  ) W^⁻$. Plot (c) shows the relative difference in the normalised purity between the two processes.}
\label{fig: Normalised purity tWm}
\end{figure}

Figure~\ref{fig: Normalised purity tWm} shows the normalised purity over the phase space, together with the relative difference in normalised purity between the two processes. The two processes exhibit markedly different behaviours in terms of normalised purity. However, when taking into account the fact that cross-sections are actually peaked (for both processes) in regions, i.e.\ at low $\beta$ and large values of $\cos\theta$, where the difference is not so striking, it is worth investigating other possibilities.  

\begin{figure}[!htp]\centering
\subfloat[]{\label{}\includegraphics[width=.33\linewidth]{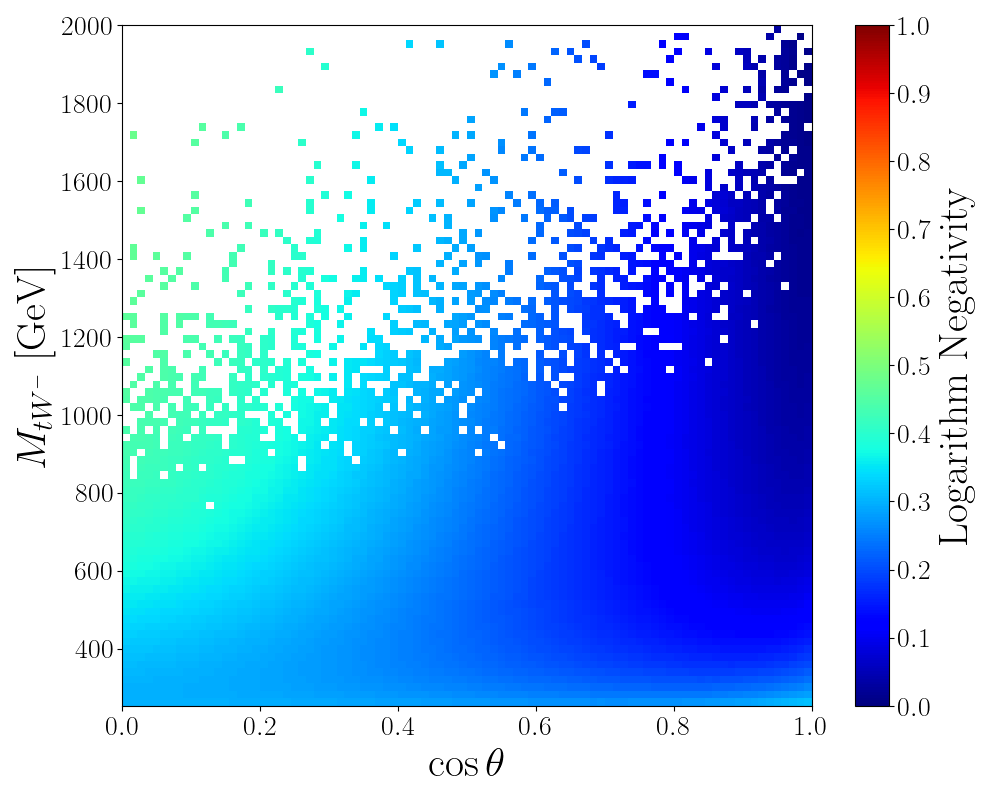}}
\subfloat[]{\label{}\includegraphics[width=.33\linewidth]{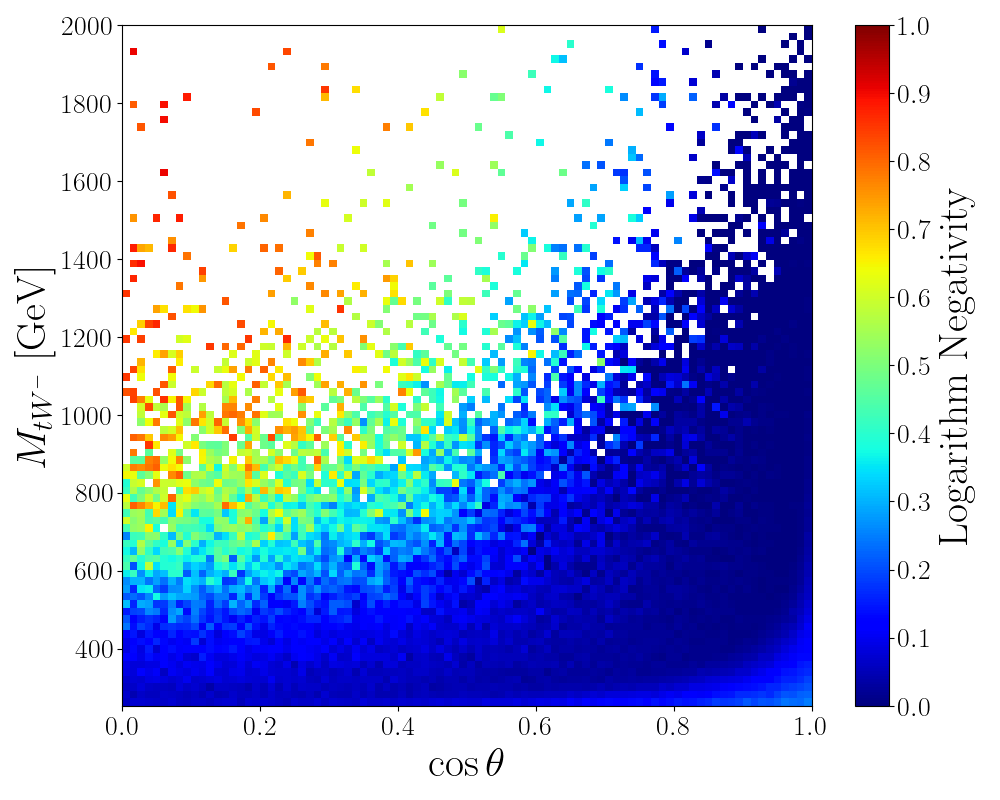}}
\subfloat[]{\label{}\includegraphics[width=.33\linewidth]{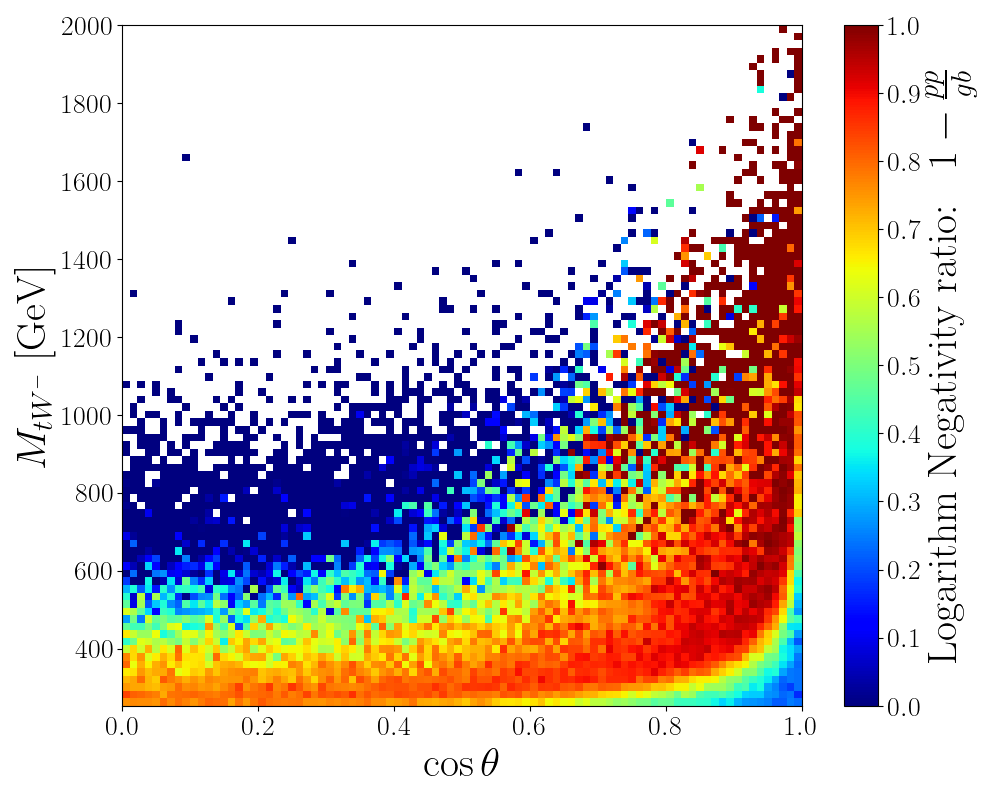}}

\caption{Logarithm negativity over the phase space for $gb \to t  W^-$ (a) and of $p  p \to t \bar t, \bar t \to \bar b  W^⁻$ (b). Plot (c) shows the relative difference in the logarithm negativity between the two processes.}
\label{fig:LogN_tWm}
\end{figure}

Let us consider the level of entanglement of the $tW^-$ system, comparing the distributions of the logarithmic negativity, defined in App.~\ref{app:QO}. Figure~\ref{fig:LogN_tWm} shows the logarithmic negativity over the phase space for the two processes, together with their relative difference.  

The two processes exhibit markedly different behaviours in the entanglement structure of the $ tW^- $ subsystem. In particular, the process $ p p \to t \bar t $, with $ \bar t \to \bar b W^- $, features an intermediate region where the logarithmic negativity nearly vanishes, indicating an almost separable state. Such a region is absent in the $ g b \to t W^- $ process. More importantly, in the threshold region around $ \cos\theta = 1 $, where both processes display a large cross-section, the difference in the logarithmic negativity becomes substantial, exceeding $ 25\% $ in every bin. These results suggest that measuring the degree of entanglement of the $ tW^- $ system, via the logarithmic negativity, could provide a powerful discriminator between the two production mechanisms, provided the experimental precision on this observable is sufficiently high.

\subsection{Same-sign top-quark correlations:  $t\bar t t$ vs. $t\bar t  t\bar t$}
\label{sec:ttt}

In this section, we illustrate the potential of our automated approach by comparing the quantum (spin) correlations in final states for which results of the corresponding amplitudes are not available (and in any case would not be practical). 

We consider like-sign top pairs, $tt$ (or $\bar t\bar t$), in multi-top-quark final states, $3t$ production, $t\bar t t/\bar t t \bar t$ ($3t$), a mixed electroweak-QCD channel which effectively amounts to single-top production accompanied by a gluon splitting $g \to t\bar t$;  and $4t$ production $t\bar t t \bar t $ ($4t$). Current measurements of multi-top-quark production at the LHC have limited power to disentangle these mechanisms~\cite{CMS:2023zdh,ATLAS:2023ajo}. Given the distinct dynamics of $3t$ versus $4t$ production, one expects sizeable differences in their spin correlations, which we now compare for the first time.~\footnote{Spin-correlation effects in $4t$ production have recently been studied in Ref.~\cite{Alsairafi:2025rjd}.}

\begin{figure}
    \centering
    \includegraphics[width=0.9\linewidth]{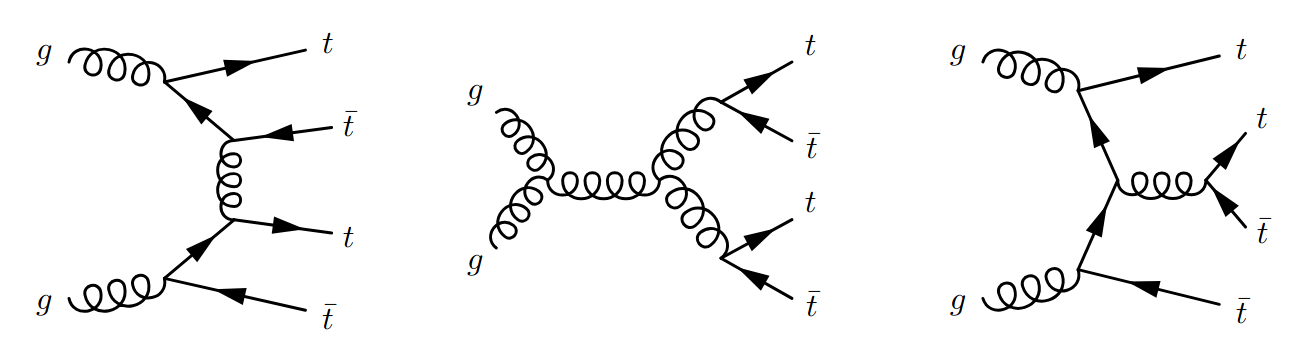}
    \caption{Representative diagrams of the process $p p \to t \bar t t \bar t$ in the Standard Model at LO.}
    \label{fig:diagrams_tttbartbar}
\end{figure}

\begin{figure}[!htb]
    \centering
    \includegraphics[width=0.8\linewidth]{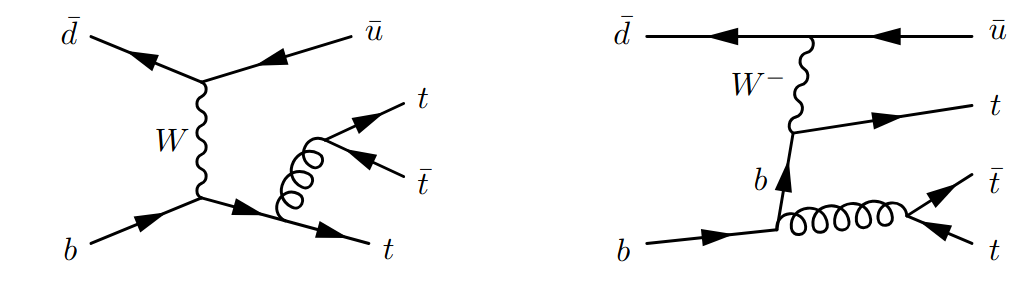}
    \caption{Example of diagrams of the process $p\;p\to t\;\bar t \; t \; q$ with $q$ any light quark in the Standard Model at LO.}
    \label{fig:diagrams_3_tops}
\end{figure}

\par
Figure~\ref{fig:diagrams_tttbartbar} shows a few representative Feynman diagrams for $4t$ production at leading order (LO) in the Standard Model, considering only QCD interactions (other processes contribute if EW interactions are included, which we do not consider here for simplicity). Figure~\ref{fig:diagrams_3_tops} shows representative diagrams for $3t$ production at LO in the Standard Model; in this case, EW interactions are needed and we only include the dominant QCD contributions to the final state. For $3t$ production, we consider only the final state $t \bar t t q$, with $q$ denoting a light quark (excluding $b$). Other processes, such as $t \bar t t W$ or $t \bar t t \bar b$, contribute whose analysis we defer to a more detailed study. 

\par
Here, we focus on the systems composed of top quarks with equal electric charge, namely $tt$ or $\bar t \bar t$. This choice is motivated by the fact that a unique combination of final-state particles gives rise to each of these pairs. One could alternatively study the quantum spin correlations of the $t \bar t$ system; however, in that case the observables would correspond to an average over the two possible pairings. Note that our code can select the particles forming the system based on any kinematic variable, such as transverse momentum, rapidity, and so on, although this selection would still mix contributions from different combinations. Once again, as our aim is illustrative, we just focus on observables related to the $tt$ system and leave the full analysis to a dedicated study. 

\par
Events are generated with \mg at a centre--of--mass energy of $13~\text{TeV}$, corresponding to LHC conditions. A five--flavour scheme is employed, treating all quarks except the top as massless. The simulations use the \texttt{NNPDF4.0} NNLO PDF set~\cite{Ball_2022}, and the input parameters for event generation are
\begin{align*}
    m_Z &= 91.118~\text{GeV}, & G_F &= 1.16639 \times 10^{-5}~\text{GeV}^{-2}, & \alpha_{EM}^{-1} &= 1.32507 \times 10^{2}.
\end{align*}
The value of $\alpha_S$ is determined by the PDF set. The density matrix $\rho$ is computed in the ZMF of the $tt$ system, and the helicity frame is defined using either of the two top quarks as reference. The choice of reference top does not affect the result, as it corresponds to the transformation $\theta \to \theta + \pi$, under which the amplitude is invariant.

\begin{figure}[!htp]\centering
\subfloat[]{\label{}\includegraphics[width=.49\linewidth]{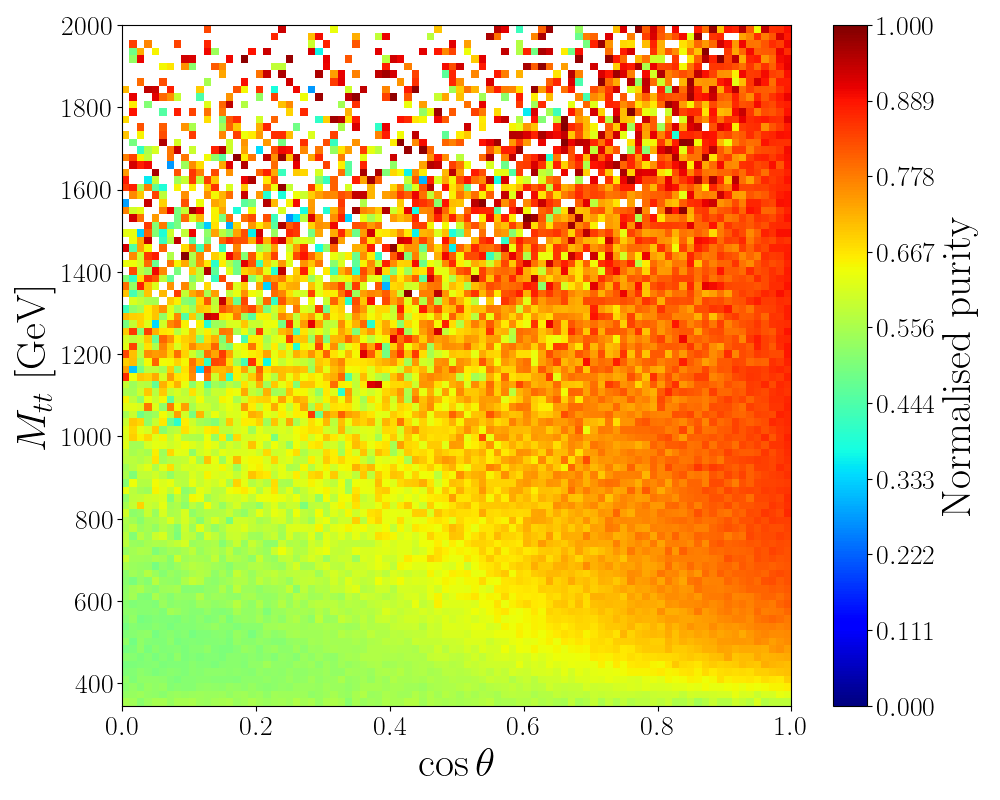}}
\subfloat[]{\label{}\includegraphics[width=.49\linewidth]{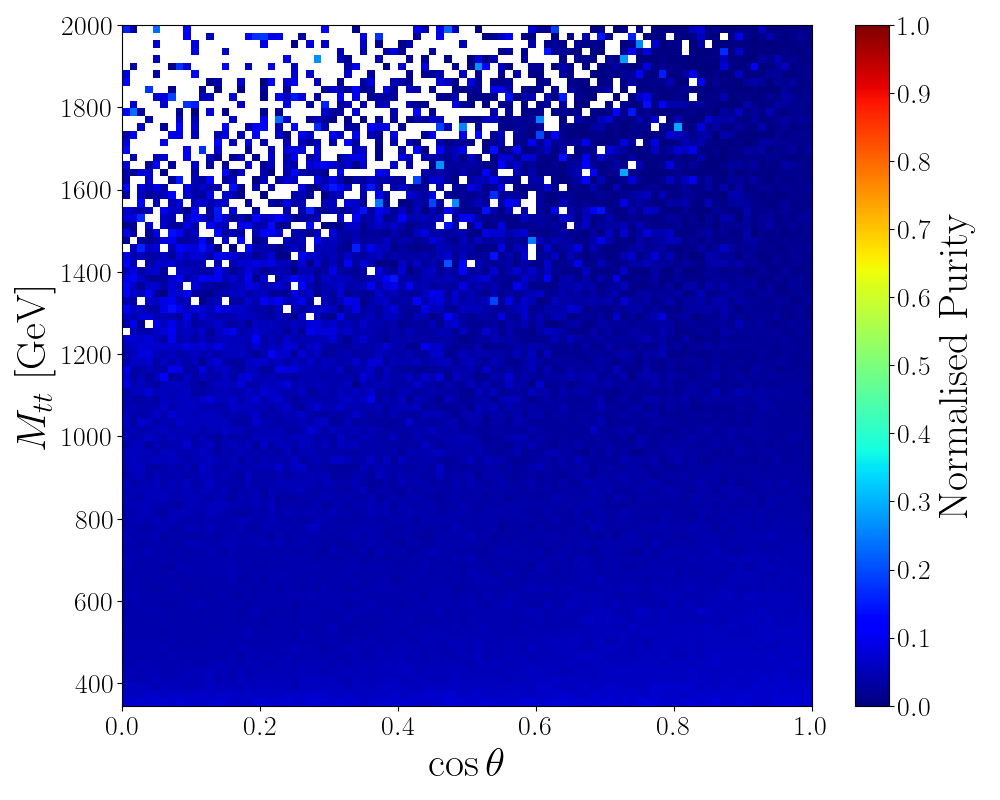}}

\caption{Normalised purity $\mu$ of the state $tt$ over the phase space spanned by $\cos\theta$ (\ref{eq: theta1}) and  $M_{tt}$ for the $3t$ production (a) and the $4t$ production (b). The difference between the two processes is stark and could help discriminate them in collider experiments}
\label{fig: Purity tttt}
\end{figure}

Figure~\ref{fig: Purity tttt} shows the normalised purity $\mu$ for $3t$ production (left) and $4t$ production (right) over the phase space $(\cos\theta, M_{tt})$. In the case of $3t$ production, the normalised purity varies between $1/2$ and $1$, with higher purity values observed in the high--energy region. In contrast, for $4t$ production, the normalised purity does not exceed $\sim 0.25$ across the entire phase space. This pronounced difference could be exploited to discriminate between the two processes, provided that the purity can be measured experimentally with sufficient precision.

\begin{figure}[!htp]\centering
\subfloat[]{\label{}\includegraphics[width=.49\linewidth]{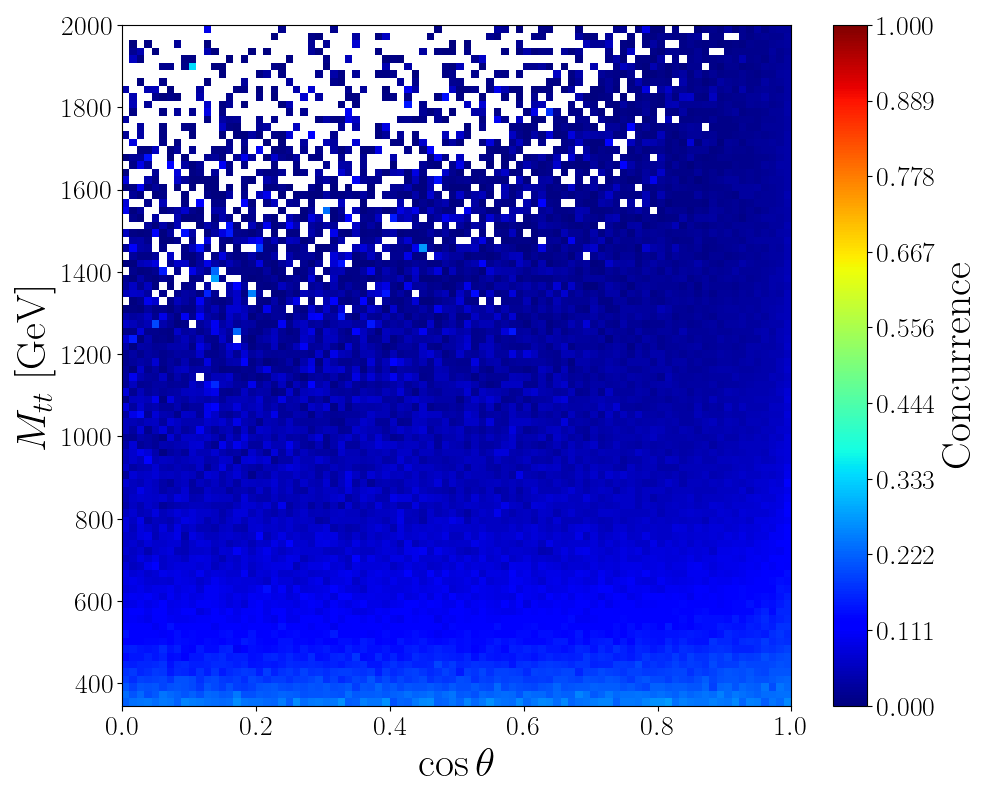}}
\subfloat[]{\label{}\includegraphics[width=.49\linewidth]{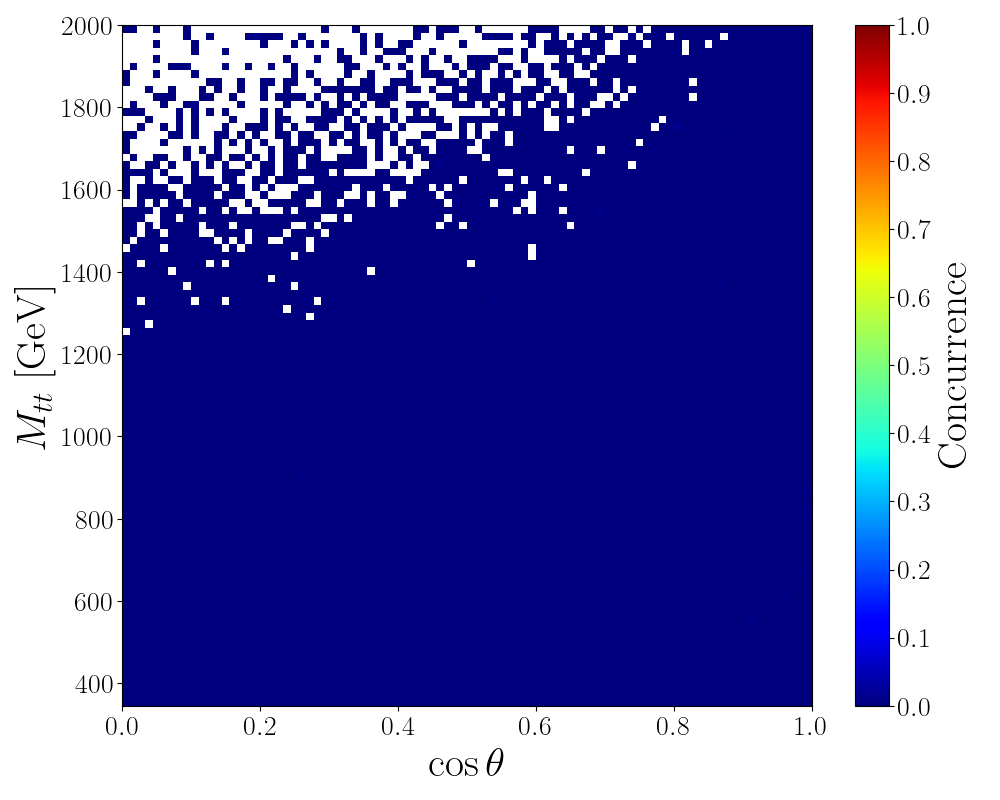}}

\caption{Concurrence of the state $tt$ over the phase space spanned by $\cos\theta$ (\ref{eq: theta1}) and $M_{tt}$ for the $3t$ production (a) and the $4t$ production (b). We observe that for the $4t$ production the state $tt$ is separable everywhere, while in the $3t$ production some entanglement can be measured in the threshold region. This difference could again be used to discriminate the two processes.}
\label{fig: Concurrence tttt}
\end{figure}

\par
Another observable that can be computed is the concurrence, which, as explained earlier in this paper, quantifies the degree of entanglement between the two particles in the $tt$ subsystem. Since the Feynman diagrams of these processes are more complex than those of simpler reactions such as $p p \to t \bar t$, it is more difficult to develop an expectation for the outcome. 

Figure~\ref{fig: Concurrence tttt} shows the concurrence over the phase space $(\cos\theta, M_{tt})$ for $3t$ production (left) and $4t$ production (right). For $4t$ production, the concurrence is nearly zero throughout the phase space, consistent with the naive expectation of weak entanglement in case particles are not ``connected'' in the Feynman diagrams. However,  $3t$ production displays a non-zero degree of entanglement, mainly in the threshold region, where the concurrence reaches values around $0.3$. This is certainly an interesting finding that will need more detailed investigations. 

\par
With our goals in mind, we emphasise that the difference in concurrence between the $3t$ and $4t$ production processes is remarkable: only the $3t$ channel exhibits a non-vanishing concurrence, albeit with a magnitude significantly smaller than in simpler systems such as $p p \to t \bar t$. This difference could allow one to distinguish between the two processes at least in the threshold region where the discrepancy is most pronounced, which is particularly relevant given that most events are produced near $\cos\theta = 1$.

\begin{figure}[!htp]\centering
\subfloat[]{\label{}\includegraphics[width=.49\linewidth]{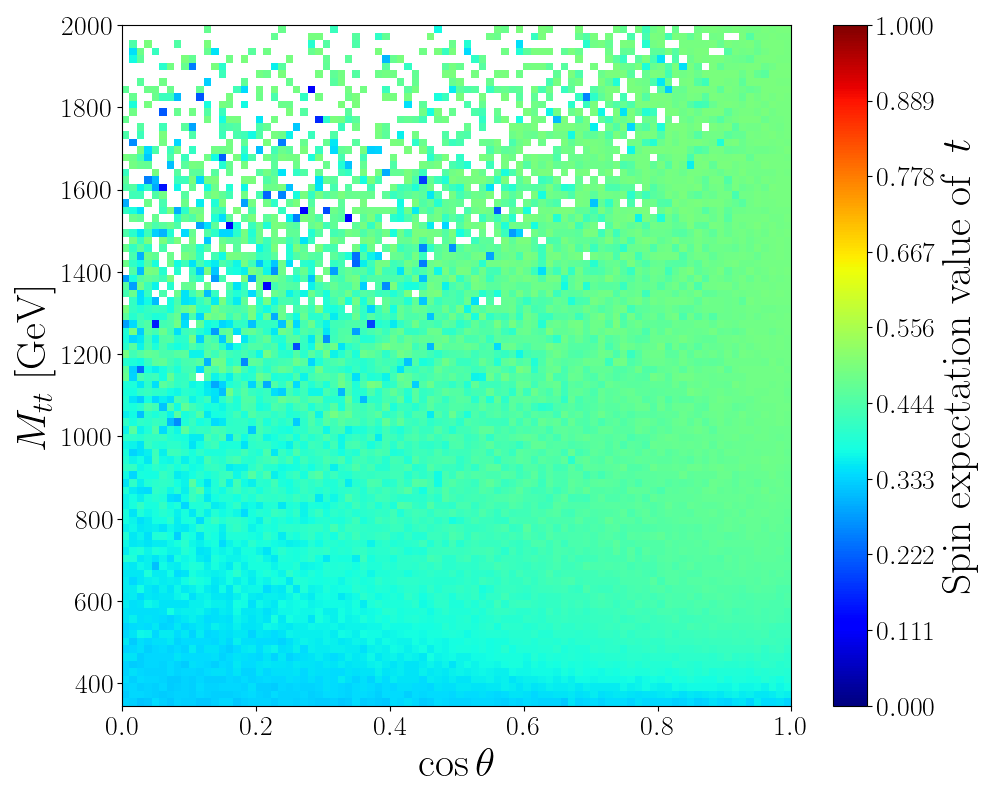}}
\subfloat[]{\label{}\includegraphics[width=.49\linewidth]{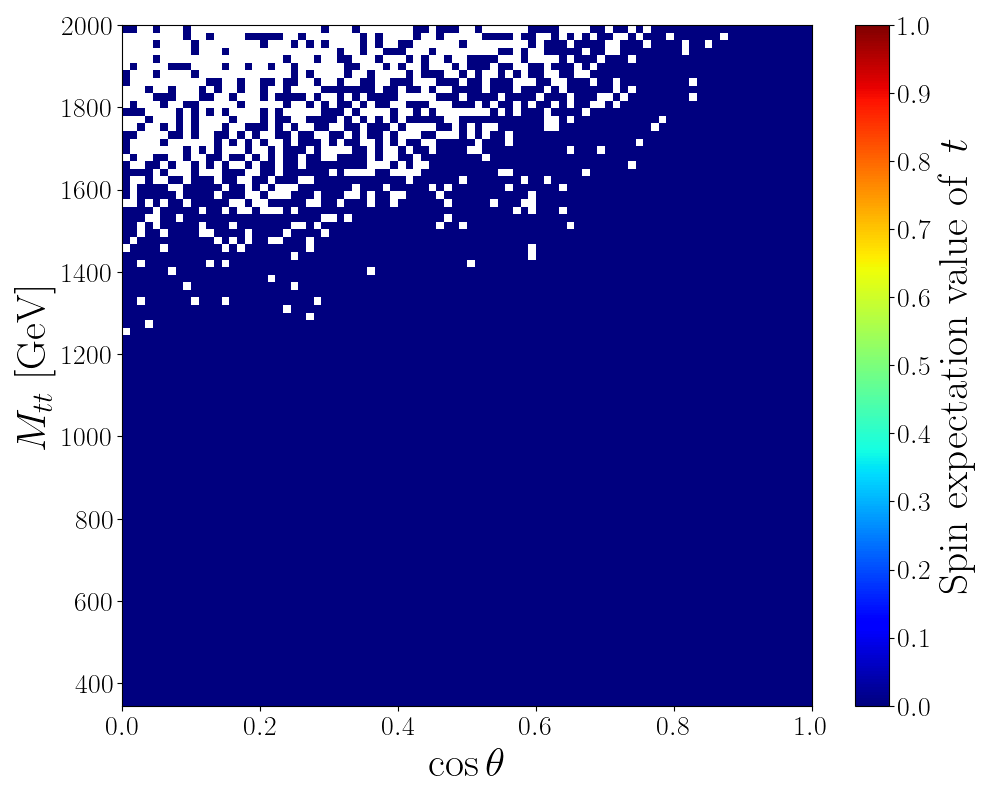}}

\caption{Norm of the spin expectation value of any of the two top quarks over the phase space $\cos\theta$ (\ref{eq: theta1}) and $M_{tt}$ for the $3t$ production (a) and the $4t$ production (b). We observe that for the $4t$ production the top quarks are non polarised, while in  $3t$ production they are. This difference could again be used to discriminate between the two processes.}
\label{fig: SeV tttt}
\end{figure}

\par
A third observable that can be used to discriminate between the two processes is the spin expectation value of either of the top quarks. Figure~\ref{fig: SeV tttt} shows the norm of the spin expectation value for one of the two top quarks in $3t$ production (left) and $4t$ production (right). As in the case of the concurrence, the norm of the spin expectation value in $4t$ production is essentially zero; in fact, all three components of the spin expectation value are very close to zero (below $10^{-4}$ in absolute value) for any event. In contrast, in $3t$ production, the norm of the spin expectation value is non-zero across the entire phase space, and each of its components is also non-zero, with magnitudes of order $10^{-1}$-$10^{-2}$. This shows that measuring the spin expectation value of the $tt$ subsystem provides a powerful discriminator between $3t$ and $4t$ production on an event--by--event basis.

\par
Another possibility to distinguish between the two processes is to define a distance between the reduced density matrices of the $tt$ pairs.  For all other observables, we computed their values on an event--by--event basis and subsequently averaged them to obtain the histograms presented in this paper. In the case of a distance, however, this approach is no longer applicable, since the events for the two processes populate different regions of phase space. Instead, we first average the density matrices within each bin and then compare the averaged density matrices bin by bin.

\begin{figure}[!htp]\centering
\subfloat[]{\label{}\includegraphics[width=.49\linewidth]{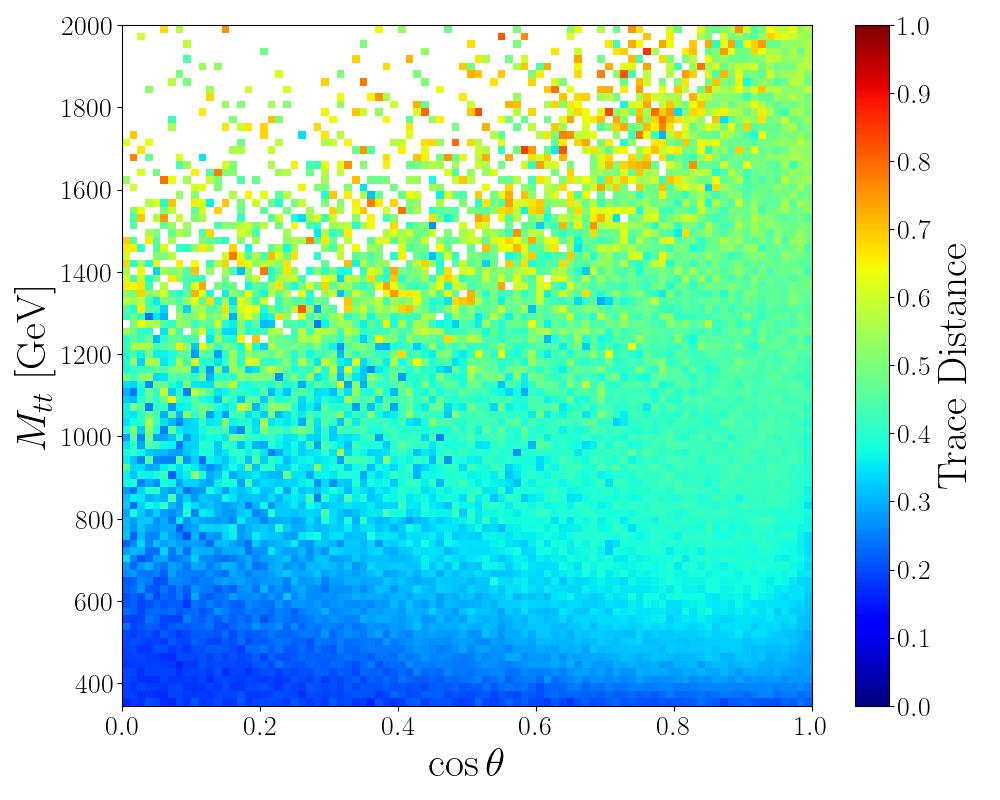}}
\subfloat[]{\label{}\includegraphics[width=.49\linewidth]{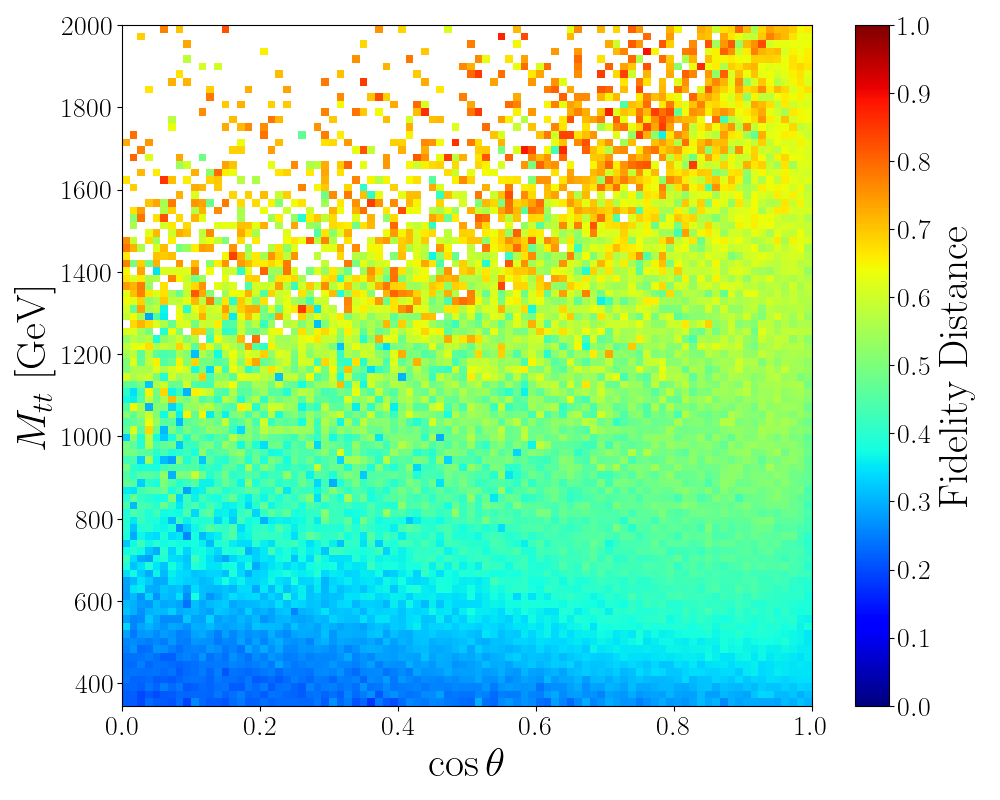}}

\caption{Trace distance (a) and fidelity distance (b) between the $3t$ and $4t$ production processes over the phase space $\cos\theta$ (\ref{eq: theta1}) and $M_{tt}$.}
\label{fig: Distances tttt}
\end{figure}

\par
We have considered and implemented two such ``distances'': the trace distance and the fidelity distance, both defined in App.~\ref{app:QO}. They share the same physical interpretation: a distance equal to zero indicates that the two density matrices describe the same physical state, while larger values correspond to increasingly dissimilar systems. Consequently, we expect the plots of these two distances to display qualitatively similar behaviour.  

Figure~\ref{fig: Distances tttt} shows the trace distance (left) and the fidelity distance (right) between the $3t$ and $4t$ production processes. As expected, both distributions exhibit very similar patterns, although the trace distance is systematically larger than the fidelity distance. The two measures consistently indicate that the density matrices of the two processes are most similar in the threshold region, particularly around $\cos\theta = 0$, and diverge progressively as the invariant mass of the $tt$ system increases. The similarity between the two plots also provides a useful validation of the implementation of these observables.

\par
This simple application of our automated code demonstrates that the measurement of quantum (spin) correlations in multi--top-quark processes can discriminate between the processes $p p \to t \bar t t q$ and $p p \to t \bar t t \bar t$, whose separation is challenging with current observables. Measurements of the concurrence, purity, and spin--expectation value each enable this discrimination and can be combined to further enhance its effectiveness.

\section{Conclusions and outlook}
\label{sec:conclusions}

We have presented a fully automated framework for computing production spin-density matrices at event level within \mg, together with a lightweight {\tt Python} library for post-processing and analysis. The implementation assembles helicity amplitudes into per-event production matrices $R$, normalises them into density matrices $\rho$, and stores the relevant objects and metadata to the LHE record. The analysis layer provides a unified interface to: (i) extract single-particle and pairwise spin polarisations $(\vec B_1,\vec B_2)$ and the correlation matrix $C$; (ii) perform frame changes; (iii) evaluate a catalogue of QI observables (purity, entanglement witnesses and monotones such as concurrence and negativity, as well as no-classicality indicators like magic); and (iv) build ensemble averages with consistent normalisation and reproducibility.

The framework was validated against known results across the three relevant classes: qubit--qubit ($t\bar t$), qubit--qutrit (associated $tV$ production), and qutrit--qutrit ($VV$ production). For each case we verified, at fixed kinematics and in fiducial regions, element-by-element agreement of $R$ and $\rho$, trace normalisation, Hermiticity and positivity.  A BSM benchmark involving a mixed scalar-pseudoscalar decay further demonstrated that the machinery naturally accommodates non-standard interactions (including CP-violating couplings) at leading order.

Building on these validations, we produced new results for quantum states arising in processes such as $q\bar q'\!\to t\bar t W^\pm$, $t$-$W$ correlations in $tW$ vs. $t \bar t$ and finally $tt$ correlations in $3t$ vs. $4t$ final states.  We showed how the event-level description enables systematic exploration of quantum observables identifying those that are suited to provide sensitivity to specific phenomenology questions.  

We believe that the range of processes considered for validation, together with the illustrative showcases presented, clearly demonstrate the potential of our implementation for phenomenological studies at present and future colliders. Many further applications can be envisaged, both within the Standard Model, such as the exploration of multi-partite correlations or decoherence phenomena, and beyond it, including investigations of SMEFT sensitivity. These directions will be the subject of further dedicated studies. 

On the code-development side, we plan to study the feasibility of constructing the production $R$ matrix at NLO QCD and electroweak accuracy, to disentangle genuine higher-order effects on the density matrix $\rho$, which can be interpreted as decoherence effects, from those originating from the tomographic reconstruction itself.

In summary, the tool provides a straightforward and standardised route from amplitudes to QI observables at colliders. By bridging precision event generation with QI methods, it enables a systematic exploitation of spin coherences and entanglement as discovery and characterisation handles for both SM precision tests and BSM searches.

\section*{Acknowledgements}
We would like to thank Quentin Herebaudt for the implementation the proof-of-concept of this work and  Spyros Argyropoulos for the help in the validation of the code and for parallel work on using density matrix for MadSpin. 
We are thankful to Rafael Aoude, Luca Mantani, and Eleni Vryonidou, for valuable comments on the manuscript and to Rikkert Frederix for feedback on the implementation. Our research is funded through the PRIN2022 grant Nr.~2022RXEZCJ, by the project ``QIHEP--Exploring the foundations of quantum information
in particle physics'', which is financed through the PNRR with NextGenerationEU funds, in the context of the extended partnership PE00000023 NQSTI (CUP J33C24001210007), by FRS-FNRS (Belgian National Scientific Research Fund) IISN projects 4.4503.16 (MaxLHC) and DR-Weave grant FNRS-DFG num\'ero T019324F (40020485).  
This article/publication is based upon work from COST Action
CA24146, supported by COST (European Cooperation in Science and Technology).
Computational resources have been provided by the supercomputing facilities of the Universit\'e catholique de Louvain (CISM/UCL) and the Consortium des \'Equipements de Calcul Intensif en F\'ed\'eration Wallonie Bruxelles (C\'ECI) funded by the Fond de la Recherche Scientifique de Belgique (F.R.S.-FNRS) under convention 2.5020.11 and by the Walloon Region.

\appendix 

\section{Quantum observables}
\label{app:QO}
In this appendix we summarise the QI observables currently implemented  in the analysis routine (\texttt{Density\_functions.py}). For each quantity we provide: (i) the definition; (ii) the class of systems it applies to (e.g.\ qubit-qubit, qubit-qutrit, general $d_A\!\times\! d_B$); and (iii) its physical interpretation (polarisation, correlations, entanglement certification/quantification, non-classicality, {etc.}). The catalogue is not exhaustive, but it covers the core diagnostics needed for baseline collider studies and can be easily extended. Where relevant, we also state normalisation conventions, invariance properties under local basis changes, and practical notes on tomographic reconstruction from decay-angle distributions.

\subsection{Purity}
\label{app:purity}
Purity is a very simple observable to compute, this simplicity allows it to be available for any quantum system when other more complex observables may not. It measures how pure/mixed a quantum system is.
\paragraph*{Systems available:} All.
\paragraph*{Theoretical formula:} Purity is defined as
\begin{equation}
    \gamma \equiv \Tr[\rho^2],
\end{equation}
This quantity indeed quantifies how pure the quantum subsystem represented by $\rho$ is, however it is not normalised in the sense that its lower bound is not $0$, we can then define the normalised purity that corrects it and is defined as
\begin{equation}
    \mu(\rho) \equiv  \frac{d}{d-1} \left( \gamma(\rho) - \frac{1}{d} \right),
\end{equation}
where $d$ is the dimension of the Hilbert space on which the density matrix is defined.
\paragraph*{Meaning:} Purity measures how mixed/pure a subsystem represented by $\rho$ is. If purity is equal to $1$ the subsystem is pure, if it is smaller than $1$, it is mixed and the closer it is to the lower bound the more mixed it is.
The bounds for purity are: $1/d \leq \gamma \leq 1$ with $d$ the dimension of the Hilbert space on which the density matrix is defined. The upper bound is reached if $\rho$ is a projector, meaning the system is in a pure state and the lower bound is reached for $\rho = \mathbb{I}_d/d$ which means that the system is in a completely mixed state. The normalised purity represents the same information but the bounds are $0 \leq \mu(\rho) \leq 1$.

\subsection{Concurrence}
\label{app:concurrence}
Concurrence is an observable that is defined for any quantum system but requires a complex minimisation procedure so it is not known how to compute it for most systems. A formula to compute concurrence is known for a system made of a pair of qubits \cite{Wootters:1997id}, also upper and lower bounds are known for a system made of a pair of qutrits but are discussed below.
\paragraph*{Systems available:} Qubit/qubit only.
\paragraph*{Theoretical formula:}
For a pair of qubits, the concurrence is:
\begin{equation}
    C (\rho) = \max(0, \lambda_1 - \lambda_2 - \lambda_3 - \lambda_4),
\end{equation}
where the $\lambda_i$ are the eigenvalues in decreasing order ($\lambda_1$ is the biggest one) of the matrix
\begin{equation}
    \sqrt{\sqrt{\rho} (\sigma_y \otimes \sigma_y) \rho^\star (\sigma_y \otimes \sigma_y) \sqrt{\rho}},
\end{equation}
where $\sigma_y$ is the Pauli matrix and $\rho^\star$ is the complex-conjugate of the density matrix $\rho$. The square root $\sqrt{\rho}$ is always defined because $\rho$ is positive semi-definite.

\paragraph*{Meaning:} For a pair of qubits the bounds on concurrence are $0 \leq \mathcal{C}(\rho) \leq 1$. The upper bound is reached if the two qubits are fully entangled (in a Bell state) and the lower bound is reached if the two qubits are in separable states.

\subsection{Entanglement of formation}
\label{app:formation}
As for concurrence, entanglement of formation is an observable that is defined for any system but requires a complicated minimisation procedure so it is only available for a pair of qubits in practice. It is another observable that is used to quantify entanglement \cite{Wootters:1997id, PhysRevLett.78.5022}.
\paragraph*{Systems available:} Qubit/qubit only.
\paragraph*{Theoretical formula:}
Entanglement of formation can be simply calculated via the concurrence $\mathcal{C}[\rho]$:
\begin{equation}
    E_F (\rho) = H\left( \frac{1 + \sqrt{1 - \mathcal{C}(\rho)^2}}{2}\right),
\end{equation}
where $H(p)$ is the Shannon entropy, defined as
\begin{equation}
    H(p) = -p \log_2(p) - (1-p) \log_2(1-p).
\end{equation}

\paragraph*{Meaning:} As expected from an entanglement witness, it is normalised such that $0 \leq E_F(\rho) \leq 1$. This quantity is a measure of entanglement, meaning that it is zero if the states are separable and the larger it is, the more entangled the two subsystems are. Specifically it the entanglement of formation is 1, it means that the two qubits are fully entangled. This observable is useful for studies where the Shannon entropy is interesting.

\subsection{Peres-Horodecki or PPT criterion}
\label{app:ppt}
The Peres-Horodecki criterion is a necessary condition for the quantum system to be separable. For $2 \times 2$ and $2\times 3$ quantum systems the criterion is necessary and sufficient, therefore it allows to distinguish between entangled and separable states for systems made of a pair of qubits or of a qubit and a qutrit. It is the main tool available to distinguish entangled and separable systems, most of the entanglement witness are derived from this criterion.
\paragraph*{Systems available:} $2\times2$  (qubit/qubit) and $2\times 3$ (qubit/qutrit) quantum systems for it to be necessary and sufficient, any bipartite system for the necessary part only.
\paragraph*{Theoretical formula:} The Peres-Horodecki criterion concerns the partial transpose of the joint density matrix defined as
\begin{equation}
    \rho^{T_B} \equiv (\mathbb{I} \otimes T) \rho,
\end{equation}
and the sign of its eigenvalues.

\paragraph*{Meaning:} The criterion states that for the system to be entangled, it is necessary for at least one eigenvalue of the partially transposed density matrix $\rho^{T_B}$ to be negative, it becomes necessary and sufficient for $2\times 2$ (qubit/qubit) and $2\times 3$ (qubit/qutrit) systems. Note that we could also transpose the density matrix of the first particle as $\rho^{T_A} = (\rho^{T_B})^T$ so their eigenvalues are identical. Derivation of this criterion and more detailed explanations can be found in \cite{Peres:1996dw, Horodecki:2009zz, Horodecki:1996nc}.

\subsection{Polarisations and spin-correlation matrix}
\label{app:bc}
The polarisation vectors from each particle in the density matrix and the correlations between their spins can be extracted from the density matrix. While the Fano coefficients of equation (\ref{eq:rho_2d}) can always be extracted from the density matrix, they can not always be linked to polarisations and spin-density matrices. In the $2\times 2$, $2\times 3$ and $3\times 3$ systems however this link is possible.
\paragraph*{Systems available:} Qubit/qubit, qubit/qutrit, qutrit/qutrit.
\paragraph*{Theoretical formula:} The Fano coefficients \cite{RevModPhys.29.74} can be extracted from the density matrix via the following formula. The different values of the prefactors come from the different definitions of the spin observables $S_i$.
\begin{align}
    &\text{For qubit/qubit:} \;\; B_{1i} = \Tr [(\sigma_i \otimes \mathbb{I}_2) \rho] && B_{2i} = \Tr [( \mathbb{I}_2 \otimes \sigma_i ) \rho] && C_{ij} = \Tr [( \sigma_i \otimes \sigma_j ) \rho] \\
    &\text{For qubit/qutrit:} \;\; B_{1i} = \Tr[(\sigma_i \otimes \mathbb{I}_{3}) \rho] && B_{2i} = \frac{1}{2} \Tr[( \mathbb{I}_{2} \otimes \lambda_i ) \rho] && C_{ij} = \frac{1}{2}\Tr [( \sigma_i \otimes \lambda_j ) \rho] \\
    &\text{For qutrit/qutrit:} \;\; B_{1i} = \frac{1}{2}\Tr [(\lambda_i \otimes \mathbb{I}_3) \rho] && B_{2i} = \frac{1}{2}\Tr [( \mathbb{I}_3 \otimes \lambda_i ) \rho] && C_{ij} = \frac{1}{4}\Tr [( \lambda_i \otimes \lambda_j ) \rho]
\end{align}
Note that in the $2\times3$ system case, we gave the formula if the qubit if the first particle or if it is the second particle but it is a convention and the physical result should not change because of it. We have seen that the expectation value of an observable $A$ is $\Tr[A \rho]$ in section \ref{sec:framework}, so the expectation values of the spin/polarisations are
\begin{align}
    \langle S_1 \rangle &\equiv \langle S_1 \otimes \mathbb{I}_{d_2}\rangle = \Tr[(S_1 \otimes \mathbb{I}_{d_2}) \rho], \\
    \langle S_2 \rangle &\equiv \langle \mathbb{I}_{d_1} \otimes S_2\rangle = \Tr[(\mathbb{I}_{d_1} \otimes S_2) \rho], \\
    \langle S_1 S_2 \rangle &\equiv \langle S_1 \otimes S_2\rangle = \Tr[(S_1 \otimes S_2) \rho],
\end{align}
where $S_1$, $S_2$ are the spin operators for the first and second particle, respectively. We also know that these spin operator for qubits are 
\begin{align}
    S_x = \frac{1}{2}\begin{pmatrix} 0 & 1 \\ 1 & 0\end{pmatrix} = \frac{1}{2}\sigma_x, &&
    S_y = \frac{1}{2}\begin{pmatrix} 0 & -i \\ i & 0\end{pmatrix}= \frac{1}{2}\sigma_y, && 
    S_z = \frac{1}{2}\begin{pmatrix} 1 & 0 \\ 0 & -1\end{pmatrix}= \frac{1}{2}\sigma_z,
\end{align}
which means that
\begin{align}
    \langle S_1\rangle = \frac{1}{2}B_1, &&  \langle S_2\rangle = \frac{1}{2}B_2, && \langle S_1 S_2\rangle = \frac{1}{4}C.
\end{align}
For a qutrit, the representation fo the spin operators read
\begin{align}
    &S_x = \frac{1}{\sqrt{2}}\begin{pmatrix} 0 & 1 & 0 \\ 1 & 0 & 1 \\ 0 & 1 & 0\end{pmatrix} = \frac{1}{\sqrt{2}} (\lambda_1 + \lambda_6), 
    &S_y = \frac{1}{\sqrt{2}}\begin{pmatrix} 0 & -i & 0 \\ i & 0 & -i \\ 0 & i & 0\end{pmatrix}= \frac{1}{\sqrt{2}} (\lambda_2 + \lambda_7), \\ 
    &S_z = \frac{1}{2}\begin{pmatrix} 1 & 0 & 0 \\ 0 & 0 &0 \\ 0 & 0 & -1\end{pmatrix} = \frac{1}{2}(\lambda_3 + \sqrt{3} \lambda_8).
\end{align}
The expectation values for the spin-operators for a system qutrit/qutrit are thus:
\begin{align}
    \langle S_1 \rangle = \begin{pmatrix} \sqrt{2}(B_{11} + B_{16}) \\ \sqrt{2}(B_{12} + B_{17}) \\
    B_{13} + \sqrt{3}B_{18}\end{pmatrix}, &&\langle S_2 \rangle = \begin{pmatrix} \sqrt{2}(B_{21} + B_{26}) \\ \sqrt{2}(B_{22} + B_{27}) \\
    B_{23} + \sqrt{3}B_{28}\end{pmatrix},
\end{align}

and 
\begin{equation}
    \footnotesize
    \langle S_1 S_2 \rangle = \begin{pmatrix}
        2 (C_{11} + C_{16} + C_{61} + C_{66}) & 2(C_{12} + C_{17} + C_{62} + C_{67}) & \sqrt{2}(C_{13} + C_{63} + \sqrt{3}(C_{18} + C_{68}) \\
        2(C_{21} + C_{71} + C_{26} + C_{76}) & 2 (C_{22} + C_{27} + C_{72} + C_{77}) & \sqrt{2}(C_{23} + C_{73} + \sqrt{3}(C_{28} + C_{78}) \\
        \sqrt{2}(C_{31} + C_{36} + \sqrt{3}(C_{81} + C_{86}) & \sqrt{2}(C_{32} + C_{37} + \sqrt{3}(C_{82} + C_{87})  & C_{33} + \sqrt{3} (C_{38} + C_{83}) + 3 C_{88}
    \end{pmatrix}.
\end{equation}

Finally, for a system made of a qubit and a qutrit, expectation values for the spin operators are
\begin{align}
    \langle S_1 \rangle = B_1, && \langle S_2 \rangle = \begin{pmatrix} \sqrt{2}(B_{21} + B_{26}) \\ \sqrt{2}(B_{22} + B_{27}) \\
    B_{23} + \sqrt{3}B_{28}\end{pmatrix},
\end{align}

\begin{equation}
    \langle S_1 S_2 \rangle = \begin{pmatrix}
        \frac{1}{\sqrt{2}} (C_{11} + C_{16}) & \frac{1}{\sqrt{2}} (C_{12} + C_{17}) & \frac{1}{2} (C_{13} + \sqrt{3}C_{18}) \\
        \frac{1}{\sqrt{2}} (C_{21} + C_{26}) & \frac{1}{\sqrt{2}} (C_{22} + C_{27}) & \frac{1}{2} (C_{23} + \sqrt{3}C_{28}) \\
        \frac{1}{\sqrt{2}} (C_{31} + C_{36}) & \frac{1}{\sqrt{2}} (C_{32} + C_{37}) & \frac{1}{2}(C_{33} + \sqrt{3}C_{38})
    \end{pmatrix}.
\end{equation}

If instead the qutrit is the first particle and the qubit the second, the spin expectations become
\begin{align}
    \langle S_1 \rangle = \begin{pmatrix} \sqrt{2}(B_{11} + B_{16}) \\ \sqrt{2}(B_{12} + B_{17}) \\
    B_{13} + \sqrt{3}B_{18}\end{pmatrix} && \langle S_2 \rangle = B_2,
\end{align}
\begin{equation}
    \langle S_1 S_2 \rangle = \begin{pmatrix}
        \frac{1}{\sqrt{2}} (C_{11} + C_{61}) & \frac{1}{\sqrt{2}} (C_{12} + C_{62}) & \frac{1}{\sqrt{2}} (C_{13} + C_{63}) \\
        \frac{1}{\sqrt{2}} (C_{21} + C_{71}) & \frac{1}{\sqrt{2}} (C_{22} + C_{72}) & \frac{1}{\sqrt{2}} (C_{23} + C_{73}) \\
        \frac{1}{2} (C_{31} + \sqrt{3} C_{81}) & \frac{1}{2} (C_{32} + C_{82}) & \frac{1}{2}(C_{33} + \sqrt{3}C_{83})
    \end{pmatrix}.
\end{equation}

Note that the structure of these matrices are not unique, but we will use these conventions for our calculations.

\paragraph*{Meaning:} The quantities $\langle S_i \rangle$ represent the average of the spin of the particle $i$ in each direction. The quantity $\langle S_1 S_2 \rangle$ represents the average spin-spin correlations between the two particles, for instance $\langle S_{11} S_{22} \rangle$ (or $\langle S_{1x} S_{2y} \rangle$) is the average of the correlation between the spin of the first particle along the $\hat{x}$ axis and the spin of the second particle along the $\hat{y}$ particle.

\subsection{Upper and lower bounds for the concurrence}
\label{app:ul}
We have seen earlier that concurrence was computable for a $2 \times 2$ quantum system only, however for the case of $3 \times 3$ systems it is possible to put bound on it \cite{Aoude:2023hxv}. These bounds allow us to have some estimation of the concurrence. 
\paragraph*{Systems available:} Qutrit/qutrit.
\paragraph*{Theoretical formula:}
The concurrence is defined for a pair of qudits but is not useful for purposes. Indeed if the density matrix (of any mixed state system) is defined as 
\begin{align}
    \rho = \sum_i p_i \ket{\psi_i} \bra{\psi_i},  && \sum_i p_i = 1, && p_i \geq 0,
\end{align}
the concurrence can be computed as:
\begin{equation}
    \mathcal{C}(\rho) = \inf_{p_i, \psi_i} \left[\sum_i p_i c(\ket{\psi_i}) \right],
\end{equation}
which is quite complicated to compute. A easier way of estimating the concurrence for a pair of qutrits is to compute its bounds, the formula for these bounds are \cite{Aoude:2023hxv}:
\begin{align}
    \mathcal{C}_{\text{LB}}^2 &\equiv 2 \max (0, \Tr[\rho^2] - \Tr[\rho_A^2], \Tr[\rho^2] - \Tr[\rho_B^2]), \\
    \mathcal{C}_{\text{UB}}^2 &\equiv 2 \min(1 - \Tr[\rho_A^2], 1 - \Tr[\rho_B^2]),
\end{align}
where $\rho_A = \Tr_B[\rho] \equiv (\mathbb{I} \otimes \Tr)\rho$ and $\rho_B = \Tr_A[\rho] \equiv (\Tr \otimes \mathbb{I})\rho$ are the partial density matrices. For a pair of qutrits, we know the maximum of concurrence is reached for a totally symmetric and entangled pure state, this state has $\mathcal{C} = 2/\sqrt{3}$ so we have:
\begin{equation}
    0 \leq \mathcal{C}_{\text{LB}}(\rho) \leq \mathcal{C}(\rho) \leq \mathcal{C}_{\text{UB}}(\rho) \leq \frac{2}{\sqrt{3}}
\end{equation}

If we consider a Hilbert space $\mathcal{H} = \mathcal{H}_A \otimes \mathcal{H}_B$ and a density matrix $\rho_{AB}$ defined on it, both partial traces are defined and calculated through:
\begin{align}
    \rho_A &\equiv \Tr_B[\rho_{AB}] \equiv \sum_i (\mathbb{I_A} \otimes \bra{j}_B) \rho_{AB} (\mathbb{I_A} \otimes \ket{j}_B), \\
    \rho_B &\equiv \Tr_A[\rho_{AB}] \equiv \sum_i (\bra{i}_A \otimes \mathbb{I}_B) \rho_{AB} (\ket{i}_A \otimes \mathbb{I}_B),
\end{align}
where $\ket{i}_A$ and $\ket{j}_B$ are any basis for $\mathcal{H}_A$ and $\mathcal{H}_B$, respectively.

\paragraph*{Meaning:} The interpretation is the same as for the concurrence of a pair of qubits, it is just an approximation because we cannot easily calculate it. If the maximum value is 0 then the system is separable, if the minimum value is strictly positive, the system is partly entangled and if the minimum value is $2/\sqrt{3}$ then we know that the system is fully entangled.

\subsection{Coefficients \texorpdfstring{$D$}{D}}
\label{app:d}
We implemented the coefficients $D^{(i)}$, which are standard observables in top--quark pair production at the LHC~\cite{PhysRevD.53.4886}. These quantities enable one to test for entanglement and to discriminate between singlet and triplet spin states.

\paragraph*{Systems available:} Qubit/qubit where the polarisations are zero and the spin-correlation matrix is diagonal.
\paragraph*{Theoretical formula:}
There are four $D$ coefficients, they can be calculated with the coefficients of the spin-correlation matrix $C$
\begin{align}
    \label{eq:D1}
    &D^{(1)}  = \frac{1}{3} (C_{11} + C_{22} + C_{33}), \\
    &D^{(x)} = \frac{1}{3} (C_{11} - C_{22} - C_{33}), \\
    &D^{(y)} = \frac{1}{3} (-C_{11} + C_{22} - C_{33}), \\
    &D^{(z)} = \frac{1}{3} (-C_{11} - C_{22} + C_{33}),
    \label{eq:Dz}
\end{align}
where $\{x, y, z\}$ is any basis of space.
From these coefficients, one can calculate the concurrence of the quantum state:
\begin{equation}
    \mathcal{C} = \frac{1}{2} \max (0, -1 - 3D_{\min}),
    \label{eq: concurrence D coeffs}
\end{equation}
where $D_{\min}$ is the smallest coefficient out of the four.
\paragraph*{Meaning:} These coefficients are useful entanglement witnesses because one can see from equation (\ref{eq: concurrence D coeffs}) that if any of the four $D$ coefficient is smaller than $-1/3$, then the concurrence is strictly positive and the system is thus entangled. Also, according to which $D$ coefficient is smaller than $-1/3$ (if any), one can determine if the state is close to the singlet state of the triplet states \cite{Maltoni:2024csn}. If $D^{(1)}$ is inferior to $-1/3$ the quantum state will be close to the singlet state and if it is any of the other three, the quantum state will be close to one of the triplet states. 

\subsection{Magic}
\label{app:magic}
The study of entangled systems has been extensively developed in the field of quantum computing, where several additional entanglement witnesses have been proposed. One particularly interesting quantity is the so-called ``magic'' \cite{PhysRevA.71.022316}. The motivation for introducing this witness stems from the observation that, in quantum computing, entanglement alone is not sufficient to guarantee a computational advantage. Indeed, it has been shown~\cite{white2024} that certain maximally entangled states, known as stabiliser states, do not provide any speed-up over classical computation. It is therefore intriguing to investigate whether this quantum resource can also emerge in collider processes, which is the reason it has been implemented in this library. Moreover, preliminary studies already suggest that ``magical'' top quarks could be produced at the LHC~\cite{white2024}.

\paragraph*{Systems available:} Qubit/qubit.
\paragraph*{Theoretical formula:} Magic quantifies the distance of the studied quantum state to the closer stabiliser state. Let us first define what is a stabilised state, a state $\ket{\psi}$ is \textit{stabilised} by a unitary operator $U$ if
\begin{equation}
    U \ket{\psi} = \ket{\psi}
\end{equation}
For a pair of qubits the unitary operators that interest us are the one forming the \textit{Pauli group}. For a single qubit, this group is
\begin{equation}
    G_1 = \{ \pm \mathbb{I}, \pm i \mathbb{I}, \pm \sigma_1, \pm i \sigma_1, \pm \sigma_2, \pm i \sigma_2, \pm \sigma_3, \pm i \sigma_3\},
\end{equation}
where $\sigma_1 = \sigma_x$, $\sigma_2 = \sigma_y$, $\sigma_3 = \sigma_z$ and for a system composed of $n$ qubits the \textit{Pauli group} is
\begin{equation}
    G_n = \bigotimes_{i = 1}^n G_1 = \{A_1 \otimes \cdots \otimes A_n, A_i \in G_1\}.
\end{equation}

We can also define the \textit{Pauli string} $\mathcal{P}_n$ which is an element of $G_n$ defined as
\begin{equation}
    \mathcal{P}_n = P_1 \otimes \cdots P_n, \;\; P_i \in \{\mathbb{I}, \sigma_1, \sigma_2, \sigma_3\},
\end{equation} 
which are the elements of $G_n$ when deleting the prefactors. From these strings we can introduce the \textit{Pauli spectrum} which is the list of the expectation values of all the Pauli strings:
\begin{equation}
    \text{spec}(\ket{\psi}) \equiv \{ \bra{\psi} P \ket{\psi}, P \in \mathcal{P}_n \}.
\end{equation}
Finally we can introduce the \textit{stabiliser R\'enyi entropies} (SREs) which are a family of functions $\{M_q\}$ defined by:
\begin{align}
    M_q \equiv \frac{1}{1 - q} \log_2(\zeta_q), && \zeta_q \equiv \sum_{P \in \mathcal{P}_n} \frac{\bra{\psi} P \ket{\psi}^{2q}}{2^n},
\end{align} 
where $q \geq 2$. The set of values $\{M_q\}$ measures the spread of the Pauli spectrum for general states, viewed as a distribution.

Note that for a stabiliser state $\ket{\psi}$, there are $2^n$ non-zero elements in $\text{spec}(\ket{\psi})$ and the non-zero elements are $\pm1$. This means that for a stabiliser state, $\zeta_q =1$ and so $M_q = 0$, which makes of $M_q$ a correct measure of magic. Specifically for a system composed of a pair of quarks, the measure of magic used is $M_2$ also known as the \textit{Second Stabiliser R\'enyi entropy}.

Since we are working in the density matrix framework, we know that:
\begin{equation}
    \bra{\psi} P \ket{\psi} = \Tr[P \rho],
\end{equation}
so that for a pure state:
\begin{equation}
    M_2(\rho) = -\log_2 \left( \frac{\sum_{P \in \mathcal{P}_2} \Tr^4[P\rho]}{2^n} \right),
\end{equation}
which can be generalised for a mixed state:
\begin{equation}
    \tilde{M}_2(\rho) = -\log_2 \left( \frac{\sum_{P \in \mathcal{P}_2} \Tr^4[P\rho]}{\sum_{P \in \mathcal{P}_2} \Tr^2[P\rho]} \right),
    \label{eq: formula magic 2 qubits}
\end{equation}
where we took $n=2$ because we are working on a system with 2 qubits. The Pauli strings for 2 qubits are:
\begin{equation}
    \mathcal{P}_2 = \{\mathbb{I}_2 \otimes \mathbb{I}_2, \mathbb{I}_2 \otimes \sigma_i, \sigma_j \otimes \mathbb{I}_2, \sigma_i \otimes \sigma_j \;|\; i,j =1,2,3  \},
\end{equation}
so there is a total of $1 + 3 + 3 + 9 = 16$ elements.

\paragraph*{Meaning:} The formula (\ref{eq: formula magic 2 qubits}) allows computing a measure of magic for a system made of two qubits. Magic quantifies the distance of the quantum state studied and a stabiliser state or the ``non-stabiliserness''. These stabiliser states do not have a clear meaning for particle physics yet, but in the field of quantum computing, the Gottesman-Knill theorem shows that zero magic states (even when maximally entangled) do not bring computational advantages compared to classical bits. Magic thus does not quantify entanglement like all the previous observables but another property that quantum state can have and has not been very common for high-energy physics studies yet. Specifically, it is common for the magic to be zero in the region with maximum entanglement. A more in depth explanation of the concept of magic in collider settings and in quantum computing can be found in \cite{white2024}.

Note also that magic for a pair of qubit is not bound by $1$ like concurrence, the bounds are \cite{Leone_2022}
\begin{equation}
    0 \leq M_2(\rho) \leq \log_2(2^n + 1 ) - \log_2 2 =_{n=2} 1.32.
\end{equation}

\paragraph*{Example:} Let us start with a density matrix of the form (general density matrix for a pair of qubits)
\begin{equation}
    \rho^I = \frac{1}{4} \left( \mathbb{I}_4 + \sum_{i = 1}^3 B_i^{I +} (\sigma_i \otimes \mathbb{I}_2) + \sum_{j = 1}^3 B_j^{I -} (\mathbb{I}_2 \otimes \sigma_j) + \sum_{i,j = 1}^3 C_{ij} (\sigma_i \otimes \sigma_j) \right),
\end{equation}
with 
\begin{align}
    B_i^{I \pm} = \frac{\tilde{B}_i^{I \pm}}{\tilde{A}^I}, && C_{ij}^{I} = \frac{\tilde{C}_{ij}^{I}}{\tilde{A}^I}.
\end{align}
Also, as written earlier, the Pauli strings are
\begin{equation}
    \mathcal{P}_2 = \{\mathbb{I}_2 \otimes \mathbb{I}_2, \mathbb{I}_2 \otimes \sigma_i, \sigma_j \otimes \mathbb{I}_2, \sigma_i \otimes \sigma_j \;|\; i,j =1,2,3  \}.
\end{equation}
We want to calculate 
\begin{align}
    \sum_{P \in \mathcal{P}_2} \Tr^4[P\rho] = \Tr^4[\rho] + \sum_{i=1}^3 \Tr^4[(\sigma_i \otimes \mathbb{I}_2)\rho] + \sum_{j = 1}^3 \Tr^4[(\mathbb{I}_2 \otimes \sigma_j)\rho] + \sum_{i,j = 1}^3 \Tr^4[(\sigma_i\otimes \sigma_j)\rho],
\end{align}

and
\begin{align}
    &\Tr[\rho] = 1 \\
    &\Tr[(\sigma_i \otimes \mathbb{I}_2)\rho^I] = B_i^{I+} \\
    &\Tr[(\mathbb{I}_2 \otimes \sigma_j)\rho^I] = B_j^{I-}\\
    &\Tr[(\sigma_i \otimes \sigma_j)\rho] = C_{ij}
\end{align}
so
\begin{align}
     \sum_{P \in \mathcal{P}_2} \Tr^4[P\rho] &=  1+\sum_{i = 1}^3 \left((B_i^{I+})^4 + (B_i^{I-})^4\right) + \sum_{i,j = 1}^3 (C^I_{ij})^4\\
     &= \frac{1}{(\tilde{A}^I)^4}\left[(\tilde{A}^I)^4 + \sum_{i = 1}^3 \left(( \tilde{B}_i^{I+})^4 + (\tilde{B}_i^{I-})^4\right) + \sum_{i,j = 1}^3 (\tilde{C}^I_{ij})^4\right],
\end{align}
and the same way,
\begin{equation}
    \sum_{P \in \mathcal{P}_2} \Tr^2[P\rho] = \frac{1}{(\tilde{A}^I)^2}\left[(\tilde{A}^I)^2 + \sum_{i = 1}^3 \left(( \tilde{B}_i^{I+})^2 + (\tilde{B}_i^{I-})^2\right) + \sum_{i,j = 1}^3 (\tilde{C}^I_{ij})^2\right],
\end{equation}

So that
\begin{equation}
    \tilde{M}_2(\rho^I) = -\log_2 \left( \frac{(\tilde{A}^I)^4 + \sum_{i = 1}^3 \left(( \tilde{B}_i^{I+})^4 + (\tilde{B}_i^{I-})^4\right) + \sum_{i,j = 1}^3 (\tilde{C}^I_{ij})^4}{(\tilde{A}^I)^2\left[(\tilde{A}^I)^2 + \sum_{i = 1}^3 \left(( \tilde{B}_i^{I+})^2 + (\tilde{B}_i^{I-})^2\right) + \sum_{i,j = 1}^3 (\tilde{C}^I_{ij})^2\right]}\right)\,. 
\end{equation}

\subsection{Mana}
\label{app:mana}
One limitation of the observables discussed so far is that they are primarily suited for systems composed of qubit pairs. To probe more complex systems, such as pairs of qutrits or, more generally, qudits, additional observables become relevant. In this context, we introduce the concept of ``mana'', which provides an operational approximation to magic for particles with higher odd-dimensional spin representations~\cite{Jain_2020}. We first define mana for a single qudit and subsequently generalise it to bipartite systems composed of qudits of identical or different dimensions.

\paragraph*{Systems available:} any system made of qu-d-its where $d$ is odd and prime. For particle physics it means mainly qutrit/qutrit systems and even ququint/ququint systems for BSM studies.

\paragraph*{Mana for a single qu-d-it:}
The \textbf{mana} of a state $\rho$ is 
\begin{equation}
    \mathcal{M}(\rho) \equiv \log_2 \left( \sum_{p, q}| W_{(p, q)} (\rho)| \right),
\end{equation}
where $W$ is the \textbf{discrete Wigner function} defined as
\begin{equation}
    W_{(p, q)} (\rho) = \frac{1}{d} \Tr[\rho A_{(p, q)}].
\end{equation}

The quantities $A_{(p, q)}$ are called the \textbf{discrete phase point operators} and are defined as
\begin{align}
    &A_{(p, q)} \equiv D_{(p, q)} A_{(0, 0)} D^\dagger_{(p, q)}, \\
    &A_{(0, 0)} \equiv \frac{1}{d} \sum_{u = 0}^{d - 1} \sum_{v = 0}^{d - 1} D_{(u, v)}.
    \label{phase point operators}
\end{align}

The discrete phase point operators are normalised by
\begin{equation}
    \Tr[A_u A_v] = d \delta(u,v).
\end{equation}

The operators $D_{(u, v)}$ are called the \textbf{Heisenberg-Weyl displacement operators}, they are conventionally defined for one qu-d-it of odd (and prime) dimension as
\begin{equation}
    D_{(u, v)} \equiv \omega_d^{uv2^{-1}} X^u Z^v,
    \label{Displacement operator}
\end{equation}
where $\omega_d$ are the d-root of the unity, there are $d^2$ such linearly-independent operators parametrised by $(u, v)$ with $u, v \in \mathbb{Z}_d$. Be careful that $2^{-1}$ is defined on $\mathbb{Z}_d$ so in fact $2^{-1} = (d + 1)/2$

The two operators $X$ and $Z$ are defined via their effect on the Fock states:
\begin{align}
    &X \ket{n} = \ket{n + 1} \\& Z\ket{n} = \omega_d^n \ket{n} \\
    &Z X = \omega_d X Z
\end{align}

The operator $X$ is called the \textbf{shift operator} and the $Z$ operator is called the \textbf{clock operator}, they are defined as
\begin{align}
    Z = \begin{pmatrix} 1 & 0 & 0 & \cdots & 0 \\
                        0 & \omega_d & 0 & \cdots & 0 \\
                        0 & 0 & \omega_d^2 & \cdots & 0 \\
                        \vdots & \vdots & \vdots & \ddots & \vdots \\
                        0 & 0 & 0 & \cdots & \omega_d^{d-1}
        \end{pmatrix} && X = \begin{pmatrix} 0 & 0 & 0 & \cdots & 0 &  1 \\
                        1 &0 & 0 & \cdots & 0 & 0\\
                        0 & 1 & 0 & \cdots & 0 & 0\\
                        0 & 0 & 1 & \cdots & 0 & 0 \\
                        \vdots & \vdots & \vdots & \ddots & \vdots & \vdots\\
                        0 & 0 & 0 & \cdots & 1 & 0   
        \end{pmatrix}
\end{align}

With these quantities, we have the information we need to compute the \textbf{mana} for a qu-d-it of odd (and prime) dimension, we now need to explain how to compute it for a pair of particles.

\paragraph*{Mana for a system of n qu-d-its:}

To generalise mana to multi-particles systems, we need to generalise the Heisenberg-Weyl displacement operators via tensor products as well as the discrete Wigner function) \cite{Prakash_2020}.

The equation (\ref{Displacement operator}) is the definition of a Heisenberg-Weyl displacement operator for a single qudit, where $(u,v)$ is used to parametrise the $d^2$ linearly-independent operators. To generalise these operators to $n$ qudits, we simply take the tensor products of them
\begin{equation}
    D_{(\Vec{u}|\Vec{v})} \equiv  D_{(u_1 | v_1)} \otimes D_{(u_2 | v_2)} \otimes \cdots \otimes D_{(u_n | v_n)},
\end{equation}
with $\Vec{u} = (u_1, u_2, \cdots, u_n)$, $\Vec{v} = (v_1, v_2, \cdots, v_n)$.

One issue with the Heisenberg-Weyl displacement operators is that they are not Hermitian (but they are unitary). An Hermitian basis for a single qudit density matrix is formed by the phase point operators defined in equation (\ref{phase point operators}). The generalisation comes naturally:
\begin{align}
    &A_{(\Vec{0}|\Vec{0})} = \frac{1}{d^n} \sum_{\Vec{u}} \sum_{\Vec{v}} D_{(\Vec{u}|\Vec{v})} = \frac{1}{d} \sum_{u_1 = 0}^{d-1} \cdots \sum_{u_n = 0}^{d-1} \sum_{v_1 = 0}^{d-1} \cdots \sum_{v_n = 0}^{d-1} D_{(u_1, \cdots, u_n| v_1, \cdots, v_n)} \\
    &A_{(\Vec{u}|\Vec{v})} = D_{(\Vec{u}|\Vec{v})} A_{(\Vec{0}|\Vec{0})} D_{(\Vec{u}|\Vec{v})}^\dagger
\end{align}
With the definition of the discrete phase point operators, their normalisation is:
\begin{equation}
    \Tr\left[A_{(\vec{u}|\vec{v})} A_{(\vec{p}|\vec{q})}\right] = d^n \delta_{\vec{u}, \vec{p}} \delta_{\vec{v}, \vec{q}} 
\end{equation}

From this the definition of the discrete Wigner function follows for n-qudits:
\begin{equation}
        W_{(\vec{u}|\vec{v})} (\rho) = \frac{1}{d^n} \Tr[\rho A_{(\vec{u}|\vec{v})}],
        \label{eq: Discrete Wigner function}
\end{equation}
and so does the generalisation of \textbf{mana}:
\begin{equation}
    \mathcal{M} (\rho) \equiv \log_2 \left( \sum_{\vec{u}, \vec{v}} |W_{(\vec{u}, \vec{v})} (\rho)| \right).
\end{equation}

\paragraph*{Meaning:}
It is important to remember that this observable is only available for particles whose Hilbert space is of odd-dimension so it can not be used to study systems with qubits. Mana is a measure of the sum of negative entries in the discrete Wigner function (\ref{eq: Discrete Wigner function}), which  is similar to the Peres-Horodecki criterion in which we are interested in the negative eigenvalues of the partially transposed density matrix. Mana is designed to be a surrogate to magic which can only be calculated to qubit/qubit systems for more complex systems.  There are two states that are known to maximise mana for a single qutrit and they are called the strange state $\ket{S}$ and the Norell state $\ket{N}$ for which mana reaches $\log 5/3$.

\subsection{Negativity}

\label{app:neg}
The Peres-Horodecki criterion presented earlier is a very convenient entanglement witness, however it is only binary, meaning it only tells us whether the system is separable or not and not how entangled the states are. Negativity is a entanglement witness that attempts to generalise this criterion for all systems. 

\paragraph*{Systems available:} For all systems.
\paragraph*{Theoretical formula:} The Peres-Hororecki states that if the partial transpose of the density matrix has any negative eigenvalues, then the two states are entangled (for $2\times 2$ and $2\times 3$ systems), so if we calculate how negative these eigenvalues are, we can determine how entangled the states are. Negativity is defined as
\begin{equation}
    N(\rho) \equiv \frac{||\rho^{T_B}|| - 1}{2},
\end{equation}
where $||X|| \equiv \Tr[\sqrt{X^\dagger X}]$.

It is entanglement monotone \cite{plenio2006}, so the bigger the quantity is, the more entangled the states are. However this entanglement witness is not additive, \textit{ie.} 
\begin{equation}
    N(\rho_1 \rho_2) \neq N(\rho_1) + N(\rho_2).
\end{equation}
If one wants an additive negativity, one can use the logarithmic negativity which is additive by construction
\begin{equation}
    E_N(\rho) \equiv \log_2 \left(||\rho^{T_B}||\right),
\end{equation}
which is also entanglement monotone \cite{plenio2006}.

\paragraph*{Meaning:} The main advantage of this entanglement witness is that it is easily computable for any system as it only requires to compute the trace norm of the density matrix. If the states are separable, then negativity $N$ vanishes \cite{Vidal:2002zz}. The maximum value for negativity is reached by fully entangled states, for $2\times 2$ systems Bell states are fully entangled, their negativity is $1/2$ and is the maximum value, the maximum value of logarithmic negativity is then $1$.

\subsection{Trace distance and fidelity}
\label{app:dist}
All the observables presented earlier consider a process calculated in a specific model and allow to determine whether the system considered is entangled. Another type of study that can be done is comparing the density matrix calculated by two different models, for instance between the SM and any BSM model, and to calculate a ``distance'' between the two. This allows to quantify any deviation from the SM through the density matrix. It can also be used on two processes that are related to determine how ``far'' they are from each other. Two of these observables are available in the code currently: trace distance and fidelity.
\paragraph*{Systems available:} For all systems.
\paragraph*{Theoretical formula:} 
For any two density matrices representing the same process, one can define the trace distance as
\begin{equation}
    D^T (\rho_1, \rho_2) \equiv \frac{1}{2} \Tr \sqrt{(\rho_1 - \rho_2)^\dagger (\rho_1 - \rho_2)} \geq 0.
\end{equation}
And one can define the fidelity as 
\begin{equation}
    F(\rho_1, \rho_2) \equiv \Tr \sqrt{\sqrt{\rho_1} \rho_2 \sqrt{\rho_1}},
\end{equation}
where $0 \leq F(\rho_1, \rho_2) \leq 1$.

To get a distance from the fidelity, one can define
\begin{equation}
    D^F(\rho_1, \rho_2) \equiv \sqrt{1 - F(\rho_1, \rho_2)^2} \geq 0,
\end{equation}
called the fidelity distance.

\paragraph*{Meaning:}
The trace distance is invariant under unitary transformations and is zero if the two density matrices commute (they represent the same physics) and does not have a upper bound. Fidelity quantify the similarity between the two different density matrices, its upper bound is $1$ and is reached only if both matrices are identical and its lower bound is $0$ and is reached only if both matrices have orthogonal supports. Note that $F(\rho_1, \rho_2) = F(\rho_2, \rho_1)$ even if that is not manifest \cite{Fabbrichesi:2025ywl}.

\subsection{Other forms of entanglement}
\noindent Concurrence is not the only quantifier of quantum correlations; different observables capture stronger or weaker forms of non-classicality. In particular, {Einstein--Podolsky--Rosen (EPR) steering} is strictly stronger than entanglement as quantified by concurrence: any steerable state is entangled, whereas some entangled states are unsteerable (e.g.\ certain Werner states). By contrast, {quantum discord} is weaker than entanglement: it can be non-zero even for separable states, revealing non-classical correlations beyond entanglement. In summary, these notions obey the hierarchy,
\begin{equation}
\text{steering}\ \Rightarrow\ \text{entanglement (non-zero concurrence)}\ \Rightarrow\ \text{non-zero discord},
\end{equation}
where none of the converse implications holds in general. These measures (and related ones) are not yet implemented in the code.

\section{Manual}
\label{app:code}
\par

In {\sc MG5aMC}, two new modes of operation have been added for the computation of QI observables. The first mode is based on the standalone output of \mg\ and computes the production matrix $R$ for a single phase-space point. This interface is useful for debugging and for interfacing with external tools or programs. The second mode is based on the reweighting module of \mg\cite{Mattelaer:2016gcx} and performs the computation of the density matrix point by point in phase space over the events of a LHE file, as well as the associated analysis.\footnote{The new functionalities are planned to be released in version \mg~\texttt{3.8.0} and, in the meantime, can be obtained by contacting the authors.}  

\subsection{Standalone mode}

In this section, we review the new option that allows the user to compute the production matrix $R$ for a single phase-space point.
The following snippet generates (and runs) an example code that first creates a sample phase-space point (using \textsc{Rambo}~\cite{Kleiss:164736}), 
then evaluates the squared amplitudes, and finally computes the density matrix for the particles requested.

\begin{mdframed}
   \begin{addmargin}[0cm]{0cm}
    \texttt{generate p p > a t t\textasciitilde}\\
    \texttt{output standalone -\,-density=4,5}\\
    \texttt{launch}       
   \end{addmargin} 
\end{mdframed}

In this example, the production matrix computed will correspond to the top and anti-top particles, since those correspond to the particle numbered 4 and 5.
We do not offer, in this mode, any easy way to change the reference frame, axis of quantisation, etc. But,  experienced user in Fortran should be able 
to easily implement such options as well as interfacing such computation with external tool/library. 

\subsection{Reweighting mode}

In addition to the standalone mode, we provide a fully integrated implementation within the \mg framework,  which allows, in a single run, the generation of events, the computation of the corresponding density matrices, and the evaluation of quantum observables.

The first step is to create a folder for the process to study, this is classically done via the standard \mg command:

\begin{mdframed}
\begin{center}
\begin{addmargin}[0cm]{0cm}
    \texttt{generate YOUR\_PROCESS}\\
    \texttt{output}\\
    \texttt{launch}
 \end{addmargin}
\end{center}
\end{mdframed}

The code then prompts the user to select the type of run to be performed. 
Since the computation of the density matrix is carried out at the parton level, 
the inclusion or omission of parton-shower and detector simulations does not affect 
the quantum-information observables. 

\newpage

\begin{mdframed}
\begin{verbatim}
The following switches determine which programs are run:
/=========================================================================\
| 1. Choose the shower/hadronisation program     shower = OFF             |
| 2. Choose the detector simulation program    detector = Not Avail.      |
| 3. Choose an analysis package (plot/convert) analysis = MadAnalysis5    |
| 4. Decay onshell particles                    madspin = OFF             |
| 5. Add weights to events for new hypp.       reweight = OFF             |
\=========================================================================/
\end{verbatim}
\end{mdframed}

Here, a new option is available for the reweight option called ``density'', one can set it by typing:
\begin{mdframed}
\begin{addmargin}[0cm]{0cm}
    \texttt{reweight = density}
\end{addmargin}
\end{mdframed}

The code then offers the possibility to edit the various configuration files controlling event generation. 
The user specifies the desired settings for the density-matrix construction directly in the \texttt{reweight\_card.dat}. 
All particle selections are provided as lists of PDG identifiers; when multiple particles share the same PDG code, 
their instances are taken according to the order established during process generation, 
that is, following the internal particle ordering of the generated subprocess. The full list of supported parameters is provided below:

\begin{itemize}
\item \texttt{boost\_choice [list[int], str, list[int]]}:
Selects the Lorentz frame by specifying which four-momenta to use for the event-level boost. The option accepts a triple
\[
\texttt{([{\rm PDG\ codes}],\ observable\_name,\ [\text{indices}])},
\]
interpreted as follows:
\begin{enumerate}
  \item {Momentum selection} (\texttt{list[int]}): list of PDG codes whose four-momenta are summed to define the boost vector. If the list contains several entries, the total momentum $P^\mu=\sum_k p_k^\mu$ is used. This allows, e.g.\ boosting to the c.m.\ frame of a pair or a subsystem.
  \item {Disambiguation key} (\texttt{str}, optional): when multiple particles share the same PDG code, they are ranked in {descending} order of the specified observable, as implemented in \texttt{FourMomentum} (see \texttt{lhe\_parser.py}). Valid keys include any attribute/method exposed by that class (e.g.\ \texttt{"pt"}, \texttt{"eta"}, \texttt{"E"}, \texttt{"mass"}, \texttt{"rapidity"}).
  \item {Index filter} (\texttt{list[int]}, optional): zero-based indices selecting which instances (within the ranked list) to include. Here \texttt{0} denotes the particle with the highest value of the chosen observable, \texttt{1} the next, etc. If omitted or empty, all matching particles are considered in the order given by the ranking.
\end{enumerate}
To {skip} any boost and keep the lab frame, set the first element to \texttt{-1}. The default is \texttt{([-1], "", [])}.

\noindent\textbf{Examples:}
\begin{itemize}
  \item \texttt{boost\_choice = ([6, -6], "", [])}: boost to the $t\bar t$ pair rest frame using the sum of the top and antitop four-momenta; no disambiguation needed.
  \item \texttt{boost\_choice = ([23], "pt", [0])}: for multiple $Z$ bosons, pick the {hardest} $Z$ (\texttt{pt}-ranked, index \texttt{0}) and boost to its rest frame.
  \item \texttt{boost\_choice = ([5, -5], "E", [0,1])}: if several $b$ and $\bar b$ are present, select the two most energetic $b$-flavoured jets by energy and boost to the rest frame of their sum.
  \item \texttt{boost\_choice = ([-1], "", [])}: no boost applied (use lab frame).
\end{itemize}

    \item \texttt{helicity\_direction [list[int], str, list[int]]}: this options allows the user to specify if and how the system should be rotated. The first option is a list of PDG codes that allow the user to specify which particle momenta should be used as reference for the rotation into the helicity as  defined in Appendix \ref{app:hel}. If the list contains several elements, the sum of their momenta is used as reference. The two other arguments are optional and serve the exact same purpose as in the boost case, please refer to the previous point for these options. The value \texttt{-1} allows specifying that no rotation should be done. \textbf{Caution:} because of the way the helicity frame is defined, even with $(\varphi, \theta) = (0,0)$, the components of the 4-momenta will change, this is done to keep the same convention used in Appendix \ref{app:hel}. Indeed, $(p_x, p_y, p_y) \to_{\varphi =0, \theta = 0} (p_y, -p_x, p_z)$ in the helicity frame. The default value for this option is: \texttt{([-1], '', [])}.

    \item \texttt{particle\_in\_density\_matrix [list[int], str, list[int]]}: this option is the only one that is required for the computation of the density matrix of each event because it specifies which particles are to be put in the density matrix. The first option is a list of PDG codes that allows the user to specify which particle should be put in the density matrix and in which order, for now the code only allows selecting up to 2 particles for this option. The two other arguments are optional and serve the exact same purpose as in the boost case, please refer to the previous point for these options. There is no default value for this option, as it is mandatory to complete it for the density mode to function.

    \item \texttt{order\_helicities [list[int]]}: this option allows the user to specify in which helicity basis the density matrix should be written. The first and only argument for this option is a list of integers contained in $\{-1, 0, +1\}$ that are the helicity values each of the particles can take. For instance, if we consider a system made of two fermions and that we want the density matrix to be written in the basis $\{\ket{--}, \ket{-+}, \ket{+-}, \ket{++}\}$, the option should be \texttt{[-1, -1, -1, +1, +1, -1, +1, +1]}. The default value of this option depends on the system considered but are always ordered with the helicities \texttt{+1} in first, then the helicities \texttt{0} and finally the helicities \texttt{-1}.

    \item \texttt{axis\_referential [list[int]]}: as explained in Appendix \ref{app:hel}, the axis used as reference for the definition of the angle $\theta$ is the $+\hat z$ axis. This option allows the user to change this reference by using one of the initial particle direction as reference, the input is thus a list of PDG codes of the selected particles. For instance if the user chooses \texttt{[-1]}, the angle $\theta$ will be using the anti-down quark of the initial state as reference for $\theta$ if it exists. The default value is \texttt{[-1]}
    \item \texttt{symmetrise\_initial\_state [bool}: if this option is set to \texttt{True}, the production matrix event will be calculated as the sum of the production matrix calculated using $\theta$ (using $+ \hat z$ as reference) and using $\theta  + \pi$ (using $- \hat z$) as reference. This option is used in some analysis (for instance \cite{Aoude:2023hxv}) and can be interesting if the initial state is not symmetric. The default value is \texttt{False}.
\end{itemize}
Note also that the user can modify the \texttt{reweight\_card} by using inline commands in \mg using the two following syntaxes, in the same way the user would do it for other cards.:

\begin{mdframed}
\begin{addmargin}[3cm]{0cm}
    \texttt{set reweight\_card input\_name input\_value} \\ 
    \texttt{set input\_name input\_value}
\end{addmargin}
\end{mdframed}

After the computation is completed, the density matrix for each event is stored in the 
\texttt{.lhe} file, and the average density matrix is also printed. 
The script \texttt{lhe\_parser.py}, together with the set of functions provided in 
\texttt{Density\_functions.py}, allows one to extract the data from the \texttt{.lhe} file 
and compute various quantum-information observables, such as the purity or the concurrence. 
A detailed technical description of this {\tt Python} library can be found in the on-line documentation, available here: \url{https://cp3.irmp.ucl.ac.be/projects/madgraph/wiki/Density}.

\subsection{Examples}
\label{app:ex}
We now list a few examples of inputs that can be put in the \texttt{reweight\_card} to study a variety of processes. 
If one aims to study spin correlations between bosons, for instance in the process 
$e^+ \, e^- \to W^+ \, W^-$, the \texttt{reweight\_card} can be set up as follows:

\begin{mdframed}
\begin{addmargin}[0cm]{0cm}
        \texttt{change boost\_choice [24, -24] \\
            change helicity\_direction [24] \\
            change particle\_in\_density\_matrix [24, -24] \\
            change order\_helicities [+1, -1, +1, 0, +1, +1, 0, -1, 0, 0, 0, +1, -1, -1, -1, 0, -1, +1]}
\end{addmargin}
\end{mdframed}

In this way, the density matrix is computed in the centre-of-mass frame of the $W$-boson pair, taking the $W^+$ as the reference for defining the helicity frame. The matrix is expressed in the basis $\{\ket{+-}, \ket{+0}, \ket{++}, \ket{0-}, \ket{00}, \ket{0+}, \ket{--}, \ket{-0}, \ket{-+}\}$.

The next example is chosen to show how to use an observable to distinguish an identical particle, for instance the process $p \, p \to Z \, Z\, Z$ contains three identical particles in the final states. If we want to study the spin correlations between two of the $Z$ bosons, one  can not just use the first two $Z$ appearing within the event record. The following syntax allows identifying which pair of $Z$ to include by ordering them by transverse momentum:

\begin{mdframed}
\begin{addmargin}[0cm]{0cm}
        \texttt{change particle\_in\_density\_matrix [23, 23] pt [0, 1]\\
            change boost\_choice [23, 23] pt [0, 1]\\
            change helicity\_direction [23] pt [2]\\
            change order\_helicities [+1, -1, +1, 0, +1, +1, 0, -1, 0, 0, 0, +1, -1, -1, -1, 0, -1, +1]}
\end{addmargin}
\end{mdframed}

With these settings, the particles included in the density matrix are, for each event, 
the two $Z$ bosons with the highest transverse momentum. 
The reference frame is then boosted to the centre-of-mass frame of these two particles, 
while the helicity reference direction is defined by the $Z$ boson with the smallest 
transverse momentum, i.e.\ the one not included in the density matrix. 
Although this particular configuration may not be of direct physical interest, 
it serves to illustrate some of the flexibility and capabilities offered by the code.

Lastly, if one wishes to study spin correlations between a massive boson and a massive fermion, 
for instance in the process $p \, p \to t\, \bar t\, W^+$, the following parameters can be used:
\begin{mdframed}
\begin{addmargin}[1cm]{0cm}
         \texttt{change particle\_in\_density\_matrix [6, 24] \\
            change boost\_choice [6, 24] \\
            change helicity\_direction [6] \\
            change order\_helicities [1, -1, 1, 0, 1, 1, -1, -1, -1, 0, -1, 1]}
\end{addmargin}
\end{mdframed}    
With these settings, the particles included in the density matrix are the top quark and the $W^+$ boson,  and the matrix is computed in the centre-of-mass frame of this pair. 
The top quark is chosen as the reference particle for the helicity frame.

\section{Conventions}
\label{app:conventions}
\subsection{Definition of the helicity frame}
\label{app:hel}
The density matrix is an object that strongly depends on various conventions 
(such as the choice of reference frame, spinor definitions, etc.). 
The choice of reference frame is left to the user. 
In this appendix, we describe the basis adopted in the code for the computation of the density matrix, 
which corresponds to the conventionally used one.

In \mg  the default reference frame is the laboratory frame 
$\{ \hat{x}, \hat{y}, \hat{z} \}$, where $\hat{z}$ is the beam axis, 
and the first incoming particle is defined to have $p_z > 0$, 
while $\{ \hat{x}, \hat{y} \}$ span the plane perpendicular to $\hat{z}$. 
However, in most studies analysing high-energy processes from a quantum-information perspective, 
the reference frame employed is the helicity frame 
$\{ \hat{n}, \hat{r}, \hat{k} \}$. 
Figures~\ref{fig:spatial_basis} and~\ref{fig:definition_phi} illustrate this basis 
for the case of top-quark pair production.
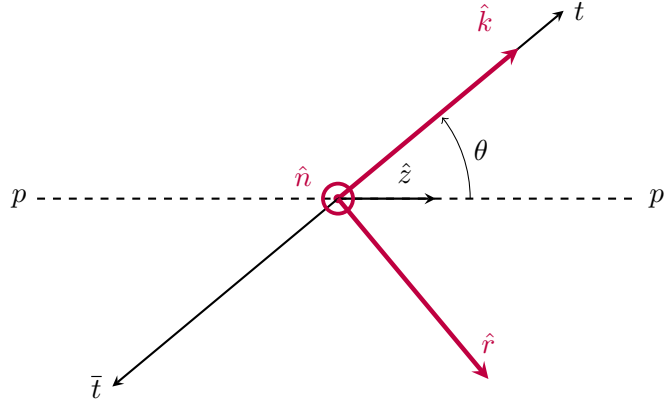
\begin{figure}[!htb]
    \centering
    \begin{tikzpicture}
    \coordinate (origo) at (0,0);
        \fill[purple] (origo) circle (0.06);
        \draw [dashed, thick] (-4,0) node[left] {$p$} -- (4,0) node (mary) [right] {$p$};
        \draw [stealth-stealth, solid, thick] (-3,-2.5) node[left] {$\overline{t}$} -- (3,2.5) node (bob) [right] {$t$};
        \draw [-stealth, solid, line width = 0.6mm, purple] (0, 0) -- (2.4, 2) node [label={[label distance=0.3mm]160:$\hat{k}$}] {} ;
         \draw [-stealth, solid, line width = 0.3mm, black] (0, 0) -- (1.3, 0) node [label={[label distance=0.3mm]160:$\hat{z}$}] {} ;
        \draw [-stealth, solid, line width = 0.6mm, purple] (0, 0) -- (2, -2.4) node [label={[label distance=0.3mm]90:$\hat{r}$}] {} ;
        \path[draw=purple, line width = 0.5 mm] (0, 0) coordinate circle[radius=2mm];
        \node (origo) [label={[purple, label distance=0.8mm]170:$\hat{n}$}] {};
        \pic [draw, ->, angle radius = 50] {angle = mary--origo--bob};
        \node[] at (1.9,0.65) {$\theta$};
    \end{tikzpicture}
\caption{Spin-helicity basis for a $2\to 2$ process. The only boost needed to go to the ZMF of the $t\bar t$ pair is along the $\hat z$ direction.}
    \label{fig:spatial_basis}
\end{figure}
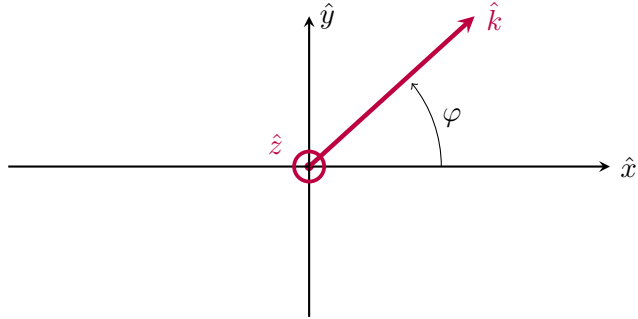
\begin{figure}[!htb]
    \centering
    \begin{tikzpicture}
    \coordinate (origo) at (0,0);
        \fill[purple] (origo) circle (0.06);
        \draw [-stealth, solid, thick] (-4,0)  -- (4,0) node (mary) [right] {$\hat{x}$};
        \draw [-stealth, solid, line width = 0.6mm, purple] (0, 0) -- (2.2,2) node (bob) [right] {$\hat{k}$};
        \draw [-stealth, solid, thick] (0, -2) -- (0, 2) node [right] {$\hat{y}$};
        \pic [draw, ->, angle radius = 50] {angle = mary--origo--bob};
        \path[draw=purple, line width = 0.5 mm] (0, 0) coordinate circle[radius=2mm];
        \node (origo) [label={[purple, label distance=0.8mm]170:$\hat{z}$}] {};
        \node[] at (1.9,0.65) {$\varphi$};
    \end{tikzpicture}
    \caption{Definition of the angle $\varphi$ to put $\hat{k}$ the plan $(O, \hat{x}, \hat{y})$.}
    \label{fig:definition_phi}
\end{figure}
This new basis is defined as:
\begin{align*}
    &\hat{k} \;\;\; \textrm{the direction of fly of the chosen particle} \\
    &\hat{r} = \frac{\hat{z} - \cos\theta \,\hat{k}}{\sin\theta} \\
    &\hat{n} =  \hat{r} \times \hat{k}
\end{align*}
\noindent One can show that with these definitions, the new vector basis are:
\begin{align*}
    \hat{k} &= \cos\varphi\sin\theta\, \hat{x} +\sin\varphi\sin\theta \,\hat{y} + \cos\theta \,\hat{z},\\
    \hat{r} &= -\cos\theta\cos\varphi\,\hat{x} - \cos\theta\sin\varphi\,\hat{y} + \sin\theta\,\hat{z},\\
    \hat{n} &= -\sin\varphi\,\hat{x} + \cos\varphi\,\hat{y}.
\end{align*}
And the momenta in this reference frame immediately are:
\begin{align}
    p_{\hat{n}} &= -\sin\varphi \, p_{\hat{x}} + \cos\varphi \, p_{\hat{y}},\\
    p_{\hat{r}} &= -\cos\varphi \cos\theta \,p_{\hat{x}} - \sin\varphi\cos\theta \,p_{\hat{y}} + \sin\theta p_{\hat{z}},\\
    p_{\hat{k}} &= \cos\varphi \sin\theta \,p_{\hat{x}} + \sin\varphi \sin\theta \,p_{\hat{y}} + \cos\theta \,p_{\hat{z}}.
\end{align}
The angles $\theta, \varphi$ are defined as:
\begin{align}
    \theta &= \arccos\frac{p_{\hat{z}}}{\sqrt{p_{\hat{x}}^2 + p_{\hat{y}}^2 + p_{\hat{z}}^2}}\,,\\
    \varphi &= \text{sign}(p_{\hat{y}})\arccos\frac{p_{\hat{x}}}{\sqrt{p_{\hat{x}}^2+p_{\hat{y}}^2}}\,. 
\end{align}
Note the $\sign(p_{\hat{y}})$ factor in the definition of $\varphi$, it ensures that $\hat{k}$ is defined in the direction of the particle and not in the opposite direction, as can happen depending on the signs of the components of $\Vec{p}$.
\par
Note that at LO for $2 \to 2$ processes, the centre-of-mass frame of the final particles is the same as the one for the initial particles, this means that the boost is only in the $\hat z$.
\par
The helicity frame presented above is the one implemented in \mg and conventionally used for quantum spin correlation studies for $2 \to 2$ processes. However, for processes with $3$ or more particles in the final state, the definition of $\theta$ becomes convention dependent. In all conventions is assumed that the centre-of-mass frame of the two particles is reached from the lab frame via a pure boost transformation (no rotations). This is called the Zero-Momentum-Frame (ZMF). Now, the $\theta$ angle can be defined in various ways, of which we consider two:
\begin{itemize}
    \item The angle $\theta$ is defined between the direction of the chosen particle (e.g.\ the top) in the ZMF and the direction of the beam axis boosted in the same frame $\vec p_1$:
    \begin{equation}
    \cos\theta \equiv \frac{\vec p_{\rm top} \cdot \vec p_1}{||\vec p_{\rm top}|| \cdot||\vec p_1||}\,. 
    \end{equation}
    \item The angle $\theta$ is defined between the direction of the chosen particle (e.g.\ the top) in the ZMF and the beam axis in the lab frame:
    \begin{equation}
    \cos\theta \equiv \frac{\vec p_{\rm top} \cdot \hat z}{||\vec p_{\rm top}||},
    \label{eq: theta1}
\end{equation}
with $\hat z = (0, 0, 1)$,
\end{itemize}
\par
In this paper we have chosen the second definition, i.e.\ to use $\hat z$ as reference for the definition of $\theta$. We have compared the values of the observables across the different processes studied for each choice of definition. While differences between the cases are visible, they do not alter the overall behaviour of the observables and therefore do not affect the interpretation of the results. 

\subsection{Conventions used for the external polarisations}
\label{app:pol}

The density matrix is an object whose specific form and structure depend strongly on the adopted conventions: the definition of spin states, the reference frame, the choice of the helicity basis, etc. The code employs a specific convention for the choice of frame, described in Appendix~\ref{app:hel}, where the user can select any external particle of the process to define the angles $\theta$ and $\varphi$. 
The spinor convention used in \mg are based on the {\sc HELAS} library, you can find them in Appendix A of Ref.~\cite{aMurayama}. The reader can note that these spinors are defined as the helicity eigenstates, meaning that the spin of the fermion is projected on the direction of its momenta. The spin quantisation axis is thus chosen by the rotation applied to the system \textit{via} the option \texttt{helicity\_direction}.

\subsection{\texttt{FeynCalc} conventions}
\label{app:diff}

\begin{align}
    \label{eq:basis_qubit_transformation}
    &\{\ket{+-}, \ket{++}, \ket{--}, \ket{-+}\},\\
    &\{\ket{+-}, \ket{+0}, \ket{++}, \ket{0-}, \ket{00}, \ket{0-}, \ket{--}, \ket{-0}, \ket{-+}\}.
    \label{eq:basis_qutrit_transformation}
\end{align} 

The difference between {\sc FeynCalc} and our code arises from distinct phase conventions. Indeed, while the moduli of all elements of the density matrix coincide, their complex phases differ. In general, one is free to multiply each component of a polarisation vector or spinor by a global phase without altering any physical observable. However, since the density matrix is not Lorentz--invariant, this phase conventions do affect its coefficients. The difference in helicity phase choices can thus be interpreted as a change of basis for the density matrix, which can be expressed through a unitary transformation $U$ such that
\begin{equation}
    \rho^\prime = U^\dagger \rho U.
\end{equation}

For the qubit--qubit processes, we find that the transformation required to recover the same convention corresponds to adding a relative phase of $\pi$ between the $+$ and $-$ helicity states of the first particle in the density matrix. If $\rho$ is expressed in basis (\ref{eq:basis_qubit_transformation}), this results in the following transformation matrix:

\begin{equation}
    U_{\textrm{2qubits}} = \begin{pmatrix}
        1 & 0 & 0 & 0 \\ 0 & 1 & 0 & 0 \\ 0 & 0 & -1 & 0 \\ 0 & 0 & 0 & -1
    \end{pmatrix}.
\end{equation}

In the qutrit-qutrit case it is also needed to dephase one polarisation vector compared to the other two by a phase of $-\pi/2$. If $\rho$ is expressed in the basis (\ref{eq:basis_qutrit_transformation}), this results in the following transformation matrix
\begin{equation}
    U_{\textrm{2qutrits}} = \begin{pmatrix}
        1 & 0 & 0 & 0 & 0 & 0 & 0 & 0 & 0 \\
        0 & -i & 0 & 0 & 0 & 0 & 0 & 0 & 0 \\
        0 & 0 & 1 &0 & 0 & 0 & 0 & 0 & 0 \\
        0 & 0 & 0 & -i & 0 & 0 & 0 & 0 & 0 \\
        0 & 0 & 0 &0 & -1 & 0 & 0 & 0 & 0 \\
        0 & 0 & 0 &0 & 0 & -i & 0 & 0 & 0 \\
        0 & 0 & 0 &0 & 0 & 0 & 1 & 0 & 0 \\
        0 & 0 & 0 &0 & 0 & 0 & 0 & -i & 0 \\
        0 & 0 & 0 &0 & 0 & 0 & 0 & 0 & 1 \\
    \end{pmatrix}.
\end{equation}

\bibliographystyle{JHEP}
\bibliography{refs}

\end{document}